%% file: mocanlo_users_guide.tex
\documentclass[UKenglish,parskip=half-]{scrartcl}
\usepackage[utf8]{inputenc}
\usepackage[T1]{fontenc}
\usepackage{courier}
\usepackage{amsmath,amssymb}
\usepackage{mathtools}
\usepackage[text={6.5in,9in},centering]{geometry}

\usepackage{tabularx}
\usepackage{subcaption}
\usepackage{multicol}
\usepackage{booktabs,xspace}
\usepackage{dirtree}
\usepackage{textcomp}
\usepackage{enumitem}
\usepackage{csquotes}
\usepackage{comment}
\usepackage[sort&compress,numbers]{natbib}

\usepackage[pdftex,table,svgnames,hyperref]{xcolor}

\definecolor{darkred}{rgb}{0.5 0 0}
\definecolor{darkgreen}{rgb}{0.5 .5 0}
\definecolor{darkblue}{rgb}{0 0 .5}
\usepackage[pdftex,bookmarksopen]{hyperref}
\hypersetup{%
        pdftitle={MoCaNLO - User's Guide}, 
        pdfauthor={Ansgar Denner, Daniele Lombardi, Santiago Lopez Portillo Chavez, Mathieu Pellen, and Giovanni Pelliccioli},%
        pdfsubject={},%
        pdfcreator={pdfTeX},%
        pdfproducer={Ansgar Denner, Daniele Lombardi, Santiago Lopez Portillo Chavez, Mathieu Pellen, and Giovanni Pelliccioli},%
        pdfkeywords={},%
        colorlinks=true,%
}
\usepackage{helvet}
\usepackage{tikz}
\usepackage[tikz]{bclogo}
\usepackage[most]{tcolorbox}
\tcbset{%
  enhanced,
lower separated=true}
\usepackage{listings} 
\usepackage{bclogo}
\input{macros}

\font\manual=manfnt
\def\dbend{{\manual\char127}} 
  {\medbreak\par}

  {\medbreak\par}

\newenvironment{info}%
{\begin{bclogo}[
    epBord=2, arrondi=0.2, logo=\bcinfo, marge=8, ombre=true, blur,
    barre=snake, 
    tailleOndu=3]}
{\end{bclogo}}%

\newenvironment{warning}%
{\begin{bclogo}[
    epBord=2, arrondi=0.2, logo=\bcattention, marge=8, ombre=true, blur,
    barre=snake, tailleOndu=3]}
{\end{bclogo}}%

  \newcommand\xmltag[1]{\lstinline[language=XMLinline]|<#1>|}
  \newcommand\bash[1]{\lstinline[language=bash]|#1|}

\usepackage{quoting,xparse}

\NewDocumentCommand{\bywhom}{m}{
  {\nobreak\hfill\penalty50\hskip1em\null\nobreak
   \hfill\mbox{\normalfont(#1)}%
   \parfillskip=0pt \finalhyphendemerits=0 \par}%
}

\NewDocumentEnvironment{pquotation}{m}
  {\begin{quoting}[
     indentfirst=true,
     leftmargin=2em,
     rightmargin=2em 
      ]\itshape}
  {\bywhom{#1}\end{quoting}}

\newcommand{\Red}[1]{{\color{red}{#1}}}

\newcommand{\basename}{\texttt{run\_2025-07-29T17h38m28s928}}

\usepackage{letltxmacro}
\LetLtxMacro\oldttfamily\ttfamily
\DeclareRobustCommand{\ttfamily}{\oldttfamily\csname ttsize\endcsname}
\newcommand{\setttsize}[1]{\def\ttsize{#1}}%
\setttsize{\small}

\begin{document}

\lstset{
  basicstyle=\ttfamily\footnotesize,
  showstringspaces=false,
  xleftmargin=4ex,
  belowskip=0pt,
  commentstyle=\color{gray}\upshape
}
\lstdefinelanguage{XMLinline}
{
  basicstyle=\ttfamily\footnotesize,
  morestring=[b]",
  moredelim=[s][\color{black}]{<}{\ },
  moredelim=[s][\color{black}]{</}{>},
  moredelim=[l][\color{black}]{/>},
  moredelim=[l][\color{black}]{>},
  morecomment=[s]{<?}{?>},
  morecomment=[s]{<!--}{-->},
  commentstyle=\color{black},
  stringstyle=\color{red},
  identifierstyle=\color{black},
  keepspaces=true,
  morekeywords={runs,run,subdirectory,subprocess,run_parameters,model_parameters,recombinations,resonances,cuts,histograms}
}
\lstdefinelanguage{XML}
{
  captionpos=b,
  basicstyle=\ttfamily\scriptsize,
  morestring=[b]",
  moredelim=[s][\bfseries\color{black}]{<}{\ },
  moredelim=[s][\bfseries\color{black}]{</}{>},
  moredelim=[l][\bfseries\color{black}]{/>},
  moredelim=[l][\bfseries\color{black}]{>},
  morecomment=[s]{<?}{?>},
  morecomment=[s]{<!--}{-->},
  commentstyle=\color{black},
  stringstyle=\color{red},
  identifierstyle=\color{black},
  morekeywords={runs,run,type,subdirectory,subprocess,run_parameters,model_parameters,recombinations,resonances,cuts,histograms}
}
\lstset{escapeinside={(*@}{@*)}}

\setcounter{page}{1}

\begin{titlepage}
\def\thefootnote{\fnsymbol{footnote}}%
\setcounter{footnote}{1}%
\begin{minipage}[t]{0.55\textwidth}%
\strut\\[-3ex]
\includegraphics[width=\textwidth]{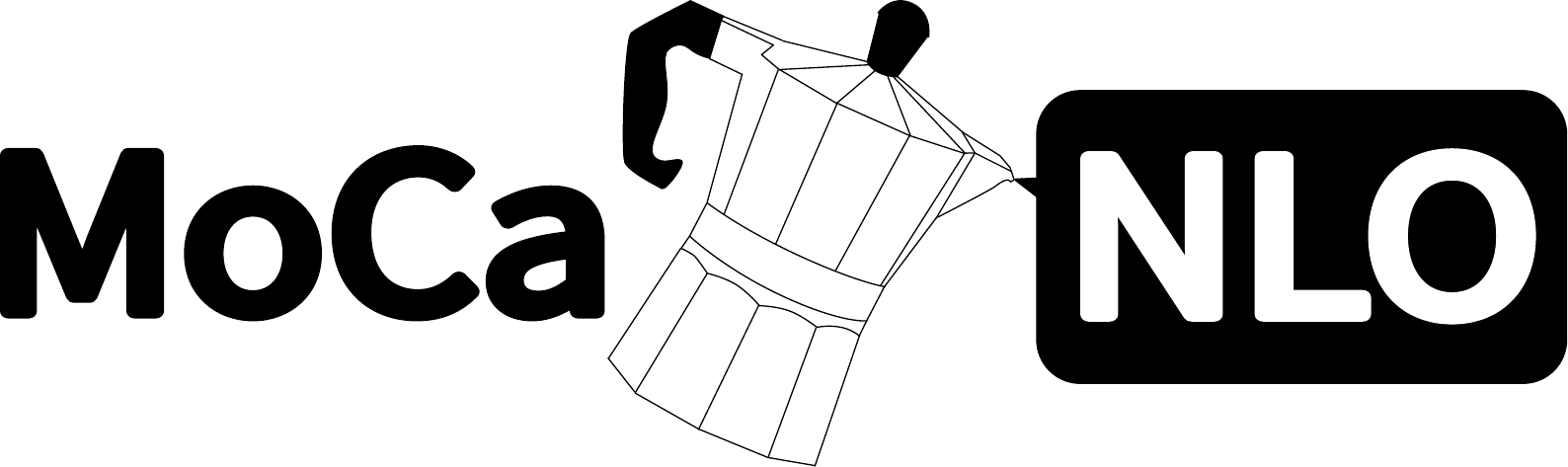}
\end{minipage}
\hfill
\begin{minipage}[t]{0.40\textwidth}
\raggedleft
COMETA-2026-03 \\
FR-PHENO-26-04\\
\end{minipage}
\vspace{1.5cm}
\begin{center}
{\Large\textbf{
MoCaNLO: a Monte Carlo integrator for NLO calculations}
\par}
\vspace{1.5cm}
{\large
{\scshape A.~Denner,$^1$ D.~Lombardi,$^2$ S.~Lopez~Portillo~Chavez,$^1$
M.~Pellen,$^3$ and G.~Pelliccioli$^4$} } \\[.5cm]
$^1$ {\itshape Institut f\"ur Theoretische Physik und Astrophysik,
  Universit\"at W\"urzburg, \\Emil-Hilb-Weg~22, 97074 W\"urzburg,
  Germany}
 \\[0.2cm]
$^2$ {\itshape
Dipartimento di Fisica, Universit\`a  di Torino and INFN, Sezione di Torino, \\
Via P. Giuria 1, 10125 Torino, Italy} 
\\[0.2cm]
$^3$ {\itshape 
Physikalisches Institut, Albert-Ludwigs-Universit\"at Freiburg, \\
Hermann-Herder-Str.~3, 79104 Freiburg, Germany}
\\[0.2cm]
$^4$ {\itshape Dipartimento di Fisica ``Giuseppe Occhialini'', Universit\`a degli Studi di Milano--Bicocca\\
  and INFN, Sezione di Milano--Bicocca,
  Piazza della Scienza 3, 20126 Milano, Italy}
\par \vskip 1em
\end{center}\par
\vfill \textbf{Abstract:}
\par
We present the Monte Carlo integration code MoCaNLO, which computes
cross sections and distributions for processes at high-energy colliders
like the LHC at leading and next-to-leading order (NLO) in the strong and
electroweak couplings. It relies on the \recola\ package for the calculation of
matrix elements and uses Catani--Seymour dipole subtraction for the
treatment of infrared singularities. It has been used for several
cutting-edge calculations of NLO QCD and electroweak
corrections over the last years, such as NLO QCD corrections to
off-shell top--antitop-quark production in association with a pair of
bottom quarks and NLO electroweak corrections to vector-boson
scattering processes.
\par
\vskip 2.5cm
\noindent
February 2026
\null

\end{titlepage}

\clearpage

{
  \hypersetup{hidelinks}
  \tableofcontents
}

\section{Introduction}
\label{se:Introduction}

The analysis of particle-physics scattering experiments relies heavily
on simulations based on first-principles calculations within the framework of the Standard
Model (SM) of particle physics. To this end, event generators
like \madgraph \cite{Alwall:2014hca}, \powheg \cite{Alioli:2010xd}, \sherpa \cite{Sherpa:2024mfk}, \herwig \cite{Bewick:2023tfi},
or \pythia \cite{Bierlich:2022pfr} are commonly used, either as stand-alone tools or
in combination, to simulate
complete events from the hard scattering to hadronisation.
On the
other hand, for processes with many particles in the final state,
these event generators are not yet able to incorporate state-of-the-art
predictions as obtained in fixed-order calculations. This kind
of predictions  can be accomplished via dedicated  Monte Carlo (MC) integration
programs that often provide results in terms of weighted events. 
One such program is \mocanlo, which is described in this manual.

\mocanlo (MOnte CArlo at NLO accuracy) is a flexible multi-purpose
MC integration program.  It computes integrated cross
sections and differential distributions for arbitrary processes at the LHC
with both NLO QCD and electroweak (EW) accuracy in the SM. Thereby, full off-shell effects
and spin correlations are taken into account. Besides, NLO corrections to processes
at lepton--lepton and lepton--hadron colliders,
as well as ultra-peripheral collisions of photons from hadrons or heavy ions are also supported.

The phase-space integration uses an elaborated
multi-channel importance sampling along the lines of
\citeres{Berends:1994pv,Denner:1999gp,Roth:1999kk,Dittmaier:2002ap}.  The required
tree-level and one-loop matrix elements are provided by the
matrix-element generator \recola \cite{Actis:2012qn,Actis:2016mpe},
which relies on the \collier library \cite{Denner:2016kdg} to
numerically evaluate the one-loop scalar and tensor integrals.  The
subtraction of the infrared (IR) divergences appearing in the real and virtual
corrections has been automated based on the Catani--Seymour dipole
formalism for both QCD and QED
\cite{Catani:1996vz,Nagy:1998bb,Dittmaier:1999mb,Catani:2002hc,Dittmaier:2008md,Basso:2015gca}.
For the separation of photons and jets in the final state both
Frixione isolation \cite{Frixione:1998jh} and the photon-to-jet
fragmentation function
\cite{Glover:1993xc,Denner:2010ia,Denner:2014bna} are
available. The IR singularities arising from the splitting of virtual
photons into quark--antiquark pairs are regularised using the
photon-to-jet conversion function \cite{Denner:2019zfp}.  Unstable
particles are treated in the complex-mass scheme
\cite{Denner:1999gp,Denner:2005fg,Denner:2006ic,Denner:2019vbn}. Alternatively, the
pole approximation, which is based on the pole expansion
\cite{Stuart:1991xk,Aeppli:1993rs,Denner:2019vbn}, can be used for the virtual corrections
following the strategy of \citere{Denner:2000bj}. The corresponding
non-factorisable virtual corrections are implemented according to
\citeres{Denner:1997ia,Accomando:2004de,Dittmaier:2015bfe}.
For a class of polarised and unpolarised processes,
the pole approximation can be employed
for all contributions to NLO cross sections, while ensuring that all
matrix elements are evaluated in the same Lorentz reference frame.
For leading-order (LO) processes, \mocanlo can deal with arbitrary
particles in the final state. However, for NLO calculations external
particles that 
emit real radiation, \ie photons or gluons, have to be massless.
While the light quarks are massless throughout, the bottom quark and the 
$\tau$~lepton can be massive, if they do not appear as radiating external
particles.
Masses for the electron and the muon only
play a role in the regularisation of collinear singularities for initial-state
leptons or as input for the lepton structure functions, \ie they are
exclusively relevant for processes at lepton colliders.

\mocanlo has demonstrated its ability to compute NLO corrections for
high-multiplicity processes up to $2\to8$. Thus, it has been used to
calculate NLO QCD and EW corrections to top-pair and associated
top-pair production
\cite{Denner:2015yca,Denner:2016jyo,Denner:2016wet,Denner:2017kzu,Denner:2020hgg,Denner:2020orv,Denner:2021hqi,Denner:2023eti,Denner:2023grl},
to the associated production of single bosons~\cite{Denner:2019zfp,Czakon:2022khx},
to the production of gauge-boson pairs
\cite{Biedermann:2017yoi,Brauer:2020kfv}, to the production of single
Higgs bosons and Higgs-boson pairs via vector-boson
fusion~\cite{Dreyer:2020xaj,Barone:2025jey}, to triple-vector-boson production
\cite{Denner:2024ufg,Denner:2024ndl}, to vector-boson scattering (VBS)
processes and their irreducible background
\cite{Biedermann:2016yds,Biedermann:2017bss,Denner:2019tmn,Pellen:2019ywl,Chiesa:2019ulk,Denner:2020zit,Denner:2021hsa,Denner:2022pwc,Denner:2024xul}, and to single-top processes \cite{Denner:2022fhu}.
It has also been employed to evaluate NLO QCD and EW corrections for
processes involving polarised vector bosons
\cite{Denner:2020bcz,Denner:2020eck,Denner:2021csi,Denner:2022riz,Denner:2023ehn,Grossi:2024jae,Denner:2024tlu,Carrivale:2025mjy,DelGratta:2025xjp,Pelliccioli:2025com,Denner:2025xdz}.

This manual is structured as follows: The installation of the code is
described in \refse{se:Installation}, while its basic use is illustrated in
\refse{se:QuickStart}. The following sections discuss the input
cards needed to run the program, namely
the \texttt{run\_card.xml} in \refse{se:ProcessSetup}, the
\texttt{proc\_card.xml} in \refse{se:process},  the
\texttt{param\_card.xml} in \refse{se:parameters}, the \texttt{cut\_card.xml}
in \refse{se:cuts}, and the  \texttt{plot\_card.xml} in \refse{se:histograms}.
User-defined features are described in
\refse{se:user-defined}. 
After the conclusions, \refse{se:conclusions}, we provide a list of validated
processes in \refapp{se:validated_processes}.

\section{Installation and compilation}
\label{se:Installation}

\mocanlo can be downloaded from the git repository
\begin{center}\vspace{-1ex}
\url{https://mocanlo.gitlab.io} 
\end{center}
Unpacking the tarball creates five directories.  The folder
\texttt{MoCaNLO} contains the actual code, while the directory \texttt{Tests}
offers some test runs with corresponding sample input and output.
In the folder \texttt{validated\_processes}, which is referred to as
\texttt{/path/to/processes} in the following, further example processes can be found.
The \texttt{manual} directory contains the PDF file for the code documentation.
The directory \texttt{card\_generator} contains the independent package
\MCG, which may be used to create input cards for MoCaNLO.  

The \mocanlo package consists of the MC code itself and
the additional program {\scshape MCG}.
The {\scshape MC} code is independent of the program 
{\scshape MCG}, whose use is optional. For details about the usage and 
compilation of  {\scshape MCG}, we refer to the file \texttt{card\_generator/mcg/README.md}.

The \texttt{MoCaNLO} folder, which is referred to as
\texttt{/path/to/mocanlo}, has the following subdirectory structure: 

\dirtree{%
        .1 /path/to/mocanlo/.
        .2 bin/. 
        .2 build/. 
        .2 include/.
        .2 src/.
}
The subdirectory \texttt{bin} contains binaries upon successful compilation and 
scripts for collecting results and generating plots. The \texttt{build} 
directory contains compiled object (\texttt{.o}) and Fortran module files 
(\texttt{.mod}) upon compilation. The \texttt{include} directory holds include 
files containing several preprocessor switches. 
Finally, the source code of \mocanlo as well as some supplementary tools
are placed in the subdirectory \texttt{src}.

\paragraph{Prerequisites and external packages}

%
\begin{table}
  \begin{minipage}{.5\linewidth}
    \centering
    \caption*{\label{tab:SoftwareRequirements} \textbf{Software requirements}}
    \begin{tabular}{lll}
      \toprule
       & \textbf{\small recommended}\\
      \midrule
      gfortran      & $\geq 8.3.0$        \\
      CMake         & $\geq 3.13.4$        \\
      gnuplot       & $\geq 5.4.1 $        \\
      \bottomrule
    \end{tabular}
  \end{minipage}%
  \hfill
  \begin{minipage}{.5\linewidth}
    \centering
    \caption*{\label{tab:ExternalPackages} \textbf{External packages}}
    \begin{tabular}{lll}
      \toprule
      & \textbf{\small recommended}\\
      \midrule      
      LHAPDF         & $\geq 6.0.5$      \\
      \recola        & $\geq 1.5$        \\
      \collier       & $\geq 1.2.9$      \\
      \bottomrule
    \end{tabular}
  \end{minipage}%
  \caption{\label{tab:RequirementsAndPackages} Software requirements 
    and external packages  needed to run \mocanlo.}
\end{table}
The software requirements for the MC code are listed in \refta{tab:RequirementsAndPackages}.
Additionally, some external packages 
are needed. Each package has to be compiled separately
and its installation location has to be properly added to the compilation script
\texttt{compile\_mocanlo} available in the \texttt{/path/to/mocanlo} folder.
They are:
\begin{itemize}
\item \textbf{LHAPDF}:
\mocanlo is linked to LHAPDF~\cite{Buckley:2014ana} for the evaluation of parton distribution 
functions (PDFs). For a successful build, compile LHAPDF dynamically
(\url{https://lhapdf.hepforge.org}) and set \texttt{LHAPDF\_PATH} in the 
script \texttt{compile\_mocanlo} to the location of the LHAPDF installation, i.e.\  
the folder with the subdirectories \texttt{bin},  \texttt{include},  \texttt{lib} (containing \texttt{libLHAPDF.so}), and \texttt{share}.
Set the LHAPATH environment variable to the path where PDF sets are installed,
\eg execute 
\begin{lstlisting}[language=sh]
export LHAPATH=/path/to/lhapdf_installation/share/LHAPDF
\end{lstlisting}
if these are stored in \texttt{/path/to/lhapdf\_installation/share/LHAPDF}.

\item \textbf{\recola/\collier}: \sloppy
\mocanlo relies on
\recola~\cite{Actis:2012qn,Actis:2016mpe} 
and \collier~\cite{Denner:2016kdg} for the matrix-element computation
and loop-integral reduction and evaluation. The two packages have to be compiled
separately from \mocanlo. Instructions and downloads can be found at
 \url{https://collier.hepforge.org/index.html} and \url{https://recola.gitlab.io/index.html}.
After a successful installation,
the path variables \texttt{COLLIER\_ROOT} and \texttt{RECOLA\_ROOT} in the 
script \texttt{compile\_mocanlo} have to be set to the location of the compiled
\collier and \recola libraries, respectively.

\begin{warning}{\recola version}
The present version of  \mocanlo does not support
{\scshape Recola2}, which is not simply an improved version of \recola.
\end{warning}

\item \textbf{gamma-UPC} (optional): For photon--photon ultra-peripheral collisions
(UPC), \mocanlo makes use of the {\scshape gamma-UPC} library~\cite{Shao:2022cly} for the photon PDF.
Instructions and downloads can be found at \url{https://www.lpthe.jussieu.fr/~hshao/gammaupc.html}.
After a successful installation, the path variable \texttt{UPCDIR} in the 
script \texttt{compile\_mocanlo} has to be set to the location of the compiled
{\scshape gamma-UPC} library.
In addition, in the script \texttt{compile\_mocanlo}, 
 \texttt{cmake} must be executed with the option \texttt{-DUPC=ON}.

\end{itemize}

\paragraph{Compilation}
To build \mocanlo, simply execute the script \texttt{compile\_mocanlo} as:
\begin{lstlisting}[language=bash]
./compile_mocanlo
\end{lstlisting}
The main \mocanlo executable \texttt{mocanlo} appears in 
\texttt{/path/to/mocanlo/bin} upon successful compilation of the source.
The script binds the different libraries to \texttt{mocanlo} using CMake.
To make use of all scripts and executable (see description later),
add the \mocanlo \texttt{bin} directory to \texttt{PATH}, by executing
\begin{lstlisting}[language=Bash]
export PATH=$PATH:/path/to/mocanlo/bin
\end{lstlisting}

\paragraph{Tests}
The correct functionality of the code can be verified upon
  executing the test runs in the directory \texttt{Tests}. Each subdirectory
  \texttt{\textit{test}} corresponds to a physical process and defines a test based on a
  single partonic subprocess. Available test folders are: 
  \texttt{sl-vbs} (\ensuremath{\Pp\Pp\to\mu^+{\nu}_\mu\Pj\Pj\Pj\Pj}),
  \texttt{ttxz} (\ensuremath{\Pp\Pp\to\Pe^+\nu_\Pe\mu^-\bar{\nu}_\mu\tau^+\tau^-\Pj_\Pb \Pj_{\Pb}}),
  \texttt{tzj} (\ensuremath{\Pp\Pp\to\Pe^+\Pe^-\mu^+\nu_\mu\Pj_\Pb\Pj}),
  \texttt{vbs-zz} (\ensuremath{\Pp\Pp\to\Pe^+\Pe^-\mu^+\mu^-\Pj\Pj}),
  and \texttt{wz-pol}  (\ensuremath{\Pp\Pp\to\Pe^+\nu_\Pe\mu^+\mu^-} with
  polarised \PZ and \PW$^+$ bosons), where $\Pj_\Pb$ denotes bottom jets.
The script \texttt{Tests/check\_process.sh}, which must be executed inside \texttt{Tests/},
takes one of the test directories \texttt{\textit{test}} as its argument
\begin{lstlisting}[language=bash]
./check_process.sh (*@\textit{test}@*)
\end{lstlisting}
and runs \mocanlo using \texttt{\textit{test}/cards} and
\texttt{\textit{test}/inputs} as input. The standard output of \mocanlo
is printed to the screen and saved in the file \texttt{\textit{test}/output\_local}, 
which is compared to \texttt{\textit{test}/output}. 
If the standard output is as expected, the script compares all cross sections and histograms calculated 
locally, which are written to \texttt{\textit{test}/\textit{test}/runs/.../run\_.../results}, to their counterparts 
found in \texttt{\textit{test}/results}.
After each pair of files is compared, the script warns if these differ and offers their location for manual 
comparison. The test finishes with a summary of failures. 
If no argument is passed to \texttt{check\_process.sh}, all directories are tested.

\begin{warning}{Tests need approriate PDFs installed}
Note that running the tests requires the relevant PDFs, as specified in
the corresponding input cards, to be installed.
\end{warning}

\section{Getting started}
\label{se:QuickStart}

\subsection{Citation}

If you use \mocanlo, please cite the present manual.
As mentioned above, input cards can be generated for arbitrary
processes using {\scshape MCG}.
Alternatively, example cards for a large variety of processes are provided.
If you happen to use some of the example cards, please also cite the related references as indicated
in the README.txt file of each process or as reported in \refse{se:validated_processes}.



\subsection{Running the example process}

This section outlines the procedure for running MC simulations of the process 
\fullProcess, which is one of the examples provided within the  \mocanlo package and can
be found under
\texttt{/path/to/processes/ttbar/pp\_tt/}. Before any MC run is submitted, the process directory 
\texttt{pp\_epvemumvmxbbx\_qcd} contains only one folder named \texttt{cards}, 
with four XML files that define subprocesses, standard model and MC 
parameters, cuts to apply, and plots to create: 
\texttt{proc\_card.xml}, \texttt{param\_card.xml}, \texttt{cut\_card.xml}, and
\texttt{plot\_card.xml}, explained in detail in 
\refse{se:ProcessSetup}.

One additional XML file, \ie \texttt{run\_card.xml}, defines MC runs for individual contributions 
of partonic subprocesses: In our example process one finds Born, virtual, real, and integrated-dipole 
contributions (see \refse{se:subprocesses}) labelled by integers.

\begin{info}{PDF sets}
Before performing runs for the example process 
\fullProcess download the PDF sets MSTW2008lo90cl and MSTW2008nlo90cl
 according to your installed LHAPDF version and place them where 
\texttt{\$LHAPATH} points to. 
\end{info}

\subsection{Running a single contribution to a subprocess}
\label{subse:running_single_contribution}
The run with \texttt{id="1"}, as displayed in \refli{lst:RunDefinition}, specifies the 
Born contribution of the gluon--gluon-induced subprocess \ggProcess
(defined in \texttt{proc\_card.xml}).
This contribution
constitutes the Born part of the NLO cross section.
\begin{lstlisting}[float,language=XML,caption=Run definition in
  \texttt{run\_card.xml}. ,captionpos=b,label=lst:RunDefinition]
<runs directory="ttx">
   <run id="1">
      <type>               born         </type>
      <subdirectory>       gg/born      </subdirectory>
      <subprocess          id="gg"      />
      <run_parameters      id="ttx_nlo" />
      <model_parameters    id="ttx_nlo" />
      <recombinations      id="ttx"     />
      <resonances          id="none"    />
      <cuts                id="ttx"     />
      <histograms          id="default" />
      <scales              id="ttx"     />
   </run>
   <run id="2">
      ...
</runs>
\end{lstlisting}

To perform the run, execute \mocanlo with the (relative or absolute) 
path to the process as first and 
the desired run ID as second command-line argument:
\begin{lstlisting}[language=Bash]
mocanlo /path/to/processes/ttbar/pp_tt/pp_epvemumvmxbbx_qcd 1
\end{lstlisting}
In the following, we make use of 
\bash{/path/to/process =
  /path/to/processes/ttbar/pp\_tt} to ease notations.
The run prints information about the initialisation and the input
parameters, and, once \recola 
is initialised, it outputs a list of parameters as being used for the evaluation 
of the matrix elements. After the initialisation phase, the MC run begins,
printing intermediate result summaries once 1000 accepted events have been reached and at increasing intervals thereafter.
Each time an intermediate result is printed to the terminal, 
this result as well as all plots defined in \texttt{plot\_card.xml} are 
dumped to disc. Thus, the process may be killed with \texttt{Ctrl+C} without
loosing the obtained results.

The run creates a new subdirectory structure in the process directory, e.g.:
\begin{lstlisting}[language=bash, basicstyle=\ttfamily\footnotesize]
/path/to/process/pp_epvemumvmxbbx_qcd/ttx/runs/gg/born/run_2025-06-18T15h56m45s045
\end{lstlisting}
where \texttt{ttx} is defined by \lstinline[language=XMLinline]|<runs directory="ttx">|\
in \refli{lst:RunDefinition}, grouping several runs into a \emph{runs set}.
The subdirectory \texttt{gg/born/} in 
\texttt{runs/} is set as the run's output directory by 
\xmltag{subdirectory} in the run definition. Each run with the same ID creates 
a subdirectory in the latter including a timestamp of the form 
\texttt{run\_YYYY-MM-DDTHHhMMmSSsSSS}, which contains the cross sections and 
plots in its \texttt{result} subdirectory, as well as information about the initialisation and 
the run phase of the MC in its \texttt{data} subdirectory. The example run is setup in such a way that it 
is computed for 3 variations of the renormalisation and factorisation scale (specified in
\texttt{param\_card.xml} in \refli{lst:run_parameters})
at the same time by rescaling the matrix elements and the PDFs. To this end, scale factors are
applied to the central scales. The first scale
factor, which usually equals one, is identified 
with the central scale, and the corresponding result is displayed as intermediate result on the 
terminal. The second and third scale factors, for example, correspond to a factor of 0.5 and 
2, respectively. When more than one scale factor is used, the subdirectory 
\texttt{result} contains folders of the type \texttt{scale\_factor\_\textit{i}} 
hosting the results for each scale factor.
The resulting directory structure is as follows:
\dirtree{%
        .1 run\_2025-06-18T15h56m45s045/.
        .2 data/.
        .3 init/.
        .3 run/.
        .2 result/.  
        .3 scale\_factor\_1/.
        .4 histograms/.
        .5 data/.
        .4 cross\_section.dat.
        .3 scale\_factor\_2/.
        .4 histograms/.
        .5 data/.
        .4 cross\_section.dat.
        .3 \ldots.
}%
{\sloppy
The intermediate result as displayed on the terminal is printed to the file 
\texttt{cross\_section.dat} for each scale factor: At the top of this file,
a selection of the most important information is also reported in one line,
preceded by a pipe symbol ``|'', to facilitate its extraction via the \texttt{grep} command.
At the bottom, a summary of occurred errors and exception is printed.
The folder \texttt{histograms/data} 
contains data files for each plot defined in \texttt{plot\_card.xml} of the form 
\texttt{histogram\_\textit{type}\_\textit{name}.dat}, \eg one has 
\texttt{histogram\_transverse\_momentum\_positron.dat} for the 
transverse-momentum distribution of the positron.}

In addition, the subdirectory 
\texttt{data/init/user\_input} contains a set of files
that reflect the inputs to the run. The file
\texttt{data/init/feynman\_diagrams.tex} encodes all Feynman diagrams
corresponding to the constructed integration channels in LaTeX
format. With the command sequence
\begin{lstlisting}[language=Bash]
latex feynman_diagrams.tex; dvips feynman_diagrams
\end{lstlisting}
a PS file is generated that shows all these diagrams, provided that the
package \texttt{axodraw} \cite{Vermaseren:1994je,Binosi:2003yf} is installed.

The subdirectory \texttt{data/run} contains the file
\texttt{feynman\_diagrams\_survey.tex}, which encodes again all
diagrams but now in the order of importance for the integration, which
is complemented by the information in the file 
\texttt{channel\_survey.dat}. The file
\texttt{integration\_stability} shows the evolution of the integration
error with the number of accepted events. Finally, the file
\texttt{max\_weight\_event.dat} stores information on the event with
the largest weight and the  file
\texttt{typical\_events.dat} information on 10 events of typical weight.

\begin{info}{Preprocessor flags} \sloppy
  The files \texttt{include/run\_flags.h} and
  \texttt{include/debug\_flags.h} contain preprocessor flags that
  allow to generate additional output for debugging.  The flags
  \texttt{MONITOR\_...} in \texttt{include/run\_flags.h} provide some
  extra output for the generation of dipoles, the initialisation of
  the recombination, the calculation of the minimal centre-of-mass (CM) energy, and
  the interface to \recola.  The flags in
  \texttt{include/debug\_flags.h} yield extensive output for
  debugging. Their use requires some knowledge of the code and they
  should not be used in runs with many events.
\end{info}

\subsection{Averaging of multiple runs}
\label{se:average_runs}

To speed up the integration, runs may be performed in parallel and their 
results, \ie integrated cross sections and differential distributions, averaged. 
For these runs to be independent, different seeds for the random number generator 
must be used. An optional third argument to \texttt{mocanlo} allows to pass a 
seed to the MC, while without a third argument, a fixed standard seed is chosen.
The Linux shell Bash provides the function 
\texttt{\$RANDOM} returning a pseudo-random integer on each invocation. It may be 
used to provide seeds to \mocanlo to perform independent MC runs by executing, 
possibly more than once
\begin{lstlisting}[language=Bash]
mocanlo /path/to/process/pp_epvemumvmxbbx_qcd 1 $RANDOM
\end{lstlisting}
which is equivalent to
\begin{lstlisting}[language=Bash]
mocanlo /path/to/process/pp_epvemumvmxbbx_qcd 1 11183
\end{lstlisting}
if \texttt{\$RANDOM} returns the random number 11183.

When a seed is passed to \mocanlo, it is appended to the name of the created 
run folder as \linebreak
\texttt{run\_YYYY-MM-DDTHHhMMmSSsSSS\_\textit{seed}}, \eg:
\begin{lstlisting}[language=Bash, basicstyle=\ttfamily\scriptsize]
/path/to/process/pp_epvemumvmxbbx_qcd/ttx/runs/gg/born/run_2025-06-19T10h18m48s164_11183
\end{lstlisting}
with the random seed 11183.

\mocanlo comes with a tool that averages runs, \ie integrated cross sections 
and differential distributions: the script \texttt{average\_runs} (to be found in the
\texttt{bin} folder like all scripts described in the following) averages all
runs in any subfolder of a runs set directory passed as a first argument to the 
binary:
\begin{lstlisting}[language=Bash]
average_runs /path/to/process/pp_epvemumvmxbbx_qcd/ttx
\end{lstlisting}
It creates the folder \texttt{average} in the directory hosting the run 
folders, which contains a subfolder \texttt{result} with the averaged cross 
section in \texttt{cross\_section.dat} and the averaged histograms in the folder 
\texttt{histograms/data}, individually for every scale factor. When the tool 
encounters contributions with only one run, it still creates the folder 
\texttt{average}, but as a symbolic link to the single run directory.
For each contribution the script performs (where needed) the average of all runs
with different random seeds. If runs with equal random seeds are found, the script stops. In this case
the user is required to manually intervene to decide which folders to keep and
make sure that only one run per random seed exists before launching the script again.
In case a contribution was restarted using the code's features described in \refse{par:running_mode},
only the last run folder that was created is considered. The latter corresponds to the folder
whose name contains the additional \texttt{step\textit{N}} identifier with the highest
integer \texttt{\textit{N}} within the contribution's folder.
Note that the averaging is computed as if the separate runs were one 
concatenated MC run. If one wants to only obtain an average of the
cross sections (without having to wait for all histograms for all scale factors to be averaged),
the \texttt{average\_runs} script can also be run with a second optional argument,
namely \texttt{-{}-xs-only}, that precisely skips the histogram averaging.

\subsection{Performing an NLO computation}

The \texttt{run\_card.xml} usually defines different runs that contribute to a full 
NLO calculation of the hadronic process, such as \fullProcess. 
An NLO calculation requires at least one run for Born,
virtual corrections, real corrections, and integrated dipole contributions (\refse{se:subprocesses}).

For the considered example process, after a run of a few hours one can retrieve the NLO results.
Executing
\begin{lstlisting}[language=Bash]
get_results /path/to/process/pp_epvemumvmxbbx_qcd/ttx n
\end{lstlisting}
performs the run averaging for the cross sections and prints them for the $n$th 
scale factor to the terminal. 
If the last argument is missing, the
results for all scale factors are printed. 

After averaging the runs as described in \refse{se:average_runs}, 
different averaged contributions can be merged with the command 
\begin{lstlisting}[language=Bash]
merge_runs -i process_path -o nlo -c "born,virt,real,idip"
\end{lstlisting}
where the argument \texttt{-i} provides the target folder where averaged results
for the different contributions can be found, \eg \bash{process_path =
  /path/to/process/pp\_epvemumvmxbbx\_qcd/ttx}. The argument \texttt{-o} specifies the
name of the output folder in the \texttt{results} directory, which in the example above
is located in \texttt{process\_path/results/nlo}. Finally, \texttt{-c} contains the list
of names of contributions to be merged, enclosed in double quotes and either space or comma separated.
Note that the name of contributions in the list must match the names of the folders where the results for
the different contributions for each partonic channel are written by the \mocanlo{} code, \ie as specified
in the \texttt{run\_card.xml}.
A summary of how to execute the script can be found by typing
\begin{lstlisting}[language=Bash]
merge_runs -h
\end{lstlisting}

The script sums up all indicated contributions to cross sections and
histograms into the subdirectory \texttt{nlo} of \texttt{results}. The structure
of the output folder is as follows:
\dirtree{%
        .1 ttx/.
        .2 results/.
        .3 nlo/.
        .4 hadronic\_cross\_section.dat.
        .4 histograms/.
        .5 data/.
        .6 scale\_factor\_1/.
        .6 scale\_factor\_2/.
        .6 \ldots/.
}%
The merged cross sections are listed for all scale~factors in the file
\texttt{hadronic\_cross\_section.dat} in the directory \texttt{nlo}. The merged
histogram files are written to the directories \texttt{scale\_factor\_\textit{i}}.

\subsection{Plotting results}

{\sloppy
\mocanlo comes with some scripts that allow to plot results using
\texttt{gnuplot} based on the histograms obtained with
\texttt{merge\_runs}. The script \texttt{create\_k-factor\_plots}
produces absolute distributions together with ratios of contributions,
\texttt{create\_correction\_plots}
absolute distributions along with relative contributions, and
\texttt{create\_scale\_envelope\_plots} absolute distributions and relative contributions
with scale envelopes based on the results in the different
\texttt{scale\_factor\_\textit{i}} directories. The syntax of these
commands reads:
}
\begin{lstlisting}[language=Bash]
create_k-factor_plots -i process_path -o k-factor_plots -c "lo,nlo"
\end{lstlisting}
The argument \texttt{-i} provides the runs set folder where merged
results for the different contributions are located in the
subdirectory \texttt{results}. The argument \texttt{-o} specifies the
name of the output folder in the \texttt{plots} directory, 
which in the example above is located
in \texttt{process\_path/plots/k-factor\_plots}. This is also the
default, in case the argument \texttt{-o} is omitted. Finally,
\texttt{-c} contains the list of names of contributions to be plotted,
enclosed in double quotes and either space or comma separated. These
contributions have to be created via \texttt{merge\_runs} with the
corresponding names. Thus, in the example above the subfolders
\texttt{results/lo} and \texttt{results/nlo} must exist and contain
the relevant information. The first contribution is used as reference for
the relative plots.

A folder \texttt{plots} is created in the runs set directory, which
contains the subdirectory specified in the plot command. The 
resulting directory structure is as follows
\dirtree{%
        .1 ttx/.
        .2 results/.
        .3 lo/.
        .4 histograms/.
        .5 data/.
        .6 scale\_factor\_1/.
        .6 scale\_factor\_2/.
        .6 \ldots/.
        .3 nlo/.
        .4 histograms/.
        .5 data/.
        .6 scale\_factor\_1/.
        .6 scale\_factor\_2/.
        .6 \ldots/.
        .2 plots/.
        .3 k-factor\_plots/.
}

Single EPS and PDF files for the individual histograms are placed
in \texttt{plots/k-factor\_plots} together with a PDF file named
\texttt{k-factor\_plots.pdf} that collects all histograms.
Each plot shows the different contributions, \ie \texttt{lo}
and \texttt{nlo} in this example, in an upper panel and the
contributions divided by the first one in the lower panel, \ie
\texttt{lo}/\texttt{lo} and \texttt{nlo}/\texttt{lo} in the example.

{\sloppy
The script \texttt{create\_correction\_plots} works in the
same way, but is meant for displaying relative corrections. If for
instance the virtual, real, and integrated dipole corrections have
been merged with appropriate executions of the \texttt{merge\_runs} script
and stored in the 
folders \texttt{results/virt},  \texttt{results/real}, and
\texttt{results/idip}, these could be plotted via
\begin{lstlisting}[language=Bash]
create_correction_plots -i process_path -o correction_plots \
                        -c "lo,virt,real,idip"
\end{lstlisting}
The resulting plots show the absolute contributions
\texttt{lo,virt,real}, and \texttt{idip} in the upper panel and the
ratios of \texttt{virt/lo}, \texttt{real/lo}, and \texttt{idip/lo} in
the lower panel in per cent. The
default output folder, in case the argument \texttt{-o} is omitted,
is  \texttt{process\_path/plots/correction\_plots}.}
 
The script \texttt{create\_scale\_envelope\_plots} works as the
script \texttt{create\_k-factor\_plots}, but displays in
addition bands for the scale variation obtained from the maxima and
minima in the different
\texttt{process\_path/results/.../histograms/data/scale\_factor\_\emph{i}}
directories. It creates in \texttt{.../histograms/data}
a further subdirectory called \texttt{scale\_envelope} containing the
data for the scale-envelope plots.

\begin{warning}{Remark on individual contributions}
Note that real, virtual, and integrated dipole contributions are individually
not physical. When creating plots for them, the contribution may fall outside the plot ranges
and in case of logarithmic scales negative contributions are not shown.
\end{warning}

\section[Setting up a process in the \texttt{\large run\_card.xml}]
{Setting up a process in the \texttt{\Large run\_card.xml}}
\label{se:ProcessSetup}

A run for a given process is fully specified in \mocanlo by five input cards: 
\texttt{run\_card.xml},
\texttt{proc\_card.xml}, \texttt{param\_card.xml}, \texttt{cut\_card.xml}, and \texttt{plot\_card.xml} (see \refta{tab:Cards}).
The code expects them to be located in a folder named \texttt{cards} inside the specific process directory:
\dirtree{%
        .1 pp\_epvemumvmxbbx\_qcd/.
        .2 cards/.
        .3 run\_card.xml.
        .3 proc\_card.xml.  
        .3 param\_card.xml.  
        .3 cut\_card.xml.
        .3 plot\_card.xml.  
}
\vspace{3mm}

\begin{table}
      \centering
      \rowcolors{2}{tablerowcolor}{}
      \begin{tabularx}{\textwidth}{llcXl}
            \toprule
            \textbf{Card}    & \textbf{XML section(s)}    &  & \textbf{Description} &\\
            \midrule
            run\_card.xml    & \xmltag{run}             & - & Definition of MC runs using links to other cards&\\
            \hline
            proc\_card.xml   & \xmltag{subprocess}       & \checkmark & List of partonic subprocesses and their coupling order & \refse{se:subprocesses}\\
                        & \xmltag{resonances}       & & Specification
                        of resonances restricting the generated phase-space
                        mappings and matrix-element contributions  &\refse{se:resonances}\\
                        & \xmltag{on_shell_projection}  &  & Details of the pole approximation and its on-shell projection & \refse{se:resonances}\\
            \hline  
            param\_card.xml  & \xmltag{model_parameters} &\checkmark & Parameters of the particle model such as masses and widths&\refse{se:model_parameters}\\
                        & \xmltag{run_parameters}   & \checkmark& Various MC-run parameters&\refse{subsubsection:run_parameters}\\
                        & \xmltag{weight_opts}       & & Tuning of the multi-channel weight optimisation & \refse{se:weight_opts}\\
            \hline
            cut\_card.xml    & \xmltag{recombinations} &   & Definitions of recombinations for the jet algorithm&\refse{se:recombinations}\\
                        & \xmltag{cuts}            &  & Phase-space cuts&\refse{se:cut-scheme}\\
                        & \xmltag{scales}          &  & Definition of the renormalisation/ factorisation scale & \refse{se:scale} \\
            \hline
            plot\_card.xml   & \xmltag{histograms}   &  & Histograms&\refse{se:histograms}\\
            \bottomrule
      \end{tabularx}
      \caption{\label{tab:Cards} List of cards for the process definition, together with the different XML sections they contain. 
      The symbol \checkmark marks the sections that must be linked to a process definition in the \texttt{run\_card.xml},
      while all remaining ones are optional.
      The rightmost column contains the section of this manual that describes each section.}
\end{table}

The \texttt{run\_card.xml} plays the role of the top-level card.
In it, all contributions (to be computed as individual runs), which together represent 
an independent and self-consistent calculation, are grouped by runs sets, as illustrated 
in \refli{lst:RunDefinition}. In this example, the XML group \xmltag{runs directory="ttx"} 
defines a runs set \texttt{ttx}, which means that the output of all 
runs in this group will appear in subdirectories of the folder \texttt{ttx} in 
the process directory. 

Within a runs set, the most elemental part of a calculation is represented by a 
run, which is fully specified by the tags contained in the XML group \xmltag{run  id="N"},
with \texttt{N} an integer number. In our example, \xmltag{run  id="1"}
defines the process with ID 1. 

\begin{info}{Remark on run-ID labelling}
  The labels for the XML groups may be any kind of string. While the same holds 
  true for the run ID, it is recommended to use integers only, to allow for an easy
  submission of array jobs on a computer cluster.
\end{info}

The tags that define a run can be divided in three different classes:
\begin{itemize}
\item \textbf{basic tags}: This set comprises two tags that should always be specified when defining a process.
The tag \xmltag{type} is used to define the type of contribution to be
computed and can take one of the
following values: \texttt{born},  \texttt{real}, 
\texttt{virt}, \texttt{idip}, \texttt{lpid}, \texttt{vnfc}, \texttt{idpi}, and \texttt{idpk}.
The output and results of the run are stored in the folder's runs set
in the path given by the tag \xmltag{subdirectory}.
\item \textbf{linking tags}: These tags are used to link XML sections that are defined in the other four cards. Each of
these tags has a one-to-one correspondence with these sections, whose complete list is reported in \refta{tab:Cards}. For instance,
the tag \xmltag{cuts id="ttx"} in \refli{lst:RunDefinition} refers to the homonymous XML group in \texttt{cut\_card.xml}, which
hosts a specific set of cuts. These tags are particularly useful, since they offer the flexibility to use different setups 
(\eg parameters, cuts, set of histograms, \dots) for different IDs, and simplify switching  back and forth between setups.
Note that the linking tags need to be specified for every process: the ones that \mocanlo requires
for a consistent run are marked with the symbol \checkmark in \refta{tab:Cards}. If those are not present, the code will stop with an appropriate
message.

\item \textbf{flag tags}: 
This set includes only the optional tag \xmltag{osp_type}, which controls some
features of the pole approximation as described in \refse{se:resonances}.

\end{itemize}

\begin{info}{Setting default linking tags in the \texttt{run\_card.xml}}
  All linking tags in the run card apart from \xmltag{subprocess} can
  also be defined outside the definition of any \xmltag{run id="N"}
  group. These definitions serve as a default for all following run
  definitions. The defaults can be overwritten within a run
  definition. They can also be redefined outside a run definition,
  and then fix a new default for the following run definitions.
\end{info}

\section[Definition of the process with the \texttt{\large proc\_card.xml}]
{Definition of the process with the \texttt{\Large proc\_card.xml}}
\label{se:process}

The process card (\texttt{proc\_card.xml}) contains the list of all partonic processes
that can be linked in the \texttt{run\_card.xml} to define a process with a given run ID
(see \refli{lst:RunDefinition}). This list is enclosed in 
the XML section \xmltag{subprocesses}. In there, each subprocess contains the information that uniquely
defines the \xmltag{run_type}, \ie the specific LO 
or NLO contribution to be computed, as described in \refse{se:subprocesses}. 

{\sloppy
In the process card, intermediate resonances can be specified in the section \xmltag{resonances} to enable a dedicated
phase-space integration and/or the pole-approximation algorithm, which
serves
as a basis for the definition of polarised cross sections (see \refse{se:resonances}). The section
\xmltag{on\_shell\_projection} allows to control some features of
the on-shell projection used to compute it.}

\subsection{Partonic process specification}
\label{se:subprocesses}

A subprocess definition is enclosed in an XML \xmltag{subprocess} block identified by
a unique subprocess ID (see \refli{lst:sub_a2a4}), which is used by the \texttt{run\_card.xml} to refer to it. Its definition
comprises two parts: the partonic process (\xmltag{partonic\_process}), which describes
the lowest-order contribution to the considered channel,
and the real process (\xmltag{real_process}), which specifies the contribution with one additional
final-state particle. The same subprocess ID can be referred to in the \texttt{run\_card.xml}
by multiple run IDs with different \xmltag{run_type} values. This allows to compute the various
contributions to the NLO cross section of a specific channel.
\begin{figure}
     \begin{lstlisting}[language=XML,caption=Subprocess definition for the
          $\mathcal{O}(\alphas^2\alpha^4)$ and its NLO QCD corrections
          for top-pair production.,
         captionpos=b,label=lst:sub_a2a4,xleftmargin=0pt]
<subprocesses>
    <subprocess id="uxu">
        <partonic_process>
            <incoming>      u~ u            </incoming>
            <outgoing> e+ ve b mu- vm~ b~   </outgoing>
            <pdf1_codes>      -2            </pdf1_codes>
            <pdf2_codes>       2            </pdf2_codes>
            <tree_qcd_order>   2            </tree_qcd_order>
            <loop_qcd_order>   4            </loop_qcd_order>
        </partonic_process>
        <real_process>
            <incoming>      u~ u            </incoming>
            <outgoing> e+ ve b mu- vm~ b~ g </outgoing>
            <pdf1_codes>      -2            </pdf1_codes>
            <pdf2_codes>       2            </pdf2_codes>
            <tree_qcd_order>   3            </tree_qcd_order>
        </real_process>
    </subprocess>
    ...
</subprocesses>
\end{lstlisting}
\end{figure}

The parameters of  both the \xmltag{partonic_process} and the
\xmltag{real_process} are summarised in \refta{tab:ProcParam}
\begin{table}
  \centering
  \renewcommand\arraystretch{1.2}
  \rowcolors{2}{tablerowcolor}{}
  \begin{tabular}{lll}
    \toprule\rowcolor{tableheadcolor} 
    \textbf{parameter} & \textbf{possible values} & \textbf{default value} \\
    \midrule
    \xmltag{incoming}              &  pair of particle names   & none \\
    \xmltag{outgoing}              &  list of particle names   & none \\
    \xmltag{pdf1_codes}            &  list of integers & none \\
    \xmltag{pdf2_codes}            &  list of integers & none \\
    \xmltag{tree_qcd_order}        &  integer &  $0$ \\
    \xmltag{tree_qcd_order_2}      &  integer &  $0$ \\
    \xmltag{loop_qcd_order}        &  integer &  $0$ \\
    \xmltag{interference}          &  integer &  $0$ \\
    \xmltag{tree_e0_order}         &  integer &  guessed \\
    \xmltag{loop_e0_order}         &  integer &  guessed \\
    \bottomrule
        \end{tabular}
        \caption{\label{tab:ProcParam} Parameters of the \xmltag{partonic_process}
          or \xmltag{real_process} subsections of
          \texttt{proc\_card.xml}  and their default values.}
\end{table}
For both 
the particle content of the initial and final states is defined in the tags
\xmltag{incoming} and \xmltag{outgoing}, respectively, using the
name convention for the particles reported in the second column of \refta{tab:ParticleNames}.
The tags \xmltag{pdf1\_codes} and \xmltag{pdf2\_codes} indicate the PDF codes  (using the numbering
convention in the fourth column of \refta{tab:ParticleNames}) to be used for the first and
second incoming beam, respectively.
\begin{table}
      \centering
      \renewcommand\arraystretch{1.2}
      \rowcolors{2}{tablerowcolor}{}
        \begin{tabular}{lllrrr}
        \toprule\rowcolor{tableheadcolor} 
        \textbf{Name} & \textbf{\mocanlo} & \textbf{\recola} & \textbf{PDG code} & \textbf{Jet type} & \textbf{PDF code} \\
        \midrule
        $\Pd$           & \texttt{d}      & \texttt{d}       & 1                 & 1                 & 1                \\
        $\Pu$           & \texttt{u}      & \texttt{u}       & 2                 & 1                 & 2                \\
        $\Ps$           & \texttt{s}      & \texttt{s}       & 3                 & 1                 & 3                \\
        $\Pc$           & \texttt{c}      & \texttt{c}       & 4                 & 1                 & 4                \\
        $\Pb$           & \texttt{b}      & \texttt{b}       & 5                 & 2                 & 5                \\
        $\Pt$           & \texttt{t}      & \texttt{t}       & 6                 & 7                 & 6                \\
        \midrule
        $\Pe^-$         & \texttt{e-}     & \texttt{e-}      & 11                & 5                 & 8               \\
        $\Pe^+$         & \texttt{e+}     & \texttt{e+}      & $-11$             & 4                 & $-8$             \\
        $\nu_\Pe$       & \texttt{ve}     & \texttt{nu\_e}   & 12                & 6 (21)                 & --          \\
        $\bar\nu_\Pe$   & \texttt{ve\char`\~} & \texttt{nu\_e\char`\~} & $-12$   & 6 (22)                 & --          \\
        $\mu^-$         & \texttt{mu-}    & \texttt{mu-}     & 13                & 17                & 9               \\
        $\mu^+$         & \texttt{mu+}    & \texttt{mu+}     & $-13$             & 16                & $-9$             \\
        $\nu_\mu$       & \texttt{vm}     & \texttt{nu\_mu}  & 14                & 6 (23)                 & --          \\
        $\bar\nu_\mu$   & \texttt{vm\char`\~} & \texttt{nu\_mu\char`\~} & $-14$  & 6 (24)                 & --          \\
        $\tau^-$        & \texttt{ta-}    & \texttt{tau-}    & 15                & 19                & 10               \\
        $\tau^+$        & \texttt{ta+}    & \texttt{tau+}    & $-15$             & 18                & $-10$            \\
        $\nu_\tau$      & \texttt{vt}     & \texttt{nu\_tau} & 16                & 6 (25)                & --           \\
        $\bar\nu_\tau$  & \texttt{vt\char`\~} & \texttt{nu\_tau\char`\~} &$-16$  & 6 (26)                 & --          \\
        \midrule
        $\gamma$        & \texttt{a}      & \texttt{A}       & 22                & 3                 & 7                \\
        $\Pg$           & \texttt{g}      & \texttt{g}       & 21                & 1                 & 0                \\
        $\PZ_0$         & \texttt{z}      & \texttt{Z0}      & 23                & 8                 & --               \\
        $\PW^+$         & \texttt{w+}     & \texttt{W+}      & 24                & 9                 & --               \\
        $\PW^-$         & \texttt{w-}     & \texttt{W-}      & $-24$             & 10                 & --               \\
        $\PH$           & \texttt{h}      & \texttt{H}       & 25                & 11                & --               \\
        \bottomrule
        \end{tabular}
        \caption{\label{tab:ParticleNames} The different columns in the table report:
                particle names, identifiers in \mocanlo and \recola, the particle codes according to the PDG MC particle numbering scheme,
                the jet types for the definition of recombinations and cuts, and the PDF codes according to LHAPDF. Antiparticles of non-self-conjugated particles have 
                the negative PDG and PDF code and a ``\texttt{\char`\~}'' added to the \mocanlo and \recola identifier
                (except for particles with explicit charge signs in the identifier, which change for the antiparticle). Jet types are usually the same for the antiparticle,
                except for the charged leptons (4, 16, 18) and the $\PW^-$
                boson (10). The different jet type for neutrinos in
                parenthesis can be switched on with the input option
                \xmltag{neutrino_species_tagging}.
                }
\end{table}

The orders of the amplitude in the QCD coupling $g_\mathrm{s}$ are fixed in \mocanlo 
with the help of additional tags. The orders in the EW coupling
$e$ follow for a given amplitude from the number of external particles.
We use the usual quantities $\alphas=g_\text{s}^2/(4\pi)$ and
$\alpha=e^2/(4\pi)$.

The value of the tag \xmltag{tree_qcd_order} defines the power of the
strong coupling $g_\text{s}$ at the amplitude level.
For the virtual contribution, which originates from the interference
of tree-level and one-loop amplitudes, the QCD order is fixed by
specifying separately the powers of $g_\text{s}$ entering the tree-level and one-loop amplitudes
using the \xmltag{tree_qcd_order} and \xmltag{loop_qcd_order} tags, respectively.
An example of the usage of these tags is given in \refli{lst:sub_a2a4}. There, the
$\mathcal{O}(\alphas^2\alpha^4)$ and its NLO QCD corrections to the
cross section are specified for a partonic channel
contributing to the example process \ttbarProcess.

For processes that receive contributions of different orders in the
QCD coupling at LO, the additional tags \xmltag{interference} and
\xmltag{tree_qcd_order_2} are needed. Contributions of interferences
between LO matrix elements with different coupling orders can be
obtained by setting \xmltag{interference}${}=1$ together with 
\xmltag{tree_qcd_order} and \xmltag{tree_qcd_order_2} to the power of
the strong coupling of the two interfering matrix elements.
Interferences between one-loop and tree-level matrix
elements  are obtained in an analogous
way by specifying \xmltag{loop_qcd_order} and
\xmltag{tree_qcd_order_2}, and interferences of loop-induced
amplitudes by \xmltag{tree_qcd_order} and \xmltag{tree_qcd_order_2}.
 Note that when \xmltag{interference} is not specified or set
to \texttt{0}, the tag \xmltag{tree_qcd_order_2} is ignored.
This feature is needed for our example process
\ttbarProcess, which receives at LO three different contributions from the
$\mathcal{O}(\alphas^2\alpha^4)$, $\mathcal{O}(\alphas\alpha^5)$, and $\mathcal{O}(\alpha^6)$.
To evaluate the $\mathcal{O}(\alphas\alpha^5)$, for instance, one needs to interfere amplitudes
with different powers of $g_\text{s}$, as exemplified in \refli{lst:sub_aa5}.
\begin{figure}
    \begin{lstlisting}[language=XML,caption=Subprocess definition for the
        $\mathcal{O}(\alphas\alpha^5)$
        and its NLO EW corrections to top-pair production. ,captionpos=b,
        label=lst:sub_aa5,xleftmargin=0pt]
    <subprocess id="uxu_int">
        <partonic_process>
            <incoming>      b~ b            </incoming>
            <outgoing> e+ ve b mu- vm~ b~   </outgoing>
            <pdf1_codes>      -2            </pdf1_codes>
            <pdf2_codes>       2            </pdf2_codes>
            <tree_qcd_order>   2            </tree_qcd_order>
            <tree_qcd_order_2> 0            </tree_qcd_order_2>
            <interference>     1            </interference>
            <loop_qcd_order>   2            </loop_qcd_order>
        </partonic_process>
        <real_process>
            <incoming>      b~ b            </incoming>
            <outgoing> e+ ve b mu- vm~ b~ g </outgoing>
            <pdf1_codes>      -2            </pdf1_codes>
            <pdf2_codes>       2            </pdf2_codes>
            <tree_qcd_order>   2            </tree_qcd_order>
            <tree_qcd_order_2> 0            </tree_qcd_order_2>
            <interference>     1            </interference>   
        </real_process>
    </subprocess>
\end{lstlisting}
\end{figure}

\begin{info}{Using \texttt{tree\_qcd\_order} to tune the integration-channel generation}
\mocanlo generates integration channels
on the basis of Feynman diagrams. Independently of the values of run
type and \xmltag{interference},
tree diagrams contributing to the amplitude of order \xmltag{tree_qcd_order} are always used for the
channel generation. To improve the integration of contributions originating from the interference
of amplitudes of different orders, the largest possible set of integration channels can be
chosen by setting \xmltag{tree_qcd_order} to the lower of the two interfering orders. This allows to
generate channels according to amplitudes with the largest power of $e$, which typically
comprise a larger number of Feynman diagrams.  
\end{info}

For processes with external on-shell photons, the corresponding
couplings should always be renormalised at zero momentum transfer in
order to avoid a dependence on the light quark masses \cite{Denner:2019vbn}. This can be
ensured by specifying the number of electromagnetic couplings
$\alpha(0)$ at tree and one-loop level using the tags
\xmltag{tree_e0_order} and \xmltag{loop_e0_order}, respectively, while
the remaining electromagnetic couplings are renormalised using the scheme
chosen via \xmltag{scheme_alpha} in the \texttt{param\_card.xml}.
This is illustrated in \refli{lst:sub_eea} for the partonic process
$\bar\Pu\Pu\to\Pe^+\Pe^-\gamma$ and its EW correction. Note that in this example only
the coupling of one photon is fixed to $\sqrt{\alpha(0)}$.
\begin{figure}
  \begin{lstlisting}[language=XML,
         caption=Subprocess definition for the
         $\mathcal{O}(\alpha^2\alpha(0))$ and its NLO EW corrections
         of order $\mathcal{O}(\alpha^2\alpha(0))$ for
         $\Pe^+\Pe^-$-pair production in association with a photon.,
         captionpos=b,label=lst:sub_eea,xleftmargin=0pt]
    <subprocess id="uxu">
        <partonic_process>
            <incoming>         u~ u      </incoming>
            <outgoing>         e+ e- a   </outgoing>
            <pdf1_codes>      -2 -4      </pdf1_codes>
            <pdf2_codes>       2  4      </pdf2_codes>
            <tree_qcd_order>   0         </tree_qcd_order>
            <tree_qcd_order_2> 0         </tree_qcd_order_2>
            <loop_qcd_order>   0         </loop_qcd_order>
            <tree_e0_order>    1         </tree_e0_order>
            <loop_e0_order>    1         </loop_e0_order>
        </partonic_process>
        <real_process>
            <incoming>         u~ u      </incoming>
            <outgoing>         e+ e- a a </outgoing>
            <pdf1_codes>      -2 -4      </pdf1_codes>
            <pdf2_codes>       2  4      </pdf2_codes>
            <tree_qcd_order>   0         </tree_qcd_order>
            <tree_qcd_order_2> 0         </tree_qcd_order_2>
            <loop_qcd_order>   0         </loop_qcd_order>
            <tree_e0_order>    1         </tree_e0_order>
        </real_process>
    </subprocess>
\end{lstlisting}
\end{figure}
Alternatively, also the coupling of the real-emission photon could
be set to $\sqrt{\alpha(0)}$ by setting \xmltag{tree_e0_order}${}=2$
in \xmltag{real_process}. However, in this case one has to additionally include
\xmltag{loop_e0_order}${}=2$
in \xmltag{partonic_process} in order not to spoil the
cancellation of IR singularities.

If \xmltag{tree_e0_order} and/or \xmltag{loop_e0_order} are not
specified in \texttt{proc\_card.xml}, \mocanlo uses the following
defaults. For processes without external photons, both are set to
zero. For processes with external photons, both are set to the
number of photons in the final state of \xmltag{partonic_process}. Since the treatment of a possible
bremsstrahlung photon is not unique for a real process, \mocanlo 
stops and offers a guess. If this guess corresponds to the desired choice,
it can be accepted by adding the tag
\begin{lstlisting}[language=XML]
<ignore_e0_check>  true </ignore_e0_check>
\end{lstlisting}
in the \xmltag{run_parameters} section of \texttt{param\_card.xml}, which
prevents the code from stopping.
Note that a proper subtraction of infrared singularities requires the same value for
\xmltag{tree_e0_order} for the \xmltag{real} and \xmltag{idip} runs.

\begin{info}{Setting default tags in the \texttt{proc\_card.xml}}
\sloppy
  For all tags needed within a \xmltag{partonic_process} or a
  \xmltag{real_process}, defaults can be specified upon defining subsections
  \xmltag{partonic_process} or \xmltag{real_process} outside an actual
  \xmltag{subprocess}. These defaults are valid for all following
  subprocesses in the card until they are overwritten by a new default.
  Within a \xmltag{subprocess}, the defaults can be overwritten as well.
\end{info}

  A standard computation at NLO accuracy (in either the $\alphas$ or the $\alpha$ coupling)
  requires to run contributions of Born ($B$), virtual ($V$), integrated-counterterm ($I$), and subtracted-real ($R-D$) type.
  More in detail, the differential cross section in a generic observable $\mathcal{O}$ at NLO reads,
  \begin{equation}
    \frac{\rd \sigma^\textrm{NLO}}{\rd \mathcal{O}} = \int_{\text{cuts}} \rd \Phi_n\,\delta_n({\mathcal O})\left(B + V + I\right)
    +\int_{\text{cuts}} \rd \Phi_{n+1}\bigl[\delta_{n+1}({\mathcal
        O})\,R-\delta_{n}({\mathcal O})\,D\bigr] ,
  \end{equation}
  where the labels $n$ and $n+1$ indicate that cuts and the evaluation of the observable should be applied (after possible recombination and isolation) to Born-like and real-correction-like phase-space kinematics, respectively.
  These four contributions are identified in \mocanlo by the \texttt{born}, \texttt{virt}, \texttt{idip}, and \texttt{real} run types.
  The local subtraction counterterms ($D$) and the corresponding integrated counterparts ($I$)
  enable the subtraction of IR singularities of QED and QCD type.
  In the term dubbed $I$, all integrated counterparts of the local counterterms $D$ are collected, including also initial-state collinear and PDF-renormalisation terms.
  In \mocanlo the counterterms are implemented in the Catani--Seymour dipole formalism \cite{Catani:1996vz,Nagy:1998bb,Dittmaier:1999mb,Catani:2002hc,Dittmaier:2008md,Basso:2015gca} and limited to IR-singular configurations caused by massless external particles (with the exception of NLO EW corrections to processes with intermediate $\PW$ bosons in the pole approximation, as detailed in~\refse{se:pole_approximation_nlo}).    

  Infrared-finite, loop-induced contributions must be handled with the \texttt{lpid} run type.
  Other run types can be used for the calculation of contributions 
  in the pole approximation as described 
  in \refse{se:pole_approximation_nfact}, specifically for
  the virtual non-factorisable corrections (\texttt{vnfc}),
  for the $I$-operator part of integrated counterterms (\texttt{idpi}),
  and for the $P$- and $K$-operator parts of the integrated dipole contributions (\texttt{idpk}).

It is important to note that, depending on the \texttt{run\_type} of a given run,
only some information of the \xmltag{subprocess} block is used.
\begin{itemize}
\item \textbf{\texttt{born}}:
  this type is intended for processes that receive tree-level
  contributions at LO. 
  \sloppy Only the \xmltag{partonic_process} part of the subprocess definition is used,
  with the order of the amplitudes computed using
  \texttt{tree\_qcd\_order} and if relevant \texttt{tree\_e0\_order}; when \texttt{interference} is set to \texttt{1}, 
  \texttt{tree\_qcd\_order\_2} is also used.

\item \textbf{\texttt{virt}}: this is intended for one-loop virtual corrections to processes
  that receive contributions at tree level.
  \sloppy Only the \xmltag{partonic_process} part of the subprocess definition is used,
  with the order of the amplitudes computed using
  \texttt{loop\_qcd\_order} and \texttt{tree\_qcd\_order},
  or, when \texttt{interference} is set to \texttt{1},
  \texttt{tree\_qcd\_order\_2};  if
  relevant also \texttt{tree\_e0\_order} and \sloppy \texttt{loop\_e0\_order} enter.

\item \textbf{\texttt{vnfc}}: only the \xmltag{partonic_process} part of the subprocess definition is used,  
  with the order of the amplitude computed as for the \texttt{born}
  \texttt{run\_type}.
This type is dedicated
  to the computation of virtual non-factorisable contributions in the pole approximation (see \refse{se:pole_approximation_nfact}).
  Its usage is only allowed if the section \xmltag{resonances id="XX"} is specified in the \texttt{run\_card.xml}
  (see \refse{se:resonances}). The type of non-factorisable corrections to be computed is steered
  by \texttt{tree\_qcd\_order} and \texttt{loop\_qcd\_order}: if \texttt{loop\_qcd\_order${}>{}$tree\_qcd\_order}
  non-factorisable QCD corrections are evaluated, and if \texttt{loop\_qcd\_order${}={}$tree\_qcd\_order}
  the EW ones are selected. If \texttt{interference${}={}$1}, \texttt{tree\_qcd\_order\_2},
  and \texttt{loop\_qcd\_order} are instead compared to choose between the two types of corrections.

\item \textbf{\texttt{lpid}}: \sloppy
this run type is intended for loop-induced processes. It works as a \texttt{born} one, where the orders of the loop diagrams
  are specified with \xmltag{tree_qcd_order} (and not by \texttt{loop\_qcd\_order}), and, if \texttt{interference} is set to \texttt{1},
  \xmltag{tree_qcd_order_2}, thus allowing to calculate loop-induced interference contributions.
Analogously, the orders of $\sqrt{\alpha(0)}$ are specified via \xmltag{tree\_e0\_order}.

\item \textbf{\texttt{real}}:  a real contribution comprises the real
  squared amplitude and all Catani--Seymour subtraction counterterms
  needed to have a finite result. The information to define the real squared amplitude is taken from the \xmltag{real\_process} block.
  The same applies for the construction of its subtraction counterterms. 
  Their complete list is obtained considering
  all underlying Born processes that can be reached
  from the real process by discarding a candidate emissus particle and replacing the candidate emitter particle by the splitting one.
  The underlying Born processes whose particle content is not compatible with the single-particle-cut requirements (see \refse{se:cut-scheme})
  specified in the \texttt{cut\_card.xml}
  are discarded already at this level, to prevent the computation of contributions that would eventually be cut away.  
  Note that the PDF codes in the \xmltag{real\_process} part are also the relevant ones for the computation of the subtraction terms.
  Note that, when the \texttt{real} is treated in the pole approximation (see \refse{se:pole_approximation_nlo}), the code also uses the information
  of the \xmltag{incoming} and \xmltag{outgoing} from the  \xmltag{partonic\_process} block to identify the radiated particle.
  
\item \textbf{\texttt{idip}}: for this run type both the \xmltag{partonic\_process} and the \xmltag{real\_process} parts
  are relevant. This contribution is computed using the PDF codes in the \xmltag{real\_process} part. It results from the sum of
  dipoles constructed as for the \texttt{real} case that also match the particle content (up to ordering of the final state) 
  and the amplitude order of the underlying Born process specified in the \xmltag{partonic\_process} block. 
  Only these dipoles are taken into account in the result. This means that the complete set of integrated counterterms is obtained by 
  running multiple \texttt{idip} runs whose \xmltag{subprocess} definitions have identical \xmltag{real\_process} parts, 
  but different \xmltag{partonic\_process} as underlying Born processes.
  Also note that, since integrated dipoles are defined on a Born-like kinematics, the integration channels are
  constructed from the \xmltag{partonic\_process} amplitude.

\item \textbf{\texttt{idpi}}/\textbf{\texttt{idpk}}: these two additional types can be run exactly as
  the idip \texttt{run\_type} described above. Even if they can be used both with and without
  the specification to the section \xmltag{resonances id="XX"} in the \texttt{run\_card.xml}
  (see \refse{se:resonances}), their purpose is to guarantee the IR safety of the full result
  when a pole approximation is applied only to the virtual
  contributions (as described in
  \refse{se:pole_approximation_nfact}). Note that the contribution
  \texttt{idpk} cannot be calculated in the pole approximation: If it is linked to
  a section \xmltag{resonances id="XX"} of the \texttt{proc\_card.xml}
  with \texttt{pole\_approximation=true}, the code stops.
\end{itemize}

In \refli{lst:real_idip} the usage of the run types \texttt{real} and \texttt{idip}
is further illustrated. The example shows some subprocesses needed to compute QCD corrections
of $\mathcal{O}(\alphas\alpha^6)$ to
the pure EW LO contribution to $\PW\PZ$ VBS.
All three subprocesses with IDs \texttt{real\_b1}, \texttt{real\_b2}, and \texttt{real\_b3} define the
same real squared amplitude, but differ in their \xmltag{partonic_process} block. The real contribution to
the channel \ensuremath{\Pu\Pu\to\Pe^+\nu_\Pe \mu^+\mu^-\Pu\Pd\Pg}\xspace is obtained by running
only one of the three subprocesses with run type \texttt{real}. To properly account for all
integrated dipoles, instead all three subprocesses must be run separately with run type \texttt{idip}.
Then, the subprocess with \texttt{id="real\_b1"} computes integrated QCD dipoles
with underlying Born of order \xmltag{tree_qcd_order}${}=0$, while the subprocesses with
\texttt{id="real\_b2"} and \texttt{id="real\_b3"} compute QED dipoles
with two different underlying Born processes
(resulting from a photon emitted from the incoming quark and entering the hard process)
of order \xmltag{tree_qcd_order}${}=1$.

\begin{lstlisting}[float,language=XML,float,caption=Example of three subprocess definitions in
    the \texttt{proc\_card.xml} needed to compute different integrated-dipole contributions
    to the same real channel. ,captionpos=b,
  label=lst:real_idip]
   <subprocess id="real_b1">
      <partonic_process>
         <incoming> u  u    </incoming>
         <outgoing> e+ ve  mu+ mu- u  d    </outgoing>
         <pdf1_codes>     2 </pdf1_codes>
         <pdf2_codes>     2 </pdf2_codes>
         <tree_qcd_order> 0 </tree_qcd_order>
         <loop_qcd_order> 2 </loop_qcd_order>
      </partonic_process>
      <real_process>
         <incoming> u  u    </incoming>
         <outgoing> e+ ve  mu+ mu- u  d  g </outgoing>
         <pdf1_codes>     2 </pdf1_codes>
         <pdf2_codes>     2 </pdf2_codes>
         <tree_qcd_order> 1 </tree_qcd_order>
      </real_process>
   </subprocess>
   <subprocess id="real_b2">
      <partonic_process>
         <incoming> a  u    </incoming>
         <outgoing> e+ ve  mu+ mu- d  g    </outgoing>
         <pdf1_codes>     2 </pdf1_codes>
         <pdf2_codes>     2 </pdf2_codes>
         <tree_qcd_order> 1 </tree_qcd_order>
         <loop_qcd_order> 0 </loop_qcd_order>
      </partonic_process>
      <real_process>
         <incoming> u  u    </incoming>
         <outgoing> e+ ve  mu+ mu- u  d  g </outgoing>
         <pdf1_codes>     2 </pdf1_codes>
         <pdf2_codes>     2 </pdf2_codes>
         <tree_qcd_order> 1 </tree_qcd_order>
      </real_process>
   </subprocess>
   <subprocess id="real_b3">
      <partonic_process>
         <incoming> u  a    </incoming>
         <outgoing> e+ ve  mu+ mu- d  g    </outgoing>
         <pdf1_codes>     2 </pdf1_codes>
         <pdf2_codes>     2 </pdf2_codes>
         <tree_qcd_order> 1 </tree_qcd_order>
         <loop_qcd_order> 0 </loop_qcd_order>
      </partonic_process>
      <real_process>
         <incoming> u  u    </incoming>
         <outgoing> e+ ve  mu+ mu- u  d  g </outgoing>
         <pdf1_codes>     2 </pdf1_codes>
         <pdf2_codes>     2 </pdf2_codes>
         <tree_qcd_order> 1 </tree_qcd_order>
      </real_process>
   </subprocess>
\end{lstlisting}

\subsubsection{Merging of partonic channels}
\label{se:merging}

Having separate runs for different partonic channels and NLO contributions allows \mocanlo
to tune the integration channels and improve the integration efficiency. This is crucial
for processes having a non-trivial resonance structure. Nonetheless, for high-multiplicity
final states, the number of contributions to be computed as separate runs grows very fast
and can potentially become a bottleneck.
To ameliorate this problem, \mocanlo supports two ways of computing different partonic
channels together, which may be used simultaneously.
\begin{figure}
  \begin{lstlisting}[language=XML,caption=Example of channel combination
        using PDF and final-state merging. ,captionpos=b,
        label=lst:pdf+fs_merging,xleftmargin=0pt]
    <subprocess id="uux_fs_merged">
        <partonic_process>
            <incoming>         u u~         </incoming>
            <outgoing> e+ ve mu+ mu- u~ d   </outgoing>
            <outgoing> e+ ve mu+ mu- c~ s   </outgoing>
            <pdf1_codes>       2  4         </pdf1_codes>
            <pdf2_codes>      -2 -4         </pdf2_codes>
            <tree_qcd_order>   0            </tree_qcd_order>
            <tree_qcd_order_2> 0            </tree_qcd_order_2>
            <interference>     1            </interference>
        </partonic_process>
        <real_process>
            <incoming>         u u~         </incoming>
            <outgoing> e+ ve mu+ mu- u~ d a </outgoing>
            <outgoing> e+ ve mu+ mu- c~ s a </outgoing>
            <pdf1_codes>       2  4         </pdf1_codes>
            <pdf2_codes>      -2 -4         </pdf2_codes>
            <tree_qcd_order>   0            </tree_qcd_order>
            <tree_qcd_order_2> 0            </tree_qcd_order_2>
            <interference>     1            </interference>
        </real_process>
    </subprocess>
  \end{lstlisting}
\end{figure}

\paragraph{PDF merging}
The tags \xmltag{pdf1\_codes} and \xmltag{pdf2\_codes}, which define the PDF codes for the first and second beam,
can be initialised to strings of integers, like $\{i_1,\dots,i_n\}$ and $\{j_1,\dots,j_m\}$, respectively.
Their length should be the same for both tags (\ie $n=m$), since values occupying the same positions in the string enter
the calculation in pairs, namely only the combinations $\{(i_1,\,j_1),\dots,(i_n,\,j_n)\}$ are considered and no cross terms. 
The resulting cross section will be the sum of as many contributions as the length of the strings. Each contribution
is obtained by weighting the same squared amplitude with different PDF factors. The underlying assumption is that
amplitudes for the considered process are identical when exchanging the flavours $i_1\leftrightarrow i_k$ and
$j_1\leftrightarrow j_k$ $\forall k\in\{2,\dots,n\}$ both in the initial and in the final states.

The lines containing \xmltag{pdf1_codes} and \xmltag{pdf2_codes} in
\refli{lst:pdf+fs_merging} illustrate the usage of PDF merging for
partonic processes contributing 
to $\PW^+\PZ$ VBS. In the example the partonic channels
$\Pu\bar{\Pu}\rightarrow\Pe^+\nu_e\mu^+\mu^-\bar{\Pu}\Pd (\gamma)$
and $\Pc\bar{\Pc}\rightarrow\Pe^+\nu_e\mu^+\mu^-\bar{\Pc}\Ps (\gamma)$  are computed in one run. This combination is allowed,
since the two processes only differ by PDF factors, but have identical matrix elements if the quarks are massless and
the CKM matrix is set to the identity matrix.

\paragraph{Final-state merging}
If the definition of a \xmltag{partonic_process}/\xmltag{real_process}
includes multiple \xmltag{outgoing} tags, partonic channels sharing
the same initial state and only differing in their final states are merged
into one \mocanlo run. When the process is initialised, the code generates all integration channels
for the different processes and eliminates identical ones. 
For the \texttt{born} and \texttt{virtual}
types, the list of processes in the \xmltag{partonic_process} block is combined,
while for the \texttt{real} type only the list in \xmltag{real_process} is relevant.
For the \texttt{idip} type, both \xmltag{partonic_process} and \xmltag{real_process}
are important. For each real process at a given position in the
list of \xmltag{outgoing} tags of \xmltag{real_process}, the code computes only the dipoles
which match the underlying Born at the same position
in the \xmltag{outgoing} tags of \xmltag{partonic_process}. Both when running the \texttt{real}
and the \texttt{idip}, identical dipoles and dipoles only differing in some charge/colour
factors are computed once and properly accounted for in the combined result.

\begin{warning}{Warning}
The final-state merging is limited to channels that
differ in the flavour of final-state fermions. No merging is supported
if a fermion is replaced by a gluon or a photon.
Further, it is implicitly assumed that recombination and cuts act on all merged final states in
exactly the same way. The reason is that recombination rules and cuts are only evaluated for the first
of the merged processes. Thus, normally quarks should not be merged
with charged leptons or neutrinos, since typically cuts act differently on
these final-state objects.
\end{warning}

\begin{info}{Merging final states efficiently}
\mocanlo generates integration channels
on the basis of Feynman diagrams for all merged partonic processes and
eliminates duplicate ones. In order to reduce the number of channels,
the final states of the different partonic processes should be ordered
in such a way that as many diagrams as possible become identical. This
is typically achieved by putting corresponding particles
in the same position in different final states.
\end{info}

In \refli{lst:pdf+fs_merging} the syntax for integrating together the channels $\Pu\bar{\Pu}\rightarrow\Pe^+\nu_e\mu^+\mu^-\bar{\Pu}\Pd (\gamma)$
and $\Pu\bar{\Pu}\rightarrow\Pe^+\nu_e\mu^+\mu^-\bar{\Pc}\Ps (\gamma)$ is explicitly shown. Since in this example the final-state
merging is used in combination with the PDF merging, the channels $\Pc\bar{\Pc}\rightarrow\Pe^+\nu_e\mu^+\mu^-\bar{\Pc}\Ps (\gamma)$
and $\Pc\bar{\Pc}\rightarrow\Pe^+\nu_e\mu^+\mu^-\bar{\Pu}\Pd (\gamma)$ are also implicitly accounted for.

\subsection{Treatment of resonances}
\label{se:resonances}

Many applications focus on contributions involving a specific set of intermediate 
resonances for a given process. The user can require a desired set of resonances 
by linking in the \texttt{run\_card.xml} a section \xmltag{resonances id="XX"},
where \texttt{XX} is the ID that denotes a particular resonance specification.
This optional section must be provided in the \texttt{proc\_card.xml} as shown in \refli{lst:resonances}.
\begin{lstlisting}[language=XML,float,caption=
Example of a minimal specification of resonances
corresponding to the partonic process defined in \refli{lst:sub_a2a4}
or \ref{lst:sub_aa5}. ,label=lst:resonances]
   <resonances id="ttbar">
      <resonance>
         <particle>       t     </particle>
         <external_legs>  3 4 5 </external_legs>
       </resonance>
       <resonance>
         <particle>       t~    </particle>
         <external_legs>  6 7 8 </external_legs>
      </resonance>
   </resonances>
\end{lstlisting}

In each resonances section, multiple resonances can be required. The information on 
each individual resonance is enclosed in the XML subsection \xmltag{resonance}, whose
available parameters are summarised in \refta{tab:ResParam}.
\begin{table}
        \centering
        \renewcommand\arraystretch{1.2}
        \rowcolors{2}{tablerowcolor}{}
        \begin{tabular}{lll}
                \toprule\rowcolor{tableheadcolor} 
                \textbf{parameter} & \textbf{possible values} & \textbf{default value} \\
                \midrule
    \xmltag{particle}              &  \texttt{t, t\char`\~, w+, w-, z, h}      &   --- \\
    \xmltag{external_legs}         &  list of integers (see text) &  --- \\
    \xmltag{phase_space_only}      &  \texttt{true, false}     & \texttt{true}  \\
    \xmltag{pole_approximation}    &  \texttt{true, false}     & \texttt{false}  \\
    \xmltag{polarisation_state}    &  \texttt{-1, +1, 0, T, U}      & \texttt{U}  \\
    \bottomrule
        \end{tabular}
        \caption{\label{tab:ResParam} Parameters of the \xmltag{resonance}
          section of \texttt{proc\_card.xml} and their default values.}
\end{table}
The resonance type is specified in \xmltag{particle} according to its
particle name as detailed in column 2 of \refta{tab:ParticleNames}.
The tag \xmltag{external_legs} defines the decay products of the resonance in terms of their positions in the list of \xmltag{outgoing}
(with counting starting from $3$) of the 
subprocess pointing to this \xmltag{resonances id="XX"}. As long as the particle types of the decay
products are consistent with the resonance, their position in the list of \xmltag{outgoing} is
not constrained, \ie they are not forced to be adjacent. In the example in \refli{lst:resonances},
the top and antitop resonances define the requirement of the decay chain $\Pt\to\Pe^+\nu_\Pe\Pb$ and $\bar\Pt\to\mu^-\bar{\nu}_\mu\bar\Pb$, 
respectively, for the process \ttbarProcess.

By default, the presence of a resonances section tells \mocanlo to construct only the integration channels that 
are compatible with the specified resonances, but it does not affect which contributions enter the matrix elements. 
This reduction of the amount of channels can speed up the convergence of the integration, because there are less 
channels to sample from. On the other hand, if there are significant contributions to the matrix elements 
in the considered fiducial region that are not sampled by the specified subset of channels, the convergence might be worse. 
For more information about the construction of integration channels we
refer to \citeres{Berends:1994pv,Denner:1999gp,Roth:1999kk,Dittmaier:2002ap}.

Setting to \texttt{false} the additional tag \xmltag{phasespace_only}
(default \texttt{true}) in a specific \xmltag{resonance}
subsection causes the decay chain to also affect the matrix-element computation, namely only diagrams including that resonance
will be part of the calculation. This is the starting point to define squared amplitudes including one or more on-shell resonances.
Nevertheless, such a definition only based on a resonant-diagram selection breaks gauge invariance. 
This is why \xmltag{phasespace_only} should only be set to \texttt{false} in combination with the \xmltag{pole_approximation}
tag (default \texttt{false}), where the matrix element is treated with a generic pole approximation algorithm, 
discussed in further detail in the following subsection.

\subsubsection{Pole approximation: general features}
\label{se:pole_approximation}

The pole approximation~\cite{Denner:2000bj} is based on the pole scheme~\cite{Stuart:1991xk,Aeppli:1993rs,Denner:2019vbn}, 
which provides a gauge-invariant way to separate resonant and non-resonant contributions at the level of the matrix element. 
When in \mocanlo the XML resonance section includes the \xmltag{phasespace_only} tag set to \texttt{false} in combination with
\xmltag{pole_approximation} set to \texttt{true}, the pole approximation is applied. Thus, only diagrams involving the
specified resonances are included in the matrix elements.  In
addition, an on-shell-projected kinematics is used throughout the calculation
except in the evaluation of resonances' propagators and when applying phase-space cuts, \ie an event is still accepted or rejected
according to its off-shell kinematics before on-shell projection. In \refli{lst:pole} two possible pole approximations are shown 
for the VBS process \vbsProcess, whose \xmltag{incoming} and  \xmltag{outgoing} lists for a specific
subprocess are given at the beginning of the same Listing.
\begin{lstlisting}[language=XML,float,caption=Examples of pole
  approximations for the process \vbsProcess. ,label=lst:pole]
   <subprocess id="sxs">
      <partonic_process>
         <incoming> s~ s               </incoming>
         <outgoing> e+ ve mu- vm~ u u~ </outgoing>
         ...
      </partonic_process>
      ...
   </subprocess>
   ...
   <resonances id="nested_hwm">
      <resonance>
         <particle>           h       </particle>
         <external_legs>      3 4 5 6 </external_legs>
         <phasespace_only>    false   </phasespace_only>
         <pole_approximation> true    </pole_approximation>
      </resonance>
      <resonance>
         <particle>           w-      </particle>
         <external_legs>      5 6     </external_legs>
         <phasespace_only>    false   </phasespace_only>
         <pole_approximation> true    </pole_approximation>
      </resonance>
   </resonances>
   <resonances id="tpa_wpwmz">
      <resonance>
         <particle>           w+      </particle>
         <external_legs>      3 4     </external_legs>
         <phasespace_only>    false   </phasespace_only>
         <pole_approximation> true    </pole_approximation>
      </resonance>
      <resonance>
         <particle>           z       </particle>
         <external_legs>      5 6     </external_legs>
         <phasespace_only>    false   </phasespace_only>
         <pole_approximation> true    </pole_approximation>
      </resonance>
      <resonance>
         <particle>           z       </particle>
         <external_legs>      7 8     </external_legs>
         <phasespace_only>    false   </phasespace_only>
         <pole_approximation> true    </pole_approximation>
      </resonance>
   </resonances>
\end{lstlisting}

In the block \xmltag{resonances id="nested_hwm"}, a pole approximation with a nested Higgs resonance is illustrated. 
There, two sets of external legs defining the Higgs and the W-boson resonances overlap.
In order to deal with nested resonances within the pole approximation, \mocanlo makes use of the
general on-shell projection algorithm presented in \citere{Denner:2024xul}.
The very same algorithm can also deal with single-pole approximations.
A special treatment is only required for a few cases, which are separately discussed at the
end of \refse{par:spa}.
\begin{table}
        \centering
        \renewcommand\arraystretch{1.2}
        \rowcolors{2}{tablerowcolor}{}
        \begin{tabular}{lll}
                \toprule\rowcolor{tableheadcolor} 
                \textbf{parameter} & \textbf{possible values} & \textbf{default value} \\
                \midrule
    \xmltag{preserve_system}               &  \texttt{cms, cms\_res}                & \texttt{cms}  \\
    \xmltag{poldef_system}         &  \texttt{cms, cms\_res, lab}                   & value of \xmltag{preserve_system}  \\
    \xmltag{preserve_directions}   &  list of integers (see text)                & ---  \\
    \xmltag{keep_resonance_widths}    &  $0,1$                    & $~0$  \\
    \midrule
    \xmltag{search_invariants}    &  $0,1$                    & $~0$  \\
    \xmltag{subsystem}               &    \texttt{full, res, nres, nres\_tot, }    & \texttt{full}  \\
                         &    \texttt{res+nres, res+nres\_tot}    &   \\
    \xmltag{priority}     &   list of resonance identifiers           & \texttt{h t* w* z}  \\
                          &   (as in second column of \refta{tab:ParticleNames})                  &   \\
    \midrule
    \xmltag{fudge_factor}         &  $0,1$                  & $~0$  \\
    \xmltag{fudge_factor_born}         &  $0,1$                  & $~0$  \\
    \xmltag{fudge_width_range}         &  real number                  & $~3.0$  \\
    \bottomrule
        \end{tabular}
        \caption{\label{tab:on_shell_projection} Parameters of the \xmltag{on\_shell\_projection}
          section of \texttt{proc\_card.xml} and their default values.}
\end{table}

{\sloppy
The behaviour of the on-shell projection can be tuned
via the \xmltag{on_shell_projection id="..."} block in the \texttt{proc\_card.xml}, 
which is linked in the subprocess definition in the \texttt{run\_card.xml}. 
The elements of this block are summarised in \refta{tab:on_shell_projection}.
The on-shell projection implemented in \mocanlo{} always preserves at least one invariant,
which is by default the partonic CM energy.
Using the tag \xmltag{preserve_system}, the system whose total momentum is preserved can be
changed from the default one, \ie the partonic CM system (\texttt{cms}),
to the one defined by the sum of the momenta of the on-shell-projected resonances (\texttt{cms\_res}).
Note that, when running a \texttt{real} contribution as part of an NLO calculation,
the radiated particle is always included in the preserved system if \xmltag{preserve_system} is set to \texttt{cms}.
If instead \texttt{cms\_res} is chosen, the radiation is included only for real corrections to the
decays (\texttt{osp\_type}${}>0$), while excluded for real corrections to the production
(\texttt{osp\_type}${}=0$). See discussion in \refse{se:pole_approximation_nlo}.
}

Furthermore, the user may require to preserve additional invariants via the input tags of the
sub-block \xmltag{preserve_invariants}.  When the tag \xmltag{search_invariants} (default \texttt{0})
is set to $1$, the code loops over all integration channels and selects the ones maximising
the number of resonances that are not projected on shell.
By default, Higgs bosons are given the highest priority, followed by top quarks, W~bosons, and
Z~bosons. This behaviour can be modified using the input \xmltag{priority} and providing a list
of resonances as \mocanlo{} identifiers (as defined in second column of \refta{tab:ParticleNames}). Different
priorities can be required for particles and antiparticles of the same resonance using the appropriate
identifiers, \eg \texttt{t}/\texttt{t\char`\~} and \texttt{w+}/\texttt{w-} for top quarks and W bosons, respectively. To give the same importance to particles and antiparticles, a star key can be used next to
the resonance identifier, \eg \texttt{t*} or \texttt{w*}.
Resonances in the first entries of the list receive higher priority and integration channels with the largest number of propagators
associated to them are selected. If no channel with at least one
resonance of the \xmltag{priority} list can be found, the code stops, since no invariant fulfilling the
input criteria can be constructed. Conversely, if a subset of channels is found, the code
further filters them according to its default priority criteria. Additionally, more importance is
given to those channels that involve resonances decaying into a
smaller amount of particles. Once an integration
channel is finally selected, invariants are collected from all resonant propagators
that are not set on shell.
Invariants corresponding to the partonic CM energy are excluded. Moreover, since the algorithm
tries to search for invariants that can be preserved both at LO and NLO, invariants including momenta
from all final-state photons and/or gluons are also discarded. 
The final selection of preserved invariants is printed to the standard
output of the run.

An additional filter on the invariants to
be preserved can be activated via the flag \xmltag{subsystem}. By default (\texttt{full}),
all invariants are kept. By choosing \xmltag{subsystem}${}={}$\texttt{res},
only invariants in the sub-system defined by the decay products of the on-shell-projected resonances
are kept, while the value \texttt{nres} restricts the momenta of the invariants to the
complementary system, \ie all final-state momenta that are not part of on-shell-projected resonances.
Invariants from both sets are chosen with the value \texttt{res+nres}. Moreover, using \texttt{nres\_tot}
and \texttt{res+nres\_tot} allows to preserve the invariant of the complementary system as a whole.
This latter option can be used for instance to address the invariant mass of the tag jets in the
context of VBS (see \citere{Denner:2024xul}). Note that the set of additionally preserved invariants
can only be a subset of the system defined by \xmltag{preserve_system}, meaning that when the
latter is set to \texttt{cms\_res}, \xmltag{subsystem} can only be set to \texttt{res}, otherwise
the code stops.

{\sloppy
An example of an \xmltag{on_shell_projection id="..."} block is shown in \refli{lst:on-shell-projection}.
There, the conservation of the CM energy is required explicitly using the \xmltag{preserve_system} tag. 
In the block \xmltag{preserve_invariants}, additional invariants to be preserved are searched in the
full final state, giving higher priority to the ones that can be associated to intermediate W bosons
(regardless of the charge) and, with lower priority, to intermediate Z bosons. 
}
\begin{lstlisting}[language=XML,float, caption=Additional on-shell
  projection information in \texttt{proc\_card.xml} extending \refli{lst:pole}.,label=lst:on-shell-projection]
   <on_shell_projection id="nested_hwm">
      <preserve_system>       cms  </preserve_system>
      <keep_resonance_widths> 1    </keep_resonance_widths>
      <preserve_invariants>
         <search_invariants>  1    </search_invariants>
         <subsystem>          full </subsystem>
         <priority>           w* z </priority>
      </preserve_invariants>
   </on_shell_projection>
\end{lstlisting}

In some applications one may want to preserve the directions of a subset of final-state particles.
By default, the on-shell projection described in \citere{Denner:2024xul} preserves all directions of the decay products
of a resonance in the rest frame of the latter, but not in the laboratory frame. This behaviour can be modified via
the optional \xmltag{preserve_directions} tag in the \xmltag{on_shell_projection} section. The latter must contain a list
of integers referring to the position of the final-state particles whose off-shell
directions in the laboratory frame one wants to preserve.
This feature, which allows to reproduce the on-shell projection introduced
in \citere{Denner:2000bj}, is only available for resonances decaying into two particles,
where no more than one decay-product direction per resonance can be preserved,
\ie each integer in the \xmltag{preserve_directions} list must belong to a different
resonance. Moreover, this option cannot be used for a \texttt{real} run type.

As part of the pole-approximation algorithm, the widths of the
on-shell-projected resonances must be set to zero everywhere in the
calculation except in the propagators of the specified resonances. 
Nevertheless, when a pole approximation with $n$ resonances is performed, 
a process may have a larger number of resonances than the specified
ones with the same types of particles. This is the case for our
example subprocess
\xmltag{subprocess id="sxs"}, which also admits triply-resonant contributions, as evidenced by
the second pole approximation \xmltag{resonances id="tpa_wpwmz"} in \refli{lst:pole}.
When this happens, setting the decay widths of the resonances to zero causes the phase-space 
integration to run into singularities, unless they are excluded by cuts. Even if no general solution
to the problem exists, \mocanlo{} offers two ways to handle divergent resonant propagators in the
context of the pole approximation.

A simple workaround is to keep the resonance widths non-zero \cite{Denner:2024tlu}. This can be enforced
by switching on the tag \xmltag{keep_resonance_widths} in the \xmltag{on_shell_projection} section
(see \refli{lst:on-shell-projection}).
If not set, the code works as usual and sets to zero the
decay widths in the pole-approximated squared amplitudes except in the propagators of the required resonances.
\begin{warning}{On the usage of \texttt{keep\_resonance\_widths}}
\begin{itemize}
\item 
The flag \xmltag{keep_resonance_widths} must not be used for
pole-approximated calculations with radiating charged resonances at NLO
(see \refse{se:pole_approximation_nlo}).  Keeping a non-zero width at NLO would lead to
inconsistencies when charged resonances are projected on shell, since singularities
stemming from the associated propagators in the real squared amplitude are already accounted for by
dedicated subtraction counterterms \cite{Denner:2024tlu}.
\item Keeping a non-zero width for the resonances cures singularities in the phase-space integration
  but introduces a gauge dependence of the squared amplitude within the pole approximation. Even if in most cases
  this effect is expected to be negligible, its impact on final results should be verified on a process-by-process basis.
\item This option may change the value of the EW coupling, 
  if the latter is computed using complex masses for the W and Z bosons as inputs of the calculation (see option
\xmltag{masses_in_alpha_gf} in \refse{se:model_parameters}).
\end{itemize}
\end{warning}

{\sloppy
If a process in the pole approximation is protected at LO
against singularities from extra resonant propagators by the definition of the fiducial
phase space, so that problems can only appear at NLO, \mocanlo{} offers an alternative regularisation.
Its usage is controlled via the block \xmltag{fudge_factor_regularisation}.
Within this block, if \xmltag{fudge_factor} is set to \texttt{1}, squared amplitudes for
the \texttt{real} and \texttt{idip} contributions receive correction factors that regularise
all resonant propagators that appear in a given partonic process. The propagators are found by the code
by looping over all integration channels and collecting all resonant propagators that are not
projected on shell. For each propagator $i$ a factor $\mathcal{F}_i$ is constructed,
}
\begin{align}\label{eq:fudge_factor}
\mathcal{F}_i(s_i)=
\begin{cases}
  1 & \quad\text{if}\quad |\sqrt{s_i}-M_i|>c_{\mathcal{F}}\,\Gamma_i\\
  \frac{(s_i-M^2_i)^2}{(s_i-M^2_i)^2+\Gamma^2_iM^2_i} &  \quad\text{if}\quad |\sqrt{s_i}-M_i|<c_{\mathcal{F}}\,\Gamma_i
  \end{cases}\,,
\end{align}
where $s_i$ refers to the invariant mass of the unregularised resonance $i$ having a
pole mass $M_i$ and a width $\Gamma_i$. The full fudge factor results from the products of
the $\mathcal{F}_i$ for all propagators found by the algorithm. In the previous formula, the parameter
$c_{\mathcal{F}}$ restricts the range of application of the fudge factor. Its value can be controlled by the
input \xmltag{fudge_width_range} (default value \texttt{3.0}). Finally, even if the fudge-factor
regularisation is mostly useful to deal with resonance singularities appearing at NLO, the usage
of the fudge factor to regularise LO processes can also be enforced by setting \xmltag{fudge_factor_born}
to \texttt{1} (when also \xmltag{fudge_factor} has been set to \texttt{1}).

A block \xmltag{fudge_factor_regularisation} is illustrated in \refli{lst:on-shell-projection-2}. This
example must be read together with \refli{lst:pole_nlo_run}, which defines the
pole approximation for the process \vbsProcess. In the context of an NLO calculation (see
\refse{se:pole_approximation_nlo}), the section \xmltag{on_shell_projection} of \refli{lst:on-shell-projection-2} regularises the resonant propagators other than the ones projected on shell.
\begin{lstlisting}[language=XML,float, caption=Additional on-shell
  projection information in \texttt{proc\_card.xml} for the pole approximation defined in \refli{lst:pole_nlo_run}.,label=lst:on-shell-projection-2]
   <on_shell_projection id="wpwm_reg">
      <preserve_system>       cms_res  </preserve_system>
      <fudge_factor_regularisation>
         <fudge_factor>       1        </fudge_factor>
         <fudge_width_range>  4.0      </fudge_width_range>
      </fudge_factor_regularisation>
   </on_shell_projection>
\end{lstlisting}

As a final observation, it is worth noting that, while the use of pole approximations is supported along with the final-state merging of different partonic processes, all merged partonic processes must have the same resonance structure.
If this is not the case, the code will still compute the correct results but with a very low convergence rate.

\subsubsection{Pole approximation at NLO}
\label{se:pole_approximation_nlo}

While \mocanlo can deal with resonances in the pole approximation in full generality at LO,
only a restricted set of processes is currently supported in pole approximation at NLO accuracy.
Limitations essentially come from the treatment of the soft and collinear
singularities in the real-radiation contributions, which becomes more cumbersome compared to the
full off-shell description.

In particular, a fully-fledged NLO (QCD or EW) calculation in the pole
approximation is only possible under the following conditions:
\begin{itemize}
\item The process involves \emph{one or two weak bosons} as intermediate states, treated
  in the single- or double-pole approximation, respectively. Neither processes with three or more resonances, nor those with a nested
  resonance structure can be handled at NLO. Top quarks cannot be treated either in this regard.
\item The resonances undergo \emph{two-body leptonic decays} at LO, and the real-radiation contributions lead to resonances that
  decay  at most into three particles, \ie \emph{up to three-body decays in real corrections} at NLO.
  Hadronic decays of resonances are not supported.
\item  Additional limitations arise for
\emph{charged resonances}. Since an intermediate charged resonance can also radiate,
 dedicated counterterms
are required for a complete subtraction of singularities, when such a
resonance is projected on shell (for more details see \citere{Denner:2024tlu}).
\mocanlo only supports processes at NLO with leptonically-decaying $\PW$ bosons.
Colour-charged resonances like top quarks cannot be handled at NLO within the pole approximation.

\item As a final remark, a full treatment of \emph{non-factorisable
  corrections} \cite{Accomando:2004de, Dittmaier:2015bfe} is not supported. Even if both factorisable and non-factorisable corrections can be computed
  for the virtual contribution (as discussed in \refse{se:pole_approximation_nfact}), only factorisable ones are available for the real contribution.
  In any case, the impact of these corrections is known to be small if 
  both the real and virtual contributions are treated in the pole approximation, since in their sum non-factorisable
  corrections cancel in observables that are inclusive with respect to
  the decay products of the resonances \cite{Melnikov:1993np,Fadin:1993dz,Fadin:1993kt,Denner:1997ia,Denner:1998rh}.
  
\end{itemize}

An NLO calculation in the pole approximation for a given subprocess can be performed by running all relevant contributions 
(\texttt{born}, \texttt{virt}, \texttt{idip}, and \texttt{real}) with the linking tag \xmltag{resonances id="XX"}
pointing to the same XML resonances block in the \texttt{proc\_card.xml}, which selects a particular pole approximation,
as explained in \refse{se:pole_approximation}. Additionally, for the real part one needs to compute separate contributions 
where the radiation is emitted from the production and the decay of the $N$ on-shell-projected resonances. This means that
one is required to submit $N+1$ separate real runs for a given subprocess to obtain a complete NLO pole-approximated result. 
Note that the radiated particle for the \texttt{real} in the pole approximation is identified by comparing
  the \xmltag{partonic_process} and the \xmltag{real_process}: This is sufficient for the cases supported by \mocanlo{},
  where a real configuration only admits one underlying born.
  The information on the origin of the radiation is instead provided directly in the \texttt{run\_card.xml} as part of the subprocess definition,
as illustrated in \refli{lst:pole_nlo_run}. This is done using the tag \xmltag{osp_type} (default value \texttt{0}), 
which assumes the value \texttt{0} for the production contribution, or any number from \texttt{1} to \texttt{N}
for the decay contributions of the $N$ resonances. These numbers refer to the ordering in which resonances are listed
in the shared \xmltag{resonances id="..."} block.
\begin{lstlisting}[language=XML,float, caption=Example of real
  contributions in pole approximation in the \texttt{run\_card.xml}. ,captionpos=b,label=lst:pole_nlo_run]
<runs directory="vbs_wpwm">
   <run id="1">
      <type>            real         </type>
      <subdirectory>    ud/real      </subdirectory>
      <subprocess       id="ud"      />
      ...
      <osp_type> 0 <osp_type/>
      <resonances          id="wpwm" />
      <on_shell_projection id="wpwm" />
   </run>
   <run id="2">
      <type>            real         </type>
      <subdirectory>    ud/real      </subdirectory>
      <subprocess       id="ud"      />
      ...
      <osp_type> 1 <osp_type/>
      <resonances          id="wpwm" />
      <on_shell_projection id="wpwm" />
   </run>
   <run id="3">
      <type>            real         </type>
      <subdirectory>    ud/real      </subdirectory>
      <subprocess       id="ud"      />
      ...
      <osp_type> 2 <osp_type/>
      <resonances          id="wpwm" />
      <on_shell_projection id="wpwm" />
   </run>      
   ...
</runs>
\end{lstlisting}

A possible \texttt{proc\_card.xml} for the real contributions in \refli{lst:pole_nlo_run} is shown in \refli{lst:pole_nlo},
where a double-pole approximation for the process \vbsProcess{} is considered. Note that the
position of the radiation in the \xmltag{outgoing} list of \xmltag{real_process} (a photon in our example)
does not need to be contiguous to the decay products of the on-shell-projected resonances (the leptons in our example). 
For illustrative purposes, the example in \refli{lst:pole_nlo}
requires to preserve the total momentum of the two $\PW$ bosons that are projected
on shell via the \xmltag{preserve_system} tag. This choice is usually more natural than the default one for
specific applications, for example when studying vector-boson polarisations (see \refse{se:polarisations}).
\begin{lstlisting}[language=XML,float,caption=Examples of pole
  approximations for the process \vbsProcess. ,label=lst:pole_nlo]
   <subprocess id="ud">
      <partonic_process>
         <incoming> u d                 </incoming>
         <outgoing> e+ ve mu- vm~ u d   </outgoing>
         ...
      </partonic_process>
      <real_process>
         <incoming> u d                 </incoming>
         <outgoing> e+ ve mu- vm~ u d a </outgoing>
         ...
      </real_process>
   </subprocess>
   <on_shell_projection id="wpwm">
      <preserve_system>    cms_res  </preserve_system>
   </on_shell_projection>
   <resonances id="wpwm">
      <resonance>
         <particle>           w+    </particle>
         <external_legs>      3 4   </external_legs>
         <phasespace_only>    false </phasespace_only>
         <pole_approximation> true  </pole_approximation>
      </resonance>
      <resonance>
         <particle>           w-    </particle>
         <external_legs>      5 6   </external_legs>
         <phasespace_only>    false </phasespace_only>
         <pole_approximation> true  </pole_approximation>
      </resonance>
   </resonances>
\end{lstlisting}

\begin{lstlisting}[language=XML,float,caption=Run definition in
  \texttt{run\_card.xml}. ,label=lst:non_fact]
<run id="1">
   <type>            vnfc  </type>
   <subdirectory> uxu/vnfc </subdirectory>
   <subprocess       id="uxu"/>
   <run_parameters   id="eex_nlo"/>
   <model_parameters id="eex_nlo"/>
   <cuts             id="eex"/>
   <recombinations   id="eex"/>
   <resonances       id="w_psp"/>
   <histograms       id="none"/>
</run>
\end{lstlisting}

\paragraph{Further remarks on the single-pole approximation}\label{par:spa}
The general pole-approximation algorithm implemented in \mocanlo{} \cite{Denner:2024xul} covers also calculations in the
single-pole approximation, up to special treatments needed for a few cases.
\begin{itemize}
\item Two-to-two process: at variance with the general case,
  the on-shell projection rescales initial-state momenta
  and (when computing NLO corrections) the final-state momentum of the radiation,
  if this is not part of the decay products of the resonance of interest.
  The rescaling factor is equal to the ratio of the pole mass over the  invariant mass of the resonance. 
  Since in this case the on-shell projection has to modify the incoming momenta, it cannot
  even preserve the partonic CM energy. Therefore, 
  the input provided via the tag \xmltag{preserve_system} is ignored.
\item Single resonance with at least one additional massive
  external particle: the pole approximation
  proceeds as in the general case, but when evaluating a real contribution to the production,
  \ie \texttt{osp\_type}${}={}$\texttt{0}, the momentum of a
  bremsstrahlung particle (a photon or a gluon) is never included in the preserved system
  (in contrast to the default behaviour, if \texttt{cms} was chosen), \ie the momentum of the radiation is never modified.
  For real corrections to the decays, no change in the algorithm is
  needed instead.
\end{itemize}
Note that the pole approximation for a single intermediate charged resonance produced
in association with massless particles is ill defined close to threshold, unless the resonance's invariant mass is constrained by kinematic cuts
preventing it from becoming arbitrarily close to the pole mass. Consequently, processes with a single $\PW$ boson
or a top quark in association with massless particles
are not supported by \mocanlo{}, while for instance cases with a single $\PZ$ boson decaying into charged-fermion pairs
are supported and treated as in the general case.

\subsubsection{Virtual non-factorisable corrections}
\label{se:pole_approximation_nfact}

Within the restrictions discussed in \refse{se:pole_approximation_nlo}, \mocanlo allows to compute all contributions
(\ie \texttt{born}, \texttt{virt}, \texttt{idip}, and \texttt{real}) in the pole approximation, including only factorisable corrections
for the virtual and real terms. This provides a consistent NLO result in this approximation
and is also the only way to define a polarised cross section, as described in \refse{se:polarisations}.
However, the pole approximation can also be used to speed up the
computation of virtual corrections, whose evaluation is particularly expensive. 
A solution, which was adopted for instance in \citeres{Denner:2000bj,Accomando:2004de}, is to compute all contributions
entering an NLO calculation exactly, with the exception of the virtual one, where a pole approximation
was used. This requires in \mocanlo to run for each subprocess a \texttt{virt} and a \texttt{vfnc} run type
both specifying a \xmltag{resonances id="XX"} section in the \texttt{run\_card.xml}.
If both factorisable and non-factorisable virtual corrections are accounted for, their IR singularities must be
cancelled applying the 
on-shell projection to the terms containing the $I$ operator in the integrated dipole contribution
in the same way. The $P$- and $K$-operator terms, on the other hand, must be evaluated with the off-shell kinematics like the
real corrections. This introduces a mismatch that is of the order of the intrinsic error
of the pole approximation. A proper evaluation of the $I$ operator, in line with our discussion,
can be achieved in \mocanlo by replacing each \texttt{idip}
contribution for a specific subprocess by an \texttt{idpi} and an \texttt{idpk} run pointing to
the same subprocess. The \texttt{idpi} runs should specify a \xmltag{resonances id="XX"} section
with \texttt{pole\_approximation=true},
while the \texttt{idpk} ones must be run with off-shell kinematics.
Examples of how to perform these kind of computations are provided in
the folder \texttt{validated\_processes}, in particular the processes in the
double-pole approximation in the \texttt{ttbar/pp\_tt} folder.
It is worth noting that while the implementation of the
non-factorisable corrections closely follows
\citere{Denner:1997ia,Accomando:2004de,Dittmaier:2015bfe} and is
therefore very general (arbitrary number of non-nested resonances), it
has only been validated for double-pole approximations involving two identical resonant particles~\cite{Denner:2016jyo,Denner:2016wet,Biedermann:2016yds,Denner:2020orv}.
\begin{warning}{Warning}
  The computation of the non-factorisable virtual corrections is limited to processes without nested resonances.
  Moreover, processes with a single charged resonance in association
  with massless particles in the final state are also not allowed,
  since the pole approximation is ill defined in these cases.
\end{warning}

\subsubsection{Selection of polarisations}
\label{se:polarisations}
 \mocanlo can compute cross sections for internal polarised vector bosons at NLO accuracy.
 The definition of polarised cross sections is based on the pole approximation, as described
 for instance in \citere{Denner:2020bcz}. 
 As a consequence, polarised results are only available for processes that can be calculated
 in this approximation (see discussion at the beginning of \refse{se:pole_approximation_nlo}).
 
 To run simulations for polarised vector bosons,
 one must add to each resonance block, where the pole approximation has been enabled,
 the required polarisation state via the tag \xmltag{polarization_state}.
 Allowed values are reported in \refta{tab:pol_states}. In the example in \refli{lst:pole_polarised},
 two on-shell $\PW$ bosons having transverse and longitudinal polarisations are required.
\begin{table}
      \centering
       \rowcolors{2}{tablerowcolor}{}
      \begin{tabularx}{0.7\textwidth}{cX}
            \toprule
            \xmltag{polarization_state}    & \textbf{Description} \\
            \midrule
            \texttt{-1} / \texttt{+1} &  Left- and right-handed \\
            \texttt{0} &   Longitudinal \\
            \texttt{T} &   Transverse \\
            \texttt{U} &  Unpolarised (default) \\
            \bottomrule
      \end{tabularx}
      \caption{\label{tab:pol_states} Available polarisation states.}
\end{table}
\begin{lstlisting}[language=XML,float,caption=Examples of pole
  approximations for the process \vbsProcess. ,label=lst:pole_polarised]
<on_shell_projection id="ww_osp">
   <poldef_system> cms_res  </poldef_system> 
</on_shell_projection>
<resonances id="ww_psp">
   <resonance>
      <particle>           w+    </particle>
      <external_legs>      3 4   </external_legs>
      <phasespace_only>    false </phasespace_only>
      <pole_approximation> true  </pole_approximation>
      <polarization_state> T     </polarization_state>
   </resonance>
   <resonance>
      <particle>           w-    </particle>
      <external_legs>      5 6   </external_legs>
      <phasespace_only>    false </phasespace_only>
      <pole_approximation> true  </pole_approximation>
      <polarization_state> 0     </polarization_state>
   </resonance>
</resonances>
\end{lstlisting}

The definition of polarisation states is Lorentz-frame dependent, and consequently the result
for the polarised cross sections too. By default, \mocanlo
defines polarisations in the same frame as requested in \xmltag{preserve_system} (see \refse{se:pole_approximation}).
Nonetheless, a different reference frame for the polarised states can be set in the XML section 
\xmltag{on_shell_projection id="..."} using the \xmltag{poldef_system} tag. In \refli{lst:pole_polarised}, 
the pole approximation preserves the partonic CM energy of the process (default when \xmltag{preserve_system}
is not specified), but computes polarisations in the CM
frame of the two vector bosons. The possible input values for \xmltag{poldef_system} (see \refta{tab:on_shell_projection})
are those of \xmltag{preserve_system} and the value \texttt{lab}, which allows to define
polarisations in the laboratory frame.
Note that this is the default frame used for the definition
of polarisations in case of a single-pole approximation.

\section[Specifying the simulation parameters: the \texttt{\large param\_card.xml}]
{Specifying the simulation parameters: the \texttt{\Large param\_card.xml}}
\label{se:parameters}

{
The parameter card \texttt{param\_card.xml} contains two types of
sections. These are labelled \linebreak \xmltag{model\_parameters id="..."}, 
where parameters of the particle model are given, and \linebreak
\xmltag{run\_parameters id="..."}, which specifies parameters 
for the MC run. The contents of these two section types are discussed in the following.}

{
A single \texttt{param\_card.xml} can contain several subsections of the same type, each with a unique \texttt{id} 
attribute that is used for linking in the \texttt{run\_card.xml} (see \refse{se:ProcessSetup}). This can be used to 
include different parameter sets, \eg for LO and NLO calculations. The exemplary 
\refli{lst:model_parameters} contains the NLO value of the top-quark width $\Gamma_\Pt$ for the NLO calculation of 
the process \fullProcess and is thus denoted with the attribute \texttt{id="ttx\_nlo"}. An additional section 
\xmltag{model_parameters id="ttx\_lo"} containing the LO value of $\Gamma_\Pt$ could be present in \texttt{param\_card.xml}. 
Similarly, different sections \xmltag{run_parameters id="ttx\_nlo"} and
\xmltag{run_parameters id="ttx\_lo"} could point to the usage of NLO and LO PDF sets, respectively.}

\subsection{Model parameters}
\label{se:model_parameters}

All parameters available in the \xmltag{model\_parameters} section of
\texttt{param\_card.xml} are summarised in
\reftas{tab:model_parameters} and \ref{tab:model_parameters2}
together with their default values, and
\refli{lst:model_parameters} shows an example of their use in
\texttt{param\_card.xml}. 

\begin{table}
    \centering
    \renewcommand\arraystretch{1.2}
    \rowcolors{2}{tablerowcolor}{}
    \begin{tabular}{lll}
        \toprule\rowcolor{tableheadcolor} 
        \textbf{parameter} & \textbf{XML element} & \textbf{default value} \\
        \midrule
        $m_\Pt$           & \xmltag{top_mass}                 & $172.0 $   \\
        $\Gamma_\Pt$      & \xmltag{top_width}                & $1.4   $   \\
        $m_\Pb$           & \xmltag{bottom_mass}              & $0.0   $   \\
        $\Gamma_\Pb$      & \xmltag{bottom_width}             & $0.0   $   \\
        $m_\tau$          & \xmltag{tau_mass}                 & $0.0   $   \\
        $m_\mu$           & \xmltag{muon_mass}                & $0.1056584$  \\
        $m_\Pe$           & \xmltag{electron_mass}            & $0.51099895\times10^{-3}$ \\
        $M_{\PZ}^\mathrm{OS}$    & \xmltag{ons_z0_mass}       & $91.1880  $  \\
        $\Gamma_\PZ^\mathrm{OS}$ & \xmltag{ons_z0_width}      & $2.4955   $  \\
        $M_\PW^\mathrm{OS}$      & \xmltag{ons_w_mass}        & $80.3692  $  \\
        $\Gamma_\PW^\mathrm{OS}$ & \xmltag{ons_w_width}       & $2.085    $  \\
        $M_\PZ$           & \xmltag{z0_mass}                  & calculated   \\
        $\Gamma_\PZ$      & \xmltag{z0_width}                 & calculated   \\
        $M_\PW$           & \xmltag{w_mass}                   & calculated   \\
        $\Gamma_\PW$      & \xmltag{w_width}                  & calculated   \\
        $M_\PH$           & \xmltag{higgs_mass}               & $125.0    $  \\
        $\Gamma_\PH$      & \xmltag{higgs_width}              & $4.07\times10^{-3}$\\
        $G_\text{F}$      & \xmltag{fermi_constant}           & $1.166390\times10^{-5}$\\
        --                & \xmltag{masses_in_alpha_gf}       & $0       $   \\
        --                & \xmltag{scheme_alpha}             & \text{gf}    \\
        $\alpha_{\GF}$    & \xmltag{alphagf}                  & $7.555310522369\times10^{-3}$ \\
        $\alpha(0)$       & \xmltag{alpha0}                   & $7.2973525\times10^{-3}$  \\
        $\alpha(\MZ^2)$   & \xmltag{alphaz}                   & $7.7983287\times10^{-3}$  \\
 $\alpha_{\MSbar}(Q_0^2)$ & \xmltag{alpha_starting_value}     & $7.8609176\times10^{-3}$  \\
        $Q_0$             & \xmltag{alpha_starting_scale}     & $91.188$ \\
        $\alphas$         & \xmltag{alphas}                   & none \\
        --                & \xmltag{alphas_running}           & $1$ if hadrons in initial state, $2$  otherwise  \\
        $\alphas(Q^2_0)$  & \xmltag{alphas_starting_value}    & $0.118  $   \\
        $Q_0$             & \xmltag{alphas_starting_scale}    & $91.188$    \\
        --                & \xmltag{n_loops}                  & $2      $   \\
        \mur    & \xmltag{renormalization_scale}    & \muf or $M_\PZ$ \\
        \muf    & \xmltag{factorization_scale}      & \mur or $M_\PZ$ \\
        $\mu_\text{IR}$   & \xmltag{infrared_scale}           & $100.0      $   \\
        \midrule
        --                & \xmltag{neutrino_species_tagging}  & $0      $   \\
        --                & \xmltag{redefine_jet_types}        & ---  \\        
        \bottomrule
    \end{tabular}
    \caption{\label{tab:model_parameters} Elements of the \xmltag{model\_parameters} section of 
             \texttt{param\_card.xml} and their defaults.}
\end{table}
\begin{table}
    \centering
    \renewcommand\arraystretch{1.2}
    \rowcolors{2}{tablerowcolor}{}
    \begin{tabular}{lll}
        \toprule\rowcolor{tableheadcolor} 
        \textbf{parameter} & \textbf{XML element} & \textbf{default value} \\
        \midrule
        $m_\text{reg}$    & \xmltag{massless_mapping_mass}     & $10^{-6}     $   \\
        $p$               & \xmltag{massless_mapping_power}    & $0.9         $   \\
        $q$               & \xmltag{subtraction_mapping_power} & $0.8         $   \\
        \bottomrule
    \end{tabular}
    \caption{\label{tab:model_parameters2} Additional technical
      parameters of the \xmltag{model\_parameters} section of
             \texttt{param\_card.xml} and their defaults.}
\end{table}
\begin{figure}
\begin{lstlisting}[language=XML,caption=Model parameters in
  \texttt{param\_card.xml}. ,label=lst:model_parameters]
<model_parameters id="ttx_nlo">
    <top_mass>                 172.0              </top_mass>
    <top_width>                  1.31670          </top_width>
    <z0_mass>                   91.1876           </z0_mass>
    <z0_width>                   2.5096596343     </z0_width>
    <w_mass>                    80.399            </w_mass>
    <w_width>                    2.0997360974     </w_width>
    <higgs_mass>               120.0              </higgs_mass>
    <higgs_width>                0.0              </higgs_width>
    <scheme_alpha>               GF               </scheme_alpha>
    <masses_in_alpha_gf>        -1                </masses_in_alpha_gf>
    <alphagf>                    7.555786d-03     </alphagf>
    <renormalization_scale>    172.0              </renormalization_scale>
    <factorization_scale>      172.0              </factorization_scale>
    <infrared_scale>           172.0              </infrared_scale>
    <neutrino_species_tagging>   1                </neutrino_species_tagging>
    <redefine_jet_types>
        <jet_particle>         b b~               </jet_particle>
        <jet_type>             2 27               </jet_type>
    </redefine_jet_types>
</model_parameters>
\end{lstlisting}
\end{figure}

Masses and widths of particles in \GeV are set via
\xmltag{particle\_mass} and \xmltag{particle\_width}, respectively.
The electron and muon masses are only used as regulators of
initial-state singularities for incoming electrons/positrons or muons
and are considered as massless otherwise. The four light quarks are
considered as massless throughout. While the masses of the bottom quark
and the $\tau$ lepton are zero by default, finite masses are allowed
as long as these particles do not appear as radiating particles in NLO
calculations.

\mocanlo uses consistently pole masses $M$ and widths $\Gamma$, which
are related to the complex poles $\mu$ via
$$
\mu^2 = M^2 -\ri M \Gamma.
$$
Since for the weak vector bosons usually the on-shell masses
$M^\mathrm{OS}$ and widths $\Gamma^\mathrm{OS}$
are quoted, these can be given alternatively using the input tags
\xmltag{ons_z0_mass}, \xmltag{ons_z0_width}, \xmltag{ons_w_mass}, and
\xmltag{ons_w_width}. From those, the pole masses and widths are
calculated via \cite{Bardin:1988xt} 
$$
  M = \frac{M^\mathrm{OS}}{\sqrt{1+(\Gamma^\mathrm{OS}/M^\mathrm{OS})^2}}
  \quad
  \quad\text{and}
  \quad
  \quad
  \Gamma = \frac{\Gamma^\mathrm{OS}}{\sqrt{1+(\Gamma^\mathrm{OS}/M^\mathrm{OS})^2}}.
$$
If no input values are provided, \mocanlo determines the pole masses
from default on-shell masses. Therefore, for weak vector bosons, pole masses and widths
are always recomputed by \mocanlo, unless directly specified via input. Note that
pole and on-shell values can not be provided together in the \texttt{param\_card.xml}.
If so, the code stops, asking to correct the input.

In \mocanlo, by default the $G_\text{F}$ scheme is used for the
renormalisation of the EW coupling $\alpha$. For
\xmltag{masses_in_alpha_gf}${}=-1$, the input value \xmltag{alphagf} is
used to set $\alpha$. Otherwise, two cases are possible, where the value of $\alpha$ is
derived from the value of $G_\text{F}$, set via
\xmltag{fermi_constant}, and from the masses of the W and Z bosons.
For \xmltag{masses_in_alpha_gf}${}=0$ (default), $\alpha$ is calculated with real
masses via
$$
\alpha = \frac{\sqrt{2}\GF}{\pi}\Re{\MW^2}\left(1-\frac{\Re{\MW^2}}{\Re{\MZ^2}}\right),
$$
while for \xmltag{masses_in_alpha_gf}${}=1$
complex masses are used according to
$$
\alpha = \frac{\sqrt{2}\GF}{\pi}\left|{\MW^2}\left(1-\frac{{\MW^2}}{{\MZ^2}}\right)\right|.
$$

Additional input schemes for $\alpha$ are available and can be chosen
using the input tag \xmltag{scheme_alpha}. For \xmltag{scheme_alpha}${}=1$ or
\xmltag{scheme_alpha}${}=\texttt{gf}$ the \GF~scheme is selected as
discussed above.  For \xmltag{scheme_alpha}${}=2$ or
\xmltag{scheme_alpha}${}=\texttt{alpha0}$, the input value
\xmltag{alpha0} is used to set $\alpha$, while for
\xmltag{scheme_alpha}${}=3$ or
\xmltag{scheme_alpha}${}=\texttt{alphaz}$, the input value
\xmltag{alphaz} is employed. Finally, for \xmltag{scheme_alpha}${}=4$
or \xmltag{scheme_alpha}${}=\texttt{alphamsbar}$, the \MSbar\ scheme
is chosen
with $\alpha$ set to \xmltag{alpha_starting_value} and the
corresponding scale set to \xmltag{alpha_starting_scale}.
The choice of \xmltag{scheme_alpha} fixes the input value of
$\alpha$ and its renormalisation. For instance, in the \MSbar\ scheme
the counterterm in Eq.~(427) of \citere{Denner:2019vbn} is used.
More details on the renormalisation of $\alpha$ can be found in the
\recola manual \cite{Actis:2016mpe}.
Note that the use of differently renormalised couplings for external
photons is steered via the parameters \xmltag{tree\_e0\_order} in \texttt{proc\_card.xml}.

In \mocanlo, the value of the strong coupling constant $\alphas$ can be either provided by the user or calculated using its running.
A fixed value of $\alphas$ can be set using the tag \xmltag{alphas}. The corresponding renormalisation scale \mur is then assumed to 
be fixed and must be set via the tag \xmltag{renormalization_scale}. Even when using a fixed $\alphas$ value, a running of this coupling is 
required if a scale variation is to be performed (see \refse{subsubsection:run_parameters}). For this reason, the value of 
\xmltag{renormalization_scale} corresponding to the given value
\xmltag{alphas} must be specified. 

In general, \mocanlo provides two options for the running of $\alphas$, controlled by the 
run parameter \xmltag{alphas_running} (see \refse{subsubsection:run_parameters}). If \xmltag{alphas_running} is set to 1, 
the running provided by LHAPDF is selected. If \xmltag{alphas_running} is set to 2, the running provided by \recola is instead used. 
In the latter case, a starting scale $Q_0$ and the value of the coupling $\alpha_\mathrm{s}(Q^2_0)$ at that scale are required
and can be specified via the parameters \xmltag{alphas_starting_scale} and \xmltag{alphas_starting_value}.
The running of the coupling also depends on the loop order of the running, fixed via the parameter \xmltag{n_loops}, 
which can be 1 or 2. Note that \mocanlo uses the five-flavour scheme.

As noted before, the running of the coupling is required even for a fixed scale if a scale variation is performed. Therefore, also in this
case a value has to be given for \xmltag{alphas_running}.
The default value of \xmltag{alphas_running} depends on the parameters \xmltag{beam1} and \xmltag{beam2} (see \refse{subsubsection:run_parameters}). 
If at least one of the beams is a proton beam, \xmltag{alphas_running} defaults to 1, otherwise it is set to 2.

If fixed renormalisation and factorisation scales \mur and \muf are employed, their values must be passed 
using the tags \xmltag{renormalization_scale} and \xmltag{factorization_scale}. If only one of the two scales is set by the user, \mocanlo 
will automatically set both to the given value. If none of the two is set, they will both be assigned the value $M_\PZ$. 
The choice of a fixed or dynamical scale is determined by the run parameter \xmltag{dynamical_scale_type} (see \refse{subsubsection:run_parameters}).
Note that the IR scale $\mu_\text{IR}$ used for the computation of virtual corrections and integrated dipoles
is not assumed to be equal to any of the previously defined scales in \mocanlo and can be independently set via the  \xmltag{infrared_scale} input.
Since the result must not depend on $\mu_\text{IR}$, varying this parameter offers a useful consistency check on the calculation.

The technical parameters $m_\text{reg}$, $p$, and $q$ enter the
mappings of massless propagators and singularities in the subtraction
terms in the MC integration. Their values can be adjusted via the tags \xmltag{massless_mapping_mass}, 
\xmltag{massless_mapping_power}, and \xmltag{subtraction_mapping_power}, respectively. 
These influence the integration performance and may be tuned, but $m_\text{reg}$ should be smaller than the lowest 
physical scale of the process by several orders of magnitude, and $p,q \lesssim 1$ should hold.  

Setting \xmltag{neutrino_species_tagging} to 1 allows to switch on
different jet types for neutrino species (see
\refta{tab:ParticleNames} in \refse{se:subprocesses}), which can be
useful for tagging or producing histograms based on specific neutrino
truth momenta. If this parameter is set to 0, all neutrinos have the
same jet type (6).

The subsection \xmltag{redefine_jet_types} allows to redefine the
default \xmltag{jet_type} of a subset of particles. The jet types in
the list \xmltag{jet_type} \ldots \xmltag{/jet_type} become the jet
types of the corresponding particles in the list \xmltag{jet_particle} \ldots
\xmltag{/jet_particle}. In the example in
\refli{lst:model_parameters}, the jet type of the antibottom quark is set to
27, which allows to recombine and tag them independently of
the bottom quark and to plot corresponding observables. As a further example,
the jet types of bottom and antibottom quarks or of the photon can be set to
1 if these shall be treated as light jets.

\subsection{Run parameters}
\label{subsubsection:run_parameters}

The section \xmltag{run_parameters} of \texttt{param\_card.xml}
defines several parameters controlling the MC run.  These parameters
together with their possible input values are listed in
\reftas{tab:run_parameters1} and \ref{tab:run_parameters2}.
\begin{table}
  \centering
  \renewcommand\arraystretch{1.2}
  \rowcolors{2}{tablerowcolor}{}
  \begin{tabular}{lll}
    \toprule\rowcolor{tableheadcolor} 
    \textbf{parameter} & \textbf{possible values} & \textbf{default value} \\
    \midrule
    \xmltag{energy}                     &                           & (group) \\
    \xmltag{cms_beam_energy}            &  positive real number     & none  \\
    \xmltag{energy_beam1}               &  positive real number     & none  \\
    \xmltag{energy_beam2}               &  positive real number     & none  \\
    \xmltag{scale}                      &                           & (group) \\
    \xmltag{dynamical_scale_type}       &  $0,1$                    & $~0$  \\
    \xmltag{scale_factors_f}            &  list of real numbers     & $1.0$  \\
    \xmltag{scale_factors_r}            &  list of real numbers     & $1.0$  \\
    \xmltag{n_scales_for_histograms}    &  integer                  & maximum \# of scale factors\\
    \xmltag{pdfs}                       &                           & (group)  \\
    \xmltag{beam1}                      &  \texttt{p,e+,e-,mu+,mu-} & ~\texttt{p}  \\
    \xmltag{beam2}                      &  \texttt{p,e+,e-,mu+,mu-} & ~\texttt{p}  \\
    \xmltag{pdf_set}                    &  see text                 & ---     \\
    \xmltag{pdf_set2}                   &  see text                 & ---     \\
    \xmltag{alphas_running}             &  $1,2$                    & see text  \\
    \xmltag{pdf_scheme}                 &  string                   & see text  \\
    \xmltag{pdf_scheme2}                &  string                   & see text  \\
    \xmltag{pdf_starting_scale}         &  positive real number     & $0.511\times10^{-3}$ \\
    \xmltag{helicity_em}                &  $-1,0,1$                 & $~0$ \\
    \xmltag{helicity_ep}                &  $-1,0,1$                 & $~0$ \\
    \xmltag{dipoles}                    &                           & (group)  \\
    \xmltag{alpha_fs_fs}                &  real $\in(0, 1]$         & $1.0$ \\
    \xmltag{alpha_fs_is}                &  real $\in(0, 1]$         & $1.0$ \\
    \xmltag{alpha_is_fs}                &  real $\in(0, 1]$         & $1.0$ \\
    \xmltag{alpha_is_is}                &  real $\in(0, 1]$         & $1.0$ \\
    \xmltag{DIS_scheme}                 &  $0,1$                    & $~0$  \\
    \xmltag{integration_termination}    &                           & (group)  \\
    \xmltag{n_target_accepted_events}   &  integer $\ge$ 1000       & $10^{15}$ \\
    \xmltag{max_generated_events}       &  integer                  & $10^{15}$ \\
    \xmltag{max_vanishing_events}       &  integer                  & $100 000$ \\
    \xmltag{target_relative_precision}  &  integer                  & $~3$  \\
    \xmltag{target_absolute_precision}  &  integer                  & $12$  \\
    \xmltag{time_wall}                  &  \small\texttt{days-hours:minutes}& ---  \\
    \bottomrule
  \end{tabular}
  \caption{\label{tab:run_parameters1} 
 Parameters of the \xmltag{run\_parameters} section of
 \texttt{param\_card.xml} that fix the kind of calculation and their defaults.}
\end{table}
In \refli{lst:run_parameters}, run parameters for the NLO calculation
of \fullProcess are displayed as an example.
\begin{lstlisting}[language=XML,caption=MC run parameters in
  \texttt{param\_card.xml}. ,label=lst:run_parameters]
<run_parameters id="ttx_nlo">
    <output_level>    4     </output_level>
    <cms_beam_energy> 8000  </cms_beam_energy>
    <scale>
        <dynamical_scale_type> 1       </dynamical_scale_type>
        <scale_factors_f>      1 0.5 2 </scale_factors_f>
    </scale>
    <pdfs>
        <pdf_set>          MSTW2008lo90cl </pdf_set>
        <pdf_mapping_emin> 120d0          </pdf_mapping_emin> 
    </pdfs>
    <integration_termination>
        <n_target_accepted_events>  1 000 000 000 </n_target_accepted_events>
        <target_relative_precision> 5             </target_relative_precision>
        <target_absolute_precision> 5             </target_absolute_precision>
        <time_wall>                 1-07:30       </time_wall>
    </integration_termination>
    <dipoles>
        <alpha_fs_fs> 1d-2 </alpha_fs_fs>
        <alpha_fs_is> 1d-2 </alpha_fs_is>
        <alpha_is_fs> 1d-2 </alpha_is_fs>
        <alpha_is_is> 1d-2 </alpha_is_is>
    </dipoles>
</run_parameters>
\end{lstlisting}

{\sloppy
The run parameters can be grouped in \texttt{param\_card.xml} into the
subsections labelled \xmltag{energy}, \xmltag{scale}, \xmltag{pdfs}, \xmltag{dipoles}, \xmltag{integration_termination},
\xmltag{output}, \xmltag{intermediate_output}, \linebreak
\xmltag{phase_space_generator}, and \xmltag{running_options}, but
these tags are ignored by \mocanlo.
}

Table \ref{tab:run_parameters1} summarises the parameters that are
needed to define the run. 
\begin{description}
\item[\texttt{<energy>}] The collider CMS energy is defined in
    \xmltag{cms_beam_energy} in GeV.  If this parameter is given, this
    energy is assumed to be equally distributed between both beams. For
    asymmetric colliders, the energies of the beams can be defined
    with \xmltag{energy_beam1} and \xmltag{energy_beam2} in GeV.
\item[\texttt{<scale>}] \sloppy This subsection contains parameters
  related to the renormalisation and factorisation scales \mur and
  \muf.  The parameter \xmltag{dynamical_scale_type} fixes the type of
  scales to be used. For \xmltag{dynamical_scale_type}${}=0$, fixed
  scales are chosen, whose values are passed via
  \xmltag{renormalization_scale} and \xmltag{factorization_scale} in
  the section \xmltag{model\_parameters}.
  Dynamical scales can be chosen by setting
  \xmltag{dynamical_scale_type}${}=1$.  The definition of the scales
  itself must be specified in the file \texttt{cut\_card.xml} (see
  \refse{se:cut-scheme} for details).

  The scales \muf and \mur can be independently re-scaled to 
  perform an estimation of the scale uncertainty. The scaling factors are passed as 
  lists of space-separated real numbers via the tags \xmltag{scale_factors_f} and \xmltag{scale_factors_r} 
  for \muf and \mur, respectively. The matrix elements and the PDFs are computed using the
  values of \muf and \mur as specified above times the first entry of 
  \xmltag{scale_factors_f} and \xmltag{scale_factors_r},
  respectively. The first entry of each list corresponds to the central scale. 
  Note that the runtime does not grow proportional to the number of scale factors.
  If the tags \xmltag{scale_factors_f} and \xmltag{scale_factors_r} are not provided, 
  a single scaling factor equal to 1 is used for both scales. If only one list of factors 
  is provided or if the lists have different lengths, the shorter list is extended to the length 
  of the longer list using additional entries equal to 1.
  When a scale variation is performed, the results of each scale
  factor are written to different directories 
  (see \refse{se:QuickStart}), whose names correspond to the position of the factors in the lists 
  \xmltag{scale_factors_f} and \xmltag{scale_factors_r}. 

  The integer \xmltag{n_scales_for_histograms} 
  defines for how many scale factors the histograms are computed at
  most (see \refse{se:histograms}),  and its default value is the larger
  of the number of scales given in the lists \xmltag{scale_factors_f} and \xmltag{scale_factors_r}.
  Histograms are produced for the first \xmltag{n_scales_for_histograms} factors in the lists \xmltag{scale_factors_f} 
  and \xmltag{scale_factors_r}. If \xmltag{n_scales_for_histograms} is larger than the length of the lists, 
  the run stops.

\item[\texttt{<pdfs>}]  \sloppy This subsection contains parameters related to the incoming particles and the employed PDFs. 
 The type of the beams is fixed via the input tags \xmltag{beam1}  and \xmltag{beam2}, which can take the values 
 \texttt{p}, \texttt{e+}, \texttt{e-}, \texttt{mu+} or \texttt{mu-}.
  By default, proton (\texttt{p}) beams are assumed.

  The PDF set is chosen via the tag \xmltag{pdf_set}.
  For proton--proton collisions, the LHAPDF library is used (see \refse{se:Installation}). 
  Different PDF sets for the second beam can be used by setting
  \xmltag{pdf_set2} to a value different from \xmltag{pdf_set}.  
  This way, proton--lepton collisions can be simulated. 
  If \xmltag{pdf_set2} is not given, it is assumed to be equal to \xmltag{pdf_set}.
    
\begin{warning}{Warning}
\mocanlo has not been thoroughly tested for proton--lepton collisions.
\end{warning}

  The tag \xmltag{pdf_scheme} fixes the subtraction scheme for the IR
  singularities in the collinear counterterms and should match the IR subtraction
  of the PDFs. Different schemes imply different finite contributions in the
  collinear counterterms. By default \xmltag{pdf_scheme} is set to \texttt{MSbar}, which is the
  only available setting for proton PDFs in \mocanlo. 
  Different settings are instead available for
  lepton PDFs, as discussed below.

  The strong coupling constant can either be taken from the PDFs or be calculated by \recola:
  This behaviour is controlled by the parameter \xmltag{alphas_running} in the \xmltag{model\_parameters}
  of the \texttt{param\_card.xml} (see \refse{se:model_parameters}).
  If \xmltag{alphas_running} is set to 1, $\alpha_\mathrm{s}$ is obtained from the PDF set,
  if it is set to 2, $\alpha_\mathrm{s}$ is calculated by \recola. 
  The default value of \xmltag{alphas_running} depends on the parameters \xmltag{beam1} 
  and \xmltag{beam2}. If at least one of the beams is a proton beam, \xmltag{alphas_running} 
  defaults to 1, otherwise it is set to 2.
  
  To simulate lepton--lepton collisions, the tag \xmltag{pdf_set} can be set to
  \texttt{none}. In this case, no PDFs are used and the initial-state
  collinear singularities are regularised with the electron or muon
  masses (see \refse{se:model_parameters}).  Alternatively, \mocanlo
  provides an internal implementation of LO lepton PDFs according to Appendix A of
  \citere{Bertone:2022ktl} including the $\order{\alpha^2}$ and
  $\order{\alpha^3}$ contributions as given in Appendix A of
  \citere{Beenakker:1996kt}. Following the nomenclature in \citere{Bertone:2022ktl},
  \xmltag{pdf_set} can be set to \texttt{LO\_beta}, \texttt{LO\_eta},
  \texttt{LO\_mixed}, and \texttt{LO\_collinear}. The value of
  \xmltag{pdf_scheme} must be chosen accordingly and set to
  \texttt{beta}, \texttt{eta}, \texttt{mixed}, and
  \texttt{collinear}, respectively.
  The starting scale $\mu_0$ on which the  \texttt{LO\_collinear} 
  case depends can be set via the input tag \xmltag{pdf_starting_scale}, which 
  has the default value $0.511\times10^{-3}$.

\begin{warning}{Warning}
  We note that lepton PDFs have not been thoroughly tested.
\end{warning}
  
  The parameters \xmltag{helicity_em} and \xmltag{helicity_ep} define
  the helicities of the incoming electrons and positrons,
  respectively. They can take the values $\pm1$ for polarised beams. A
  value $0$ corresponds to an unpolarised beam.

\item[\texttt{<dipoles>}] This section is only relevant for the real and integrated-dipole contributions. Here, the so-called $\alpha$-parameters 
  that control the phase-space volume of the subtraction and integrated dipoles can be set (see \citere{Nagy:2003tz}). 
  The $\alpha$-parameters can be independently chosen using different input tags
  for some classes of emitter--emissus dipoles, specifically  for
  final-state--final-state (\xmltag{alpha_fs_fs}), 
  final-state--initial-state (\xmltag{alpha_fs_is}), 
  initial-state--final-state (\xmltag{alpha_is_fs}), and 
  initial-state--initial-state (\xmltag{alpha_is_is}) dipoles.
  The $\alpha$-parameters can take values in the interval $(0,1]$ and their common default value is 1. 
  Note that no $\alpha$-parameters are available for subtraction dipoles
  related to internal resonances in the pole approximation.
  
  Corresponding real and integrated-dipole contributions must be calculated using the same value for the $\alpha$-parameter, and the sum of these 
  contributions must be independent of the specific value. Thus, the correctness of the subtraction procedure can be tested by varying the $\alpha$-parameters.
  
A further parameter controls the subtraction of collinear singularities.
By choosing \xmltag{DIS_scheme}${}=1$, the DIS factorisation scheme is
switched on for photon PDFs, while the default is to use the \MSbar\ 
factorisation scheme throughout. 
The flag \xmltag{DIS_scheme} has no effect for strongly interacting particles.

\item[\texttt{<integration\_termination>}] This section allows to set conditions for a regular termination of the integration. 
  The integration terminates upon fulfilling at least one of the
  following conditions. 
  The integration stops once the number of accumulated accepted,
  generated, or vanishing events exceeds the values given via
  the parameters \xmltag{n_target_accepted_events}, \xmltag{max_generated_events}, and \xmltag{max_vanishing_events}. 
  Here, an event is vanishing when its weight is 0,
  while we refer to accepted events as those that pass the cuts (see \refse{se:cut-scheme}). 
  If the input value for \xmltag{n_target_accepted_events}
  is lower than 1000, it is replaced by 1000 and a corresponding message appears on the standard output. 
  To ease legibility for the values of these three parameters, spaces can be used to separate groups of digits.
  
  A run is also terminated after a certain relative or absolute precision has been reached. These are specified  
  with the tags \xmltag{target_relative_precision} and
  \xmltag{target_absolute_precision}, whose values represent negative powers of ten.
In \refli{lst:run_parameters}, for example, the target relative and absolute precisions are $10^{-5}$. 
  Finally, the termination of the integration is triggered by exceeding the run time specified by the \xmltag{time_wall} parameter, 
  whose value must have the format
  \texttt{days-hours:minutes}.
This option can be useful when running on computer clusters with
CPU-time limits. 
  If no value is given to \xmltag{time_wall}, this termination condition is not employed.   
\end{description}

Table \ref{tab:run_parameters2} summarises the parameters that control
the MC integration and should only affect the results within
integration errors.
\begin{table}
  \centering
  \renewcommand\arraystretch{1.2}
  \rowcolors{2}{tablerowcolor}{}
  \begin{tabular}{lll}
    \toprule\rowcolor{tableheadcolor} 
    \textbf{parameter} & \textbf{possible values} & \textbf{default value} \\
    \midrule
    \xmltag{cms_matrixelement}          &  $0,1$                    & $~1$  \\
    \xmltag{output}                     &                           & (group) \\
    \xmltag{output_level}               &  $2,4,6,8$                & $~2$  \\
    \xmltag{print_feynman_diagrams}     &  \texttt{true,false}      & \texttt{true}    \\
    \xmltag{intermediate_output}        &                           & (group)  \\
    \xmltag{intermediate_result_start}  &  integer                  & $1000$  \\
    \xmltag{intermediate_result_stride} &  integer                  & $1000000$  \\
    \xmltag{intermediate_result_factor} &  positive real number     & $1.1$  \\
    \xmltag{phase_space_generator}      &                           & (group) \\
    \xmltag{use_weight_optimization}    &  \texttt{true,false}      & \texttt{true}  \\
    \xmltag{channel_permutation}        &  $0,1$                    & $~0$  \\
    \xmltag{channel_elimination}        &  $0,1$                    & $~0$  \\
    \xmltag{subtraction_channels}       &  $0,1$                    & $~0$  \\
    \xmltag{me_virt_veto_threshold}     &  positive real number     &  none \\
    \xmltag{pdf_mapping_emin}           &  positive real number     & $0.0$  \\
    \xmltag{minimum_relative_invariant} &  positive real number     & $10^{-10}$ \\
    \xmltag{cos_theta_tolerance}        &  positive real number     & $10^{-10}$ \\
    \xmltag{tmax_tolerance}             &  positive real number     &
    $6.4\times 10^{-3}$ \\
    \xmltag{momentum_conservation_tolerance}&  positive real number & $10^{-12} $\\
    \xmltag{onshellness_tolerance}      &  positive real number     & $10^{-12}$ \\
    \xmltag{running_options}            &                           & (group)  \\
    \xmltag{store_run_resuming_info}    &  $0,1$                    & $~0$  \\
    \xmltag{writing_mode}               &  $0,1$                    & $~0$  \\
    \xmltag{load_run_resuming_info}     &  $0,1$                    & $~0$  \\
    \xmltag{folder_overwriting}         &  $0,1,2$                  & $~0$  \\
    \bottomrule
  \end{tabular}
  \caption{\label{tab:run_parameters2} 
 Elements of the \xmltag{run\_parameters} section of
 \texttt{param\_card.xml}  affecting the details of the MC integration  and their defaults.}
\end{table}

{\sloppy The matrix elements can be calculated either in the
  laboratory system or with momenta boosted to the CM frame to improve
  numerical accuracy. This behaviour is controlled by the parameter
  \xmltag{cms_matrixelement}. By default,
  \xmltag{cms_matrixelement}${}=1$ and the CM-frame momenta are
  employed. By setting the \xmltag{cms_matrixelement}${}=0$, 
  the use of laboratory frame momenta can be enforced.}

  \begin{warning}{Warning}
  Note that the parameter \xmltag{cms_matrixelement} has no effect when calculating
  cross sections with polarised vector bosons (see \refse{se:polarisations}). In this case,
  \mocanlo makes sure that the system where the momenta used for the evaluation
  of the matrix elements are defined corresponds to the frame
  specified by \xmltag{poldef_system},
  which is the only correct choice.
  \end{warning}

\begin{description}
\item[\texttt{<output>}]
The parameter \xmltag{output_level} controls the amount of information for
the standard output. The values
\xmltag{output_level}${}=2,4,6,8$ correspond to minimal, intermediate,
maximal relevant, and complete output, respectively.  

{\sloppy
By default, \mocanlo prints two files collecting all Feynman diagrams
that were used to construct the integration channels. The file
\texttt{feynman\_diagrams.tex} in the subdirectory
\texttt{run\ldots/data/init} contains the channels sorted 
in the order they were generated, while the file
\texttt{feynman\_diagrams\_survey.tex} in the subdirectory
\texttt{run\ldots/data/run} contains them sorted by their importance
in the integration. For complicated processes,
these files can become very large and their writing very
time consuming. The writing of these files can be switched off by
issuing \xmltag{print_feynman_diagrams}${}={}$\texttt{false}.
}

\item[\texttt{<intermediate\_output>}] The parameters of this section control when intermediate results of the integration are written out. 
  The parameter \xmltag{intermediate_result_start} specifies after how many accepted events this occurs for the first time. Its default value is 1000. 
  After the first time, intermediate results are written whenever the number of accepted events has increased by the value of  
  \xmltag{intermediate_result_stride}, with default 1000000, or by a factor corresponding to the value of \xmltag{intermediate_result_factor}, 
  with default $1.1$.
  
\item[\texttt{<phase\_space\_generator>}] This section includes
  parameters that are relevant for the phase-space generation.

The parameter \xmltag{use_weight_optimization} turns on or off the use
of \mocanlo's adaptive multi-channel weight-optimisation algorithm and
should normally be set to \texttt{true}.
  
Several flags exist to control the selection of integration channels:
\begin{itemize}
\item If \xmltag{channel_permutation}${}=1$, extra integration
channels with permuted resonances are included. This is needed for
processes with nested resonances, for example.
\item If \xmltag{channel_elimination}${}=1$, integration
channels that differ from others only by internal non-resonant lines
are eliminated.  This can be useful to reduce the number of channels,
in particular if \xmltag{channel_permutation}${}=1$.
\item The flag \xmltag{subtraction_channels} is relevant only for real
contributions. If its value is 1, extra integration channels
corresponding to Catani--Seymour dipole subtraction contributions are
taken into account. This can significantly increase the total number
of channels.  Note that no extra channels are available for
subtraction dipoles related to internal resonances in the pole
approximation.
\end{itemize}
The real parameter \xmltag{me_virt_veto_threshold} can be set to
discard events with unusually large virtual matrix elements, \ie
$2\Re\,\mathcal{M}_\text{virt}\mathcal{M}_\text{Born}^* >{}$
\xmltag{me_virt_veto_threshold}${}\cdot
|\mathcal{M}_\text{Born}|^2$. Such events might result from
phase-space points that are problematic for \collier,
but whose removal does not impact the final result, as long as
their number is limited. It is the user's responsibility to ensure
 that only a few events are discarded when this parameter is activated,
for instance by inspecting the output file \texttt{cross\_section.dat}
in the \texttt{result/} folder of the run.
  
A minimal partonic CM energy can be required to avoid the generation
of events which would be cut away anyhow. This value can be specified
via the parameter \xmltag{pdf_mapping_emin}.  We note that \mocanlo
also computes a minimal partonic CM energy from the transverse-momentum
cuts provided via the \texttt{cut\_card.xml} file (see
\refse{se:cut-scheme}). The value that is actually employed is the
larger of the two. It needs to be larger than zero, otherwise the code
stops with an error message.
  
The parameters \xmltag{minimum_relative_invariant},
\xmltag{cos_theta_tolerance}, \xmltag{tmax_tolerance},
\xmltag{momentum_conservation_tolerance}, and
\xmltag{onshellness_tolerance} are technical cut parameters of the MC
integration that avoid the appearance of very large unphysical weights.
They can be varied (typically by factors of 10) about their
default values to verify that the integration results are independent
of these parameters.

\item[\texttt{<running\_options>}] \sloppy This subsection contains a set of parameters
  that control the storing and loading of information during a computation, 
  which can be used to resume a run after a walltime has been reached, for example. 
  A description of the parameters in this subsection  is given in \refse{par:running_mode}.

\end{description}

\subsubsection{How to perform sequential runs}
\label{par:running_mode}

\mocanlo is capable of loading the results of a previous run and 
restarting a calculation that has met a termination criterion, 
provided the necessary information has been stored. In combination with the \texttt{time\_wall}
parameter  (part of the \xmltag{integration\_termination} subsection of \texttt{param\_card.xml}),
this feature can be used 
to perform a resumption chain in which each step is within the walltime of a cluster, but which effectively 
corresponds to a longer integration runtime. 
By using this method, a smaller integration error can be achieved with the same CPU time compared to parallel job execution.
This is because the optimised channel weights are inherited by subsequent steps of the resumption chain, whereas parallel runs 
perform the same channel weight optimisation independently. We remark that parallel and sequential running can be combined by 
simply computing  parallel resumption chains.

The \xmltag{running\_options} subsection of the
\xmltag{run\_parameters} section in the file \texttt{param\_card.xml}
contains a set of parameters that control the storing and loading of 
information during a computation.
\refli{lst:running_options} displays the complete set of
parameters of this subsection.

\begin{lstlisting}[language=XML,caption=MC running options
  that can be specified in the \texttt{run\_parameters} section in
  \texttt{param\_card.xml}. ,label=lst:running_options] 
   <running_options>
      <store_run_resuming_info>  1 </store_run_resuming_info>
      <writing_mode>             0 </writing_mode>
      <load_run_resuming_info>   1 </load_run_resuming_info>
      <folder_overwriting>       1 </folder_overwriting>
   </running_options>      
\end{lstlisting}

Their possible and default values are found in the lower part of \refta{tab:run_parameters2}.

Setting \xmltag{store_run_resuming_info}${}=1$ enables the storage 
of the information necessary to resume that run later. If \xmltag{store_run_resuming_info}${}=0$, 
no storage will take place.

The parameter \xmltag{writing_mode} controls at which points of the run the storage of 
information necessary for resumption takes place. 
By default (\xmltag{writing_mode}${}=0$), the storage occurs when a \mocanlo termination 
condition is met (see subsection \xmltag{integration\_termination} in \refse{subsubsection:run_parameters}).
If \xmltag{writing_mode}${}=1$, the storage occurs each time \mocanlo writes 
intermediate results (see the beginning of \refse{subse:running_single_contribution}).

Setting \xmltag{load_run_resuming_info}${}= 1$ tells the program that the user intends to resume a run 
and that it should load the necessary information. In this case, the run directory of the run that is 
to be resumed must be passed as an argument (see the example below for further details). 
If \xmltag{load_run_resuming_info}${}= 0$ (default), no loading will take place.
By choosing both \xmltag{load_run_resuming_info}${}= 1$ and \xmltag{store_run_resuming_info}${}=1$, 
intermediate steps in a chain of resumptions can be performed. 

The parameter \xmltag{folder_overwriting} determines how the results of a resumption run are stored, 
which is relevant for subsequent run resumptions. 
The default value \xmltag{folder_overwriting}${}= 0$ makes a backup of the results of the run that will be 
resumed and stores the results of the resumption under the name of the original run directory. 
If one sets \xmltag{folder_overwriting}${}= 1$, the results of previous steps of a resumption chain are kept, 
but the results of a step are overwritten if that step is performed again. 
Finally, the value \xmltag{folder_overwriting}${}= 2$ prevents any overwriting, so that the results of all 
intermediate and parallel steps are kept. An exemplary usage of this parameter can be found
at the end of this subsection.

In the following, details on the storage and resumption procedures are given using the example introduced 
in \refse{subse:running_single_contribution}. There, a
\mocanlo run for a specific subprocess was started 
\begin{lstlisting}[language=Bash]
mocanlo  /path/to/process/pp_epvemumvmxbbx_qcd  1
\end{lstlisting}
where the two arguments are the path to the process directory, which contains the \texttt{cards}
directory, and the run ID of the subprocess. 
If this run is performed with \xmltag{store_run_resuming_info}${}=0$, the run's output directory contains
the \texttt{data} subdirectory, with information about the initialisation and the run phase, as well as 
the \texttt{result} subdirectory, where cross sections and distributions are saved. 
If the information storage is activated by setting \xmltag{store_run_resuming_info}${}=1$,
an additional subdirectory named \texttt{carbon\_copy} is created:

\dirtree{%
  .1 run\_2025-07-29T17h38m28s928/.
  .2 data/.
  .3 \ldots .
  .2 result/.  
  .3 \ldots .
  .2 run\_info.dat .
  .2 carbon\_copy/ .
  .3 channels\_info.dat .
  .3 cuts\_info.dat .
  .3 rnd\_number\_info.dat .
  .3 run\_id\_info.dat .
  .3 scale\_factor\_1/ .
  .4 cross\_section\_info.dat .
  .4 histograms/ .
  .3 scale\_factor\_2/ .
  .4 cross\_section\_info.dat .
  .4 histograms/ .
  .3 \ldots .
}%

{\sloppy
The files in \texttt{carbon\_copy} contain the required information for the run resumption. 
In the file \texttt{run\_id\_info.dat} general information on how the run
was performed is written, like its \texttt{run\_type}, \texttt{run\_parameters}, and any other tag
specified for the run ID in the \texttt{run\_card.xml}. This information is used when 
resuming a run to check that the code is restarted with the same set of parameters.
}

In \texttt{rnd\_number\_info.dat}, a set of $25$ random numbers is stored that uniquely specifies
the current point in the \texttt{RANLUX} pseudo-random number sequence~\cite{James:1993np}.
Using this information, the event generation chain can restart from the exit point of the previous run. 
In \texttt{channels\_info.dat}, information on the integration channels is stored, like the weight-optimisation 
step, together with the channels' weights and variances, the respective contribution to the integrals and  so on. 
This way, the program can benefit from a run in which an optimisation of the weights has occurred.

In order to consistently restore a run, information on the hits of the different cuts specified
in \texttt{cut\_card.xml} is saved in \texttt{cuts\_info.dat}. This allows the cumulative counting of the hits
for each cut.

{\sloppy
Depending on the number of scales specified by 
\xmltag{scale_factors_f} and \xmltag{scale_factors_r}, an equal number
of scale subdirectories is stored within the \texttt{carbon\_copy} subdirectory (if only one scale is
provided, no additional scale directory is produced). Therein, one can find all information to restart
the cross section computation at the integrated level, in \texttt{cross\_section\_info.dat}, and
even differentially, in the subdirectory \texttt{histograms}, containing information for all
required distributions in separate files named after the observable name specified by the user.
Moreover, in \texttt{cross\_section\_info.dat},
the maximum weight and the sum of the absolute weights computed so far are also contained.}

In order to resume a previous run that contains a \texttt{carbon\_copy} subdirectory, like
the one of our example, whose output has been stored in \basename,
one has to set the parameter \xmltag{load_resuming_run_info} to
\texttt{1} and start a new run as 
\begin{lstlisting}[language=Bash]
mocanlo  /path/to/process/pp_epvemumvmxbbx_qcd  1  \
         /path/to/run_2025-07-29T17h38m28s928
\end{lstlisting}
where the second argument is the run ID of the subprocess in question
and the third argument is the path to the directory that contains the 
results of the run to be resumed. 
 
{\sloppy
  An additional argument containing the random number seed before or after the name 
  of the directory to resume is tolerated but not used, since 
  all initialisation conditions for the random number chain and the integral computation
  are loaded from \basename\texttt{/carbon\_copy}. 
  If a run is restarted with a different set of relevant parameters, the code
  stops with an error message pointing to a file containing the conflicting lines of the cards 
  that triggered the error. 
  This file is located in the subdirectory  \texttt{data/init/user\_input/} 
  of the current run and named \texttt{diff\_user\_input\_\textit{section}.dat}, where
  \texttt{\textit{section}} can be \texttt{cuts}, \texttt{histograms},
  \texttt{monte\_carlo}, \linebreak \texttt{on\_shell\_projection}, \texttt{partonic\_process},
  \texttt{recombinations}, \texttt{resonances}, \texttt{scales}, \linebreak
  \texttt{standard\_model}, or \texttt{weight\_opts}
  depending on which section caused the error. These files are temporary and are removed 
  if the run resumption succeeds.
  The parameters contained in the subsections \xmltag{running\_options} and \xmltag{integration\_termination} 
  of \texttt{param\_card.xml}  are ignored when comparing  the input parameters of subsequent runs,
  so that their  values can change without triggering an error. 
}

The options \xmltag{load_run_resuming_info} and \xmltag{store_run_resuming_info} can be 
activated at the same time to perform chains of run resumptions. 
Depending on the value of \xmltag{folder_overwriting} (discussed above),
the name of the directory containing the 
results of a run resumption might not contain the time at which the directory was created, 
as opposed to the case of a fresh run.  Therefore, this information is printed into a 
\texttt{run\_info.dat} file, together with the name of the directory from which the data for the 
resumption was loaded. This file can be found directly in the directory of the resuming run, 
at the same level of the \texttt{data}, \texttt{results} and \texttt{carbon\_copy} subdirectories.

{\sloppy
The parameter \xmltag{folder_overwriting} controls the location of the output
  of a resumed run. Setting \xmltag{folder_overwriting}${}=2$ and executing
the restart command above, the results 
of the restarted run are located in \basename\texttt{\_step1\_2025-07-30T17h42m25s330}. 
Note that the name of this run directory comprises the original directory's name from which the data have 
been loaded, the number of times \texttt{N} the run has been resumed (as \texttt{stepN}), and a new time stamp 
with the time at which the run has been restarted. By executing again
the restart command above, a directory with a different time stamp, \\ 
\eg~\basename\texttt{\_step1\_2025-07-30T18h14m01s811}, 
is created, containing results from a new resumption of the same run, thus carrying a different (second) timestamp.}

{\sloppy
If the flag \xmltag{store_run_resuming\_info} was set to 1, this run
can be further resumed by 
running
\begin{lstlisting}[language=Bash]
mocanlo  /path/to/process/pp_epvemumvmxbbx_qcd  1  \
         /path/to/run_2025-07-29T17h38m28s928_step1_2025-07-30T18h14m01s811
\end{lstlisting}
which creates \basename\texttt{\_step2\_2025-07-31T18h17m28s388} directory, for instance. 
Note that the timestamp of the original run (step 0) is kept in the name at the first position, while the second
timestamp corresponds to the last performed step (\texttt{step2} in this case). The directory 
\basename\texttt{\_step1\_2025-07-30T18h14m01s811}, which contains the results of the intermediate 
\texttt{step1},  as well as the directory \\ \basename\texttt{\_step1\_2025-07-30T17h42m25s330}, which
contains the results of a \texttt{step1} that were not used for a run resumption, remain.
}

{\sloppy
In order to avoid the proliferation of intermediate-result directories, 
\xmltag{folder_overwriting} can be set 
to 1. This causes the result directory of \texttt{stepN} to be replaced if a new \texttt{stepN}, \ie~the same step, is 
performed using the same loaded information. This option also affects the name of the result directories,
in that no second timestamp is included. 
In our example, setting \xmltag{folder_overwriting}${}=1$ and executing
the first restart command 
stores the results into a folder \basename\texttt{\_step1}. Re-executing the
command again
will overwrite the \texttt{step1} directory. Resuming the \texttt{step1} of the run with the command
\begin{lstlisting}[language=Bash]
mocanlo  /path/to/process/pp_epvemumvmxbbx_qcd  1  \
         /path/to/run_2025-07-29T17h38m28s928_step1
\end{lstlisting}
creates a directory \basename\texttt{\_step2} and so on. The directory \linebreak 
\basename\texttt{\_step1} remains. Therefore, only the most recent results of each step are stored.}

{\sloppy 
By setting \xmltag{folder_overwriting}${}=0$, the storage can be restricted to the two directories containing the results 
of the last two steps. 
By executing the first restart command, the original
directory \basename{} is renamed into 
\basename\texttt{\_bckp} and the results of this first resumption are saved under the previous name of 
the original directory, \ie \basename. Issuing the restart command 
again, the second step 
is performed and its results are saved under the name \basename,
while \basename\texttt{\_bckp} now contains the results of the first step.
}

\subsubsection{Photon--photon ultra-peripheral collisions}

MoCaNLO is able to simulate initial-state photons originating from
heavy ions in ultra-peripheral collisions.
To that end, it relies on the external program gamma-UPC~\cite{Shao:2022cly}.

To steer all the possible options of the library, several flags are available in \texttt{param\_card.xml} in the \xmltag{pdfs} section.
\begin{table}
  \centering
  \renewcommand\arraystretch{1.2}
  \rowcolors{2}{tablerowcolor}{}
  \begin{tabular}{lll}
    \toprule\rowcolor{tableheadcolor} 
    \textbf{parameter} & \textbf{possible values} & \textbf{default value} \\
    \midrule
    \xmltag{alpha_UPC}                  &  positive real number     & $1/137.036$  \\
    \xmltag{UPC_photon_flux}            &  $0,1$                    & $~0$  \\
    \xmltag{nuclear_A_beam1}            &  integer                  & $208$  \\
    \xmltag{nuclear_A_beam2}            &  integer                  & $208$  \\
    \xmltag{nuclear_Z_beam1}            &  integer                  & $82$  \\
    \xmltag{nuclear_Z_beam2}            &  integer                  & $82$  \\
    \bottomrule
  \end{tabular}
  \caption{\label{tab:run_parameters_UPC} Additional parameters
    for ultra-peripheral collisions in the
    \texttt{param\_card.xml} and their default values.}
\end{table}
{\samepage
An example to run Pb--Pb collisions using
the electric-dipole form factor (EDFF)~\cite{Cahn:1990jk} (instead of the charge (ChFF)
form factor~\cite{Vidovic:1992ik}) is:\\
\begin{minipage}{\linewidth}
\begin{lstlisting}[language=XML]
   <pdfs>
      <beam1>              ion          </beam1>
      <beam2>              ion          </beam2>
      <alpha_UPC> 7.2973525205055605d-3 </alpha_UPC>       ! alpha used in UPC
      <UPC_photon_flux>    0            </UPC_photon_flux> ! (0) EDFF scheme, (1) ChFF scheme
      <nuclear_A_beam1>    208          </nuclear_A_beam1> ! Case of Pb
      <nuclear_A_beam2>    208          </nuclear_A_beam2> ! Case of Pb
      <nuclear_Z_beam1>    82           </nuclear_Z_beam1> ! Case of Pb
      <nuclear_Z_beam2>    82           </nuclear_Z_beam2> ! Case of Pb
   </pdfs>
\end{lstlisting}
\end{minipage}}
The value of $\alpha$ used in UPC can be changed independently of
the one used in the matrix element.
The default values of the UPC parameters can be found in Table~\ref{tab:run_parameters_UPC}.

\subsection{Setting a-priori weights for the Monte Carlo integration}
\label{se:weight_opts}

By default, the a-priori weights of the different MC integration channels
are all set to the same value, equal to the inverse of the
total number of integration channels for the process of interest.
They represent initialisation values for the multi-channel optimisation,
which adapts them during the integration \cite{Kleiss:1994qy}. 
In order to improve the convergence of the result for
processes dominated by diagrams containing resonances, it can
be useful to tune the a-priori weights at initialisation. This can be
achieved by linking in the
\texttt{run\_card.xml} a section \xmltag{weight_opts id="XX"}, where XX is an ID
that denotes a particular a-priori weight setting. This section contains
one or more subsections
\xmltag{weight_opt} that independently specify the information
on the resonant contributions to be enhanced. 
Therefore, the a-priori weights of the integration channels corresponding to Feynman diagrams
containing at least \xmltag{min_n_resonant_particles} and at most \xmltag{max_n_resonant_particles}
particles specified by \xmltag{particle} as an $s$-channel resonance are adjusted. They are
multiplied by a factor \xmltag{mult_factor} for each resonance found
in the associated Feynman diagram.  The global tag
\xmltag{total_max_n_resonant_particles} limits the total number of
enhancement factors to the specified number.
Thus, a channel with more resonances than \xmltag{total_max_n_resonant_particles} is not enhanced at all.

An exemplary application is shown in \refli{lst:weightopt} for $\Pt\PZ\Pj$ production, where
a-priori weights can be enhanced by up to a factor of 30, \ie by a factor of 10 if they belong to
integration channels corresponding to diagrams with exactly one top quark, and by an additional factor
of three if their integration channels also include one $\PZ$ boson. Note that in the example
there are no a-priori weights that can be multiplied by a factor
higher than 30, as the global tag
\xmltag{total_max_n_resonant_particles} limits the number of
enhancement factors to 2. Therefore, there are four distinct situations in which a-priori weights
can be enhanced: for channels
with one top quark (multiplied by 10), one Z boson (multiplied by 3), two Z bosons
(multiplied by 9), and one top quark and one Z boson (multiplied by 30).
\begin{lstlisting}[language=XML,caption=A-priori weights in
  \texttt{param\_card.xml}. ,label=lst:weightopt]
   <weight_opts id="tzj">
      <total_max_n_resonant_particles> 2   </total_max_n_resonant_particles>
      <weight_opt>
         <particle>                 t   </particle>
         <min_n_resonant_particles> 1   </min_n_resonant_particles>
         <max_n_resonant_particles> 1   </max_n_resonant_particles>
         <mult_factor>              10  </mult_factor>
      </weight_opt>
      <weight_opt>
         <particle>                 z   </particle>
         <min_n_resonant_particles> 1   </min_n_resonant_particles>
         <max_n_resonant_particles> 2   </max_n_resonant_particles>      
         <mult_factor>              3   </mult_factor>
      </weight_opt>
   </weight_opts>
\end{lstlisting}

Few additional comments on the \xmltag{particle} tag are in order. First of all,
note that particles and antiparticles are distinguished. In order to treat them on the
same footing, the \mocanlo{} identifier of the resonance (see second column of \refta{tab:ParticleNames})
must be accompanied by a star key, \ie \texttt{t*} indicates either a \texttt{t} or a \texttt{t\char`\~},
and similarly \texttt{w*} either a \texttt{w+} or a \texttt{w-}. Moreover, a list of particles
can be specified in \xmltag{particle}: particles in the list are treated on the same level as far as
the assignment of an enhancement factor of the a-priori weight is concerned.


\section[Recombination, cut, and scale definitions: the \texttt{\large cut\_card.xml}]
{Recombination, cut, and scale definitions: the \texttt{\Large cut\_card.xml}}
\label{se:cuts}

The \texttt{cut\_card.xml} defines the sections
\xmltag{recombinations}, \xmltag{cuts}, and \xmltag{scales} for setting
the recombination parameters for the clustering algorithm, defining
the acceptance function, and fixing the dynamical scales, respectively. 

After recombination, clustered objects are stored in \emph{sorted clusters},
which consist of lists of different jet types, each ordered by transverse momentum
(see \refse{se:cut-scheme}).
The cut procedure operates on these sorted clusters by modifying and
extending them. The definition of scales and of observables of
histograms is based on the sorted clusters that result from the
cut procedure.

\subsection{Recombinations}
\label{se:recombinations}

The settings in the section \xmltag{recombinations}  of the 
\texttt{cut\_card.xml} configure the jet algorithm.
The scheme allows for different steps of recombinations, which can be executed
consecutively or in parallel. The recombination may involve several levels,
at which the parallel recombinations are merged. The input of the recombination is the
final-state particles of the considered process, while the output is a list of clusters,
to be used as input for the cut routines.
Note that the final list of clusters includes only objects that pass through all recombination levels,
even if they are not subject to recombination.

The parameters, with their possible and default values, of the section
\xmltag{recombinations}, the subsections \xmltag{recombination_step},
and their subsections \xmltag{recombination} are listed in
\reftas{tab:recombination_list}, \ref{tab:recombination_step}, and
\ref{tab:recombination_rule}, respectively.
\begin{table}
      \centering
      \renewcommand\arraystretch{1.2}
      \rowcolors{2}{tablerowcolor}{}
      \begin{tabular}{lll}
        \toprule\rowcolor{tableheadcolor} 
        \textbf{Parameter} & \textbf{possible values} &
        \textbf{default value}  \\
        \midrule
      \xmltag{rapidity_maximum}    & real number                & $100.0$  \\
      \xmltag{recombination_index} & real number                & $-1$  \\
      \xmltag{recombination_step}  & ---                        & ---  \\
      \bottomrule
      \end{tabular}
      \caption{\label{tab:recombination_list} Parameters and possible values of
        the \xmltag{recombinations} section.}
\end{table}
\begin{table}
      \centering
      \renewcommand\arraystretch{1.2}
      \rowcolors{2}{tablerowcolor}{}
      \begin{tabular}{lll}
        \toprule\rowcolor{tableheadcolor} 
        \textbf{Parameter} & \textbf{possible values} &
        \textbf{default value}  \\
        \midrule
      \xmltag{name}                    & string                & none  \\
      \xmltag{recombination_index}     & real number           & $-1$  \\
      \xmltag{recombination_level_in}  & non-negative integer  & $~0$  \\
      \xmltag{recombination_level_out} & positive integer      & $~1$  \\
      \xmltag{recombination_jet_types_in}  & list of jet types & none  \\
      \xmltag{recombination_jet_types_out} & list of jet types & none  \\
      \xmltag{recombination}  & ---                        & ---  \\
      \bottomrule
      \end{tabular}
      \caption{\label{tab:recombination_step} Parameters and possible values of
        the \xmltag{recombinations_step} section.}
\end{table}
\begin{table}
      \centering
      \renewcommand\arraystretch{1.2}
      \rowcolors{2}{tablerowcolor}{}
      \begin{tabular}{lll}
        \toprule\rowcolor{tableheadcolor} 
        \textbf{Parameter} & \textbf{possible values} &
        \textbf{default value}  \\
        \midrule
      \xmltag{name}                     & string           & none  \\
      \xmltag{recombinant}              & integer          & $~0$  \\
      \xmltag{recombinendus}            & integer          & $~0$  \\
      \xmltag{recombinendus1}           & integer          & \xmltag{recombinant}  \\
      \xmltag{minimum_distance}         & real number      & $0.0$  \\
      \xmltag{jet_type_selector}        & real number      & $0.0$  \\
      \xmltag{jet_isolation_distance}   & real number      & $0.0$  \\
      \xmltag{jet_isolation_parameter}  & real number      & $0.0$  \\
      \xmltag{jet_isolation_exponent}   & real number      & $1.0$  \\
      \bottomrule
      \end{tabular}
      \caption{\label{tab:recombination_rule} Parameters and possible values of
        the \xmltag{recombination} section.}
\end{table}
A simple example with only one recombination step and thus one recombination
level is shown in \refli{lst:recombination},
\begin{lstlisting}[language=XML,caption=Example recombinations in
  \texttt{cut\_card.xml}. ,label=lst:recombination]
<recombinations id="ttx">
    <rapidity_maximum>             5   </rapidity_maximum>
    <recombination_step>
        <recombination_index>     -1   </recombination_index>
        <recombination_level_in>   0   </recombination_level_in>
        <recombination_level_out>  1   </recombination_level_out>
        <recombination type="different_type">
            <name> bjet-jet </name>
            <recombinant>          2   </recombinant>
            <recombinendus>        1   </recombinendus>
            <minimum_distance>     0.5 </minimum_distance>
        </recombination>
        <recombination type="same_type">
            <name> bjet-bjet </name>
            <recombinendus1>       2   </recombinendus1>
            <recombinant>          1   </recombinant>
            <minimum_distance>     0.5 </minimum_distance>
        </recombination>
    </recombination_step>
</recombinations>
\end{lstlisting}
while a more complicated example with three recombination steps and two recombination
levels is given in \refli{lst:recombination2}.
\begin{lstlisting}[language=XML,caption=Example recombinations in
  \texttt{cut\_card.xml} with multiple steps. ,label=lst:recombination2]
<recombinations id="vbs_semileptonic">
    <rapidity_maximum>             5   </rapidity_maximum>
    <recombination_step>
        <name> lepton-photon </name>
        <recombination_index>      0   </recombination_index>
        <recombination_level_in>   0   </recombination_level_in>
        <recombination_level_out>  1   </recombination_level_out>
        <recombination type="different_type">
            <name> antimuon-photon </name>
            <recombinant>          16  </recombinant>
            <recombinendus>        3   </recombinendus>
            <minimum_distance>     0.1 </minimum_distance>
        </recombination>
    </recombination_step>
    <recombination_step>
        <name> slim-jets </name>
        <recombination_index>     -1   </recombination_index>
        <recombination_level_in>   1   </recombination_level_in>
        <recombination_level_out>  2   </recombination_level_out>
        <recombination type="same_type">
            <name> jet-jet </name>
            <recombinant>          1   </recombinant>
            <minimum_distance>     0.4 </minimum_distance>
        </recombination>
        <recombination type="different_type">
            <name> jet-photon </name>
            <recombinant>          1   </recombinant>
            <recombinendus>        3   </recombinendus>
            <minimum_distance>     0.4 </minimum_distance>
        </recombination>
    </recombination_step>
    <recombination_step>
        <name> fat-jets </name>
        <recombination_index>     -1   </recombination_index>
        <recombination_level_in>   1   </recombination_level_in>
        <recombination_level_out>  2   </recombination_level_out>
        <recombination_jet_types_in>   1 3 </recombination_jet_types_in>
        <recombination_jet_types_out> 31 3 </recombination_jet_types_out>
        <recombination type="same_type">
            <name> jet8-jet8-jet8 </name>
            <recombinant>          31  </recombinant>
            <minimum_distance>     0.8 </minimum_distance>
        </recombination>
        <recombination type="different_type">
            <name> jet-photon </name>
            <recombinant>          31  </recombinant>
            <recombinendus>        3   </recombinendus>
            <minimum_distance>     0.8 </minimum_distance>
        </recombination>
    </recombination_step>
</recombinations>
\end{lstlisting}

{\sloppy
The section \xmltag{recombinations id="XX"}, linked via the ID in the
\texttt{run\_card.xml}, contains one or more
\xmltag{recombination_step} subsections as well as the further parameters
\xmltag{rapidity_maximum} and \xmltag{recombination_index}.  Particles
with a rapidity larger than \xmltag{rapidity_maximum} are disregarded
in the recombination algorithm. Since neutrinos do not undergo any recombination,
they are not affected by it. The parameter
\xmltag{recombination_index} defines the default clustering algorithm and
takes the values $-1$, $0$, and $1$ for the anti-$k_\mathrm{t}$~\cite{Cacciari:2008gp}, the
Cambridge--Aachen~\cite{Dokshitzer:1997in,Wobisch:1998wt}, and the $k_\mathrm{t}$ algorithm~\cite{Catani:1993hr,Ellis:1993tq}, respectively.}

The parameters of a \xmltag{recombination_step} are summarised in
\refta{tab:recombination_step}.  The optional tag \xmltag{name} is
only used for output purposes.  The optional parameter
\xmltag{recombination\_index} allows to overwrite, for the
corresponding \xmltag{recombination_step}, the value of
\xmltag{recombination\_index} defined for the \xmltag{recombinations
  id="XX"} section, discussed above.  Note that only one algorithm can
be chosen within a single \xmltag{recombination_step}.

Since multiple steps of recombination are allowed,
one must specify the level in the recombination chain on which a
recombination step acts via
the tag \xmltag{recombination_level_in}. Recombination steps with the same
  \xmltag{recombination_level_in} values are run in parallel.
  The tags \xmltag{recombination_level_in} and \xmltag{recombination_level_out}
  define the input and output level of the recombination step, respectively,
  and by default equal 0 (all particles) and 1.
  Internally, the clusters of level
  \xmltag{recombination_level_in} are copied to clusters of level
   \xmltag{recombination_level_out} (to allow for parallel
  recombination with different jet resolution parameters for
  instance), and thereafter the recombination step is performed on
  clusters of level \xmltag{recombination_level_out}. Once
  all steps for an output level have been done, the resulting
  clusters are merged to a new list of clusters, which can serve as
  input for subsequent, not necessarily consecutive, levels.  The merged clusters of the final level
  are the output of the recombination procedure.
  
{\sloppy The list \xmltag{recombination_jet_types_in} allows to
  restrict the jet types on which the recombination step acts. If this
  list is specified, all recombined objects get the jet type
  given by the corresponding list \xmltag{recombination_jet_types_out}. 
  If \xmltag{recombination_jet_types_in} is not specified or empty, all
  visible jet types, \ie those within 
  \xmltag{rapidity_maximum}, and all neutrinos are used as input and, if
  not recombined, passed to the output.
  Note that each jet type that should appear in the final output of the recombination
  routine must also be present in the input at every recombination level.
  This requirement can be satisfied by defining a
  sequence of recombination steps without the \xmltag{recombination_jet_types_in} parameter,
  thereby propagating all jet types through all levels.
}

{\sloppy
Each recombination step necessarily contains different recombination rules,
defined either for jets of the same
type by the tag \xmltag{recombination\ type="same_type"} or for jets of different
type by the tag \xmltag{recombination\ type="different_type"}.  An arbitrary name
can be set for both types via \xmltag{name}, which is only used for
output purposes. The jet-resolution parameter is fixed by \linebreak
\xmltag{minimum\_distance}, corresponding to the
rapidity--azimuthal-angle distance, and the jet type of the result of the
clustering by \xmltag{recombinant}. 
}

For same-type-jet recombination, the type of the jets to cluster is
defined by \xmltag{recombinendus1}. If \xmltag{recombinendus1} is
not specified, it is assumed to be equal to \xmltag{recombinant}. In the example in \refli{lst:recombination}, the same-type
recombination with the name \texttt{bjet-bjet} defines the clustering
of b jets (jet type 2) with a jet resolution parameter $R=0.5$ into
flavourless jets (jet type 1).

A different-type recombination requires the definition of two
different types of jets to be recombined via \xmltag{recombinendus1}
and \xmltag{recombinendus}. While
\xmltag{recombinendus} needs to be specified, \linebreak \xmltag{recombinendus1}
is assumed to be equal to \xmltag{recombinant}, if not given.  In
\refli{lst:recombination}, the different-type recombination with the
name \texttt{bjet-jet} defines the clustering of b jets and jets (jet
type 1) into b jets with a jet resolution parameter $R=0.5$.

\begin{warning}{Valid values for \xmltag{jet_type} created in
    the recombination}   
  The jet types appearing within recombination steps must correspond
  to existing jet types of \mocanlo as given in
  \refta{tab:ParticleNames} or new ones defined via
  \xmltag{recombination_jet_types_out} or \xmltag{recombinant}
  within the range $30$--$50$.
\end{warning}

In the example \refli{lst:recombination2}, in the first recombination
step photons are clustered with antimuons, using
the Cambridge--Aachen algorithm with jet resolution parameter $R=0.1$. This step
passes all jets to level 1. At this level recombination steps 2 and 3 cluster jets of
type 1 (and photons) twice in parallel using the
anti-$k_{\mathrm{T}}$ algorithm with resolution parameters $0.4$
(slim-jets) and $0.8$ (fat-jets). While the slim jets keep their jet
types (if not clustered), the fat jets get jet type 31 via the
parameter \xmltag{recombination_jet_types_out}${}=31$ before
clustering. Note that also photons (jet type 3) have to be present in
the lists \xmltag{recombination_jet_types_in} and
\xmltag{recombination_jet_types_out} to allow their clustering with fat jets.
At level 2, which is the last one in the example, the outputs of
recombination steps 2 and 3 are internally merged. Thus, the lists of jets in the
output of the recombination involve the same objects twice. This has
to be rectified by using appropriate cuts.

\subsection{Treatment of photons}
\label{se:photons}

The treatment of photons in the final state (via isolation, fragmentation, and
conversion to jets) is determined to a large extent by the
recombination rules.

\subsubsection{Frixione isolation}
Photons can be isolated via the Frixione isolation scheme
\cite{Frixione:1998jh}. To this end, a recombination rule as shown in
\refli{lst:Frixione} should be added to the corresponding
recombination step in the run card.
\begin{lstlisting}[language=XML,caption=Frixione isolation in
  \texttt{cut\_card.xml}. ,label=lst:Frixione]
    <recombination type="different_type">
        <name>                jet-photon </name>
        <recombinant>             3      </recombinant>
        <recombinendus>           1      </recombinendus>
        <minimum_distance>        0.5d0  </minimum_distance>
        <jet_isolation_parameter> 0.11d0 </jet_isolation_parameter>
        <jet_isolation_exponent>  1d0    </jet_isolation_exponent>
        <jet_isolation_distance>  0.5d0  </jet_isolation_distance>
    </recombination>
\end{lstlisting}
If the rapidity--azimuthal-angle distance $\Delta R$ between the
photon (=\xmltag{recombinant}) and the jet (=\xmltag{recombinendus}) is
smaller than the cone size $R_0$ (=\xmltag{jet-isolation-distance}),
the two objects \xmltag{recombinant} and \xmltag{recombinendus}
are discarded, unless they respect the relation
\begin{equation}\label{eq:Frixione}
p_{\rT,\mathrm{recombinendus}} < \epsilon p_{\rT,\gamma}
\left(\frac{1-\cos(\Delta R_{\gamma,\mathrm{recombinendus}})}{1-\cos(R_0)}\right)
^n,
\end{equation}
where $p_{\rT,\mathrm{recombinendus}}$ and $p_{\rT,\gamma}$ are the
transverse momenta of the jet and the photon, respectively. 
If $\Delta R < R_0$ and \refeq{eq:Frixione} is fulfilled, the two
objects are merged into one of type \xmltag{recombinant}, \ie
a photon. For $\Delta R > R_0$, both objects are kept independently of
\refeq{eq:Frixione}.
The prefactor $\epsilon$ (=\xmltag{jet_isolation_parameter}) controls the
allowed range of the energy fraction of the recombinendus in the cone
around the photon and the exponent $n$
(=\xmltag{jet_isolation_exponent}) the approach to the limit. Both
should be positive parameters of order one and have default values
\xmltag{jet_isolation_parameter}${}=0$ and
\xmltag{jet_isolation_exponent}${}=1$.

\begin{warning}{Restriction}
\sloppy
The use of Frixione isolation via \xmltag{jet_isolation_parameter}, 
\xmltag{jet_isolation_exponent}, and \xmltag{jet_isolation_distance}
is only implemented for the case of a photon as
\xmltag{recombinant} and a jet originating from a massless parton as \xmltag{recombinendus},
more precisely for \xmltag{jet_type}${}\in\{1,2,30,\ldots,50\}$, where
jets of type $30$--$50$ must be defined appropriately.
\end{warning}

\subsubsection{Quark-to-photon fragmentation function}

{
Alternatively, isolated photons can be treated by employing a cut on
the energy fraction of the photon within a recombined quark--photon
pair.  To this end, a recombination rule as shown in
\refli{lst:fragmentation_photon} should be added to the corresponding
recombination step in the run card.
\begin{lstlisting}[language=XML,caption=Use of fragmentation function
  for final states with photons in \texttt{cut\_card.xml}. ,label=lst:fragmentation_photon]
    <recombination type="different_type">
        <name>              jet-photon </name>
        <recombinant>       3          </recombinant>
        <recombinendus>     1          </recombinendus>
        <minimum_distance>  0.5d0      </minimum_distance>
        <jet_type_selector> 0.1d0      </jet_type_selector>
    </recombination>
\end{lstlisting}
\sloppy
If \xmltag{jet_type_selector}${}=0.0$ (default),  \xmltag{recombinant}
and \xmltag{recombinendus} are recombined to \xmltag{recombinant} as usual.
For \xmltag{jet_type_selector}${}>0.0$, the recombination is modified as
follows.
If the energy fraction of the \xmltag{recombinendus}
$$
z = \frac{E_{\mathrm{recombinendus}}}{E_{\mathrm{recombinendus}}+E_{\mathrm{recombinant}}}
$$
is larger than $z_\mathrm{cut}$ (=\xmltag{jet_type_selector}), both
jets are recombined into a jet of type \linebreak \xmltag{recombinendus},
otherwise they are recombined into a jet of type \xmltag{recombinant}.
}

A recombination depending on the energy fraction of the \xmltag{recombinendus}
must be accompanied by a fragmentation-function contribution in order
to ensure IR-finite results. A quark-to-photon fragmentation function
has been introduced in \citeres{Glover:1993xc,Glover:1993he} and the
non-perturbative fit parameters entering it were obtained in
\citere{Buskulic:1995au} and read 
\begin{equation}
\mu_0 = 0.14 \GeV \qquad \textrm{and} \qquad C = -13.26 .
\end{equation}
The quark-to-photon fragmentation function has been implemented in
\mocanlo following \linebreak
\citeres{Denner:2009gj,Denner:2010ia,Denner:2011vu,Denner:2014ina,Denner:2019vbn}.

The quark-to-photon
fragmentation function is needed in two different situations.  The
first case appears in the calculation of EW corrections (real photon
radiation) to a LO process containing a final-state gluon.  An example
is the process $\Pp\Pp \to \PZ \Pj$ \cite{Denner:2011vu}, which
contains the partonic process $\Pq\bar\Pq \to \PZ \Pg$.  When
considering real photonic corrections, a hard photon can be recombined
with a soft jet.  Such a configuration leads to a QCD IR singularity
that is not cancelled by any standard dipole but needs a
quark-to-photon fragmentation function.  The appropriate recombination
rule for this case is given in \refli{lst:fragmentation_jet}.
\begin{lstlisting}[language=XML,caption=Use of fragmentation function
  for final states with gluons in \texttt{cut\_card.xml}. ,label=lst:fragmentation_jet]
   <recombination type="different_type">
       <name>              jet-photon </name>
       <recombinant>       1          </recombinant>
       <recombinendus>     3          </recombinendus>
       <minimum_distance>  0.1d0      </minimum_distance>
       <jet_type_selector> 0.7d0      </jet_type_selector>
   </recombination>
\end{lstlisting}
If the energy fraction of the photon (\xmltag{recombinendus})
$$
z_\gamma = \frac{E_\mathrm{recombinendus}}{E_\mathrm{recombinendus}+E_\mathrm{recombinant}}
$$
is larger than $z_\mathrm{cut}$ (=\xmltag{jet_type_selector}), the
two objects are recombined into an object of type photon, otherwise
they are recombined into a jet of type \xmltag{recombinant}.
The real and integrated dipole-subtraction terms inherit the splitting
cut $z_\mathrm{cut}$ from the recombination rule
(parameter \xmltag{jet_type_selector}) and the corresponding
integrated dipoles, \ie the ones with a fermionic emitter and a massless vector boson as emissus,
take the quark-to-photon
fragmentation function properly into account.

In the second case, QCD corrections are computed for a LO process
containing a final-state photon.  An example 
is the process $\Pp\Pp \to \PZ \gamma$ \cite{Denner:2015fca}.  When
considering QCD corrections, a soft photon can be recombined with a
hard quark jet, for instance in the process $\Pu\Pg\to\PZ\gamma\Pu$.
This gives rise to a collinear singularity that is compensated by a
contribution of the quark-to-photon fragmentation function in
combination with the LO process $\Pu\Pg\to\PZ\Pu$, as
discussed in \citeres{Denner:2014bna,Denner:2015fca}.  
The appropriate recombination rule for this case is given in
\refli{lst:fragmentation_photon}.
If the energy fraction of the jet (\xmltag{recombinendus})
$$
z_\Pj = \frac{E_\mathrm{recombinendus}}{E_\mathrm{recombinendus}+E_\mathrm{recombinant}}
$$
{\sloppy
is larger than $z_\mathrm{cut}$ (=\xmltag{jet_type_selector}), the objects
are recombined into a jet of the type \xmltag{recombinendus},
otherwise they are recombined into an object of type photon (\xmltag{recombinant}).
As in the previous case, the real and integrated dipole-subtraction terms inherit the splitting
cut from the recombination rule (parameter \xmltag{jet_type_selector}) and the
integrated dipoles,
\ie the ones with a fermionic emitter and a massless vector boson as emissus,
take the quark-to-photon fragmentation function
properly into account.
}

\begin{warning}{Restriction}
\sloppy
The quark-to-photon fragmentation function  is only implemented for the case of a photon as
\xmltag{recombinant} and a massless quark, specifically with
\texttt{PDG code}${}\le5$, as \xmltag{recombinendus}. 
\end{warning}

\subsubsection{Photon-to-jet conversion function}

The second type of IR singularities associated with photons concerns
the splitting of a virtual photon into a quark--antiquark pair,
$\gamma^* \to \Pq \bar \Pq$, in the final state. These singularities
are absorbed in \mocanlo\ by using a photon-to-jet conversion function
as discussed in detail in \citeres{Denner:2019zfp,Denner:2019tmn}. In
\mocanlo\ the photon-to-jet conversion function is switched on by
default in integrated dipoles for the splitting of a photon into a
quark--antiquark pair. 
The implementation is based on the 5-flavour scheme, and the finite
contributions are distributed according to their squared charges to
the five light quarks.
The value for
$\Delta\alpha^{(5)}_{\text{had}}(\MZ^2)$ is taken from
\citere{Keshavarzi:2018mgv} and hard-coded:
$$
\Delta\alpha^{(5)}_{\text{had}}(\MZ^2) = 276.11\times 10^{-4}.
$$.

\subsubsection{DIS factorisation scheme for photon PDFs}
For photon PDFs in hadron collisions, the DIS factorisation scheme can
also be used.  It can be switched on adding the entry
\begin{lstlisting}[language=XML]
                <DIS_scheme> 1 </DIS_scheme>
\end{lstlisting}
to the \xmltag{run_parameters} section of the \texttt{param\_card.xml}.
For the default value \xmltag{DIS\_scheme}${}=0$, the
\MSbar\ factorisation scheme is used, while
\xmltag{DIS\_scheme}${}=1$ turns on the DIS factorisation scheme
only for photon PDFs.

\subsection{Cuts}
\label{se:cut-scheme}

In \mocanlo, the cuts are not only used for event selection, but they
also serve to tag particles, order them, and 
create reconstructed objects.
These cuts take the form of a list of operations to be applied
sequentially to the final-state clusters resulting from the recombination.
It is important to note that the order in which the cuts are applied is crucial.

The cuts operate on {\em sorted clusters}, which contain for each jet type
a list of {\em recombinants}, resulting from the recombination. 
Recombinants, jets, and clusters are used as synonyms
in the following, even if these can be any
kind of clusters of particles or single particles like leptons. The recombinants
are objects of a specific \xmltag{jet_type}, as defined in
column 5 of \refta{tab:ParticleNames}, or of new jet types within the range
30--50 generated by the recombination or cut procedure. 
Each jet in a list of recombinants has two additional
attributes, a \xmltag{jet_tag} and a \xmltag{jet_order}, 
which might be changed by applying a cut.
By default, after recombination and before applying any cut,
\xmltag{jet_tag} and \xmltag{jet_order} are equal 
to 0, and the lists of recombinants of each \xmltag{jet_type} 
are ordered according to their transverse momenta.
The cuts do not change the entries in the list of
recombinants, but only modify their associated parameters
\xmltag{jet_tag} and \xmltag{jet_order}.
Besides the lists of recombinants, sorted clusters
contain two further lists for each \xmltag{jet_type}: the \emph{ordered recombinants} and the
\emph{reconstructed recombinants}, which are both initially empty. 
The reconstructed recombinants contain the 
reconstructed objects created by the \emph{reconstruction cuts}, and the
lists of ordered recombinants are filled subsequently whenever a cut
\emph{orders} jets.

As mentioned above, the cuts allow to tag and order the jets, to apply cuts on jets
with specific \xmltag{jet_tag} and \xmltag{jet_order}, and to create
new types of jets upon combining existing jets. Tagging a jet
sets the corresponding \xmltag{jet_tag} to one. Ordering a jet adds
this jet to the list of ordered recombinants and sets its
\xmltag{jet_order} to a value equal to the position of the
object in the list of ordered recombinants for the corresponding
\xmltag{jet_type}. 
The same jet is still present in the list of
recombinants, but now with parameter \xmltag{jet_order} different from zero.
Reconstruction cuts create new jets upon combining or cloning existing
ones and add them to the lists of reconstructed
recombinants. Upon ordering them, they also enter the corresponding
lists of ordered recombinants.

Besides eliminating events and jets, the cuts define the jets
that enter the histograms and the definition of dynamical scales.  To
this end, the usage of the lists of ordered recombinants is
recommended, because only these are uniquely defined.
Thus, all jets intended for histogram filling or scale definition
should be tagged and ordered. Once a given type of clusters has been ordered,
no further cuts must be applied that could remove or reorder any of the ordered
jets of this type.
Otherwise, subsequent access to these jets, for example by additional cuts,
scale definitions, or histogram calls, may lead to run-time errors.

{\sloppy
Even if a mechanism has been implemented to warn the user in case of
inconsistencies in the input, not all problematic cases might be catched.
In order to further check the correct implementation of the cuts in the
\texttt{cut\_card.xml}, it might be useful to compile the code with the
flag \texttt{DEBUG\_CUTS} switched on in
\texttt{include/debug\_flags.h}, to run it with a small number
of events, \eg using the tag \xmltag{max_generated_events}, 
and to check whether the cuts act as intended.}

After this introduction, we turn to a more detailed description
of the cut implementation. In a typical application, at a given
stage of the cut sequence,
tagged jets (\xmltag{jet_tag}${}=1$), which are distinguished by their
order, tagged jets without an order (\xmltag{jet_order}${}=0$), and 
untagged jets (\xmltag{jet_tag}${}=0$), which are unordered, are present simultaneously.
In addition, there can be discarded jets with \xmltag{jet_tag}${}=-1$.
Some examples of cuts are shown in \refli{lst:cuts2} and explained in detail later
in this section.
\begin{figure}
\begin{lstlisting}[language=XML,caption=Cuts in
  \texttt{cut\_card.xml}. ,label=lst:cuts2]
<cuts id="vbs">
    <cut type="transverse_momentum">
        <name>        jet_pt_cut </name>
        <jet_type>    1          </jet_type>
        <jet_tag>     0          </jet_tag>
        <jet_order>   0          </jet_order>
        <target>      1          </target>
        <action>      1          </action>
        <min_value>   30.0       </min_value>
        <n_required>  2          </n_required>
    </cut>
    ...
    <cut type="distance">
        <name> ep_tag_jet_distance_cut </name>
        <jet_type_1>  1          </jet_type_1>
        <jet_tag_1>   1          </jet_tag_1>
        <jet_order_1> 0          </jet_order_1>
        <jet_type_2>  4          </jet_type_2>
        <jet_tag_2>   0          </jet_tag_2>
        <jet_order_2> 0          </jet_order_2>
        <target>      1          </target>
        <action>      0          </action>
        <min_value>   0.4        </min_value>
        <n_required>  2          </n_required>
    </cut>
    ...
    <cut type="rapidity">
        <name> jet_rapidity_cut  </name>
        <jet_type>    1          </jet_type>
        <jet_tag>     1          </jet_tag>
        <jet_order>   0          </jet_order>
        <target>      1          </target>
        <action>      0          </action>
        <order>       1          </order>
        <max_value>   4.7        </max_value>
        <n_required>  2          </n_required>
    </cut>
    ...
    <cut type="invariant_mass ">
        <name> invariant_mass_tag_jets </name>
        <jet_type>    1 1        </jet_type>
        <jet_tag>     1 1        </jet_tag>
        <jet_order>   1 2        </jet_order>
        <min_value>   400.0      </min_value>
    </cut>
</cuts>
\end{lstlisting}
\end{figure}

All cuts are defined in the section \xmltag{cuts} in
\texttt{cut\_card.xml}. Each cut is set via the XML tag
\xmltag{cut} and can be given an arbitrary name with the tag
\xmltag{name}, which is only used for output purposes. 
Cuts always come with a type, which is defined by the tag attribute
\texttt{type}. There are four classes of cuts: single-particle, pair-particle,
multi-particle, and reconstruction cuts. All these are explained in the following.

Each cut acts on a single jet or on a set of jets, characterised by their type,
tag, and order status. If no matching jets are found, the cut either has no
effect or results in the removal of the event, depending on its definition.
The latter is to a large extent determined by three more tags: the \xmltag{target},
the \xmltag{action}, and the \xmltag{order} of a cut.

The \xmltag{target} tag specifies the object the cut
acts on. It can take the values \xmltag{target}${}=0$, if acting on the whole
event, \xmltag{target}${}=1$, if on the jet of \xmltag{jet_type} or
\xmltag{jet_type_1}, or \xmltag{target}${}=2$, if on the jet of
\xmltag{jet_type_2} in a pair-particle cut, and even larger values in
multi-particle or reconstruction cuts. By default, \mocanlo sets 
\xmltag{target}${}=0$ for all cuts apart from the reconstruction cuts,
where the default is \xmltag{target}${}=1$. In a reconstruction cut,
for \xmltag{target}${}=2$ the tagging applies to both lists
\xmltag{jet_type_1} and \xmltag{jet_type_2} (if present), while the cut acts
always on the last object (most right) in \xmltag{jet_type_1}.

The tag \xmltag{action} defines the action of the cut, with a default value \xmltag{action}${}=-1$.
For \xmltag{action}${}=-1$ and \xmltag{target}${}=0$, the event is discarded,
while for  \xmltag{action}${}=-1$ but \xmltag{target}${}>0$, the corresponding
target of the cut is eliminated, if the value of the corresponding
observable is not between \xmltag{min_value} and \xmltag{max_value}. For instance, if \xmltag{target}${}=1$, the
jet of \xmltag{jet_type} or \xmltag{jet_type_1} is removed, \ie its parameter
\xmltag{jet_tag} is set to $-1$, if the value of the corresponding
observable is outside the cut range.
For \xmltag{action}${}=1$ and
\xmltag{target}${}=0$, the cut acts as a veto, \ie if the variable is
within the cut range the
event is discarded.  For \xmltag{action}${}=1$ and
\xmltag{target}{}$=1$, the jet is tagged, \ie its parameter
\xmltag{jet_tag} is set to $1$, if it passes the cut, \ie if the value
of the corresponding observable is in the range between \xmltag{min_value} and
\xmltag{max_value}. Conversely, for \xmltag{action}${}=0$ and
\xmltag{target}${}=1$, the jet is untagged, \ie its parameter
\xmltag{jet_tag} is set to $0$, if it does not pass the cut. 
The tagging and untagging acts on all
jets with specified \xmltag{jet_type}, \xmltag{jet_tag}, and
\xmltag{jet_order}. For \xmltag{action}${}=-1$, all jets with specified
\xmltag{jet_type} and \xmltag{jet_order}, but with tag parameter
greater or equal to \xmltag{jet_tag}, are discarded.

Finally, cuts also come with an \xmltag{order} tag, whose meaning 
depends on the cut type.
All single-particle cuts with the tag \xmltag{order}${}=1$ and
pair-particle cuts with the tag \xmltag{order_1}${}=1$ or
\xmltag{order_2}${}=1$ order all respective tagged jets  after application 
of the cut by setting their parameter \xmltag{jet_order} to
consecutive values larger than zero. The default is
\xmltag{order}${}=0$, which means do not order. 
Multi-particle cuts and reconstruction cuts with \xmltag{order}${}=1$
set the order of the tagged jets to the number of previously tagged
jets plus one. 

\begin{warning}{Ordering of jets}
  Ordering via single- or pair-particle cuts must be done only once
  per jet type, since these order all tagged jets of the respective
  type. Tagging via multi-particle cuts and reconstruction cuts should
  only be done when no re-ordering via single- or pair-particle cuts
  follows thereafter. For reconstruction cuts, which should normally
  be the last cuts to be applied, \xmltag{order}${}=1$ implies
  ordering of all jets tagged by this cut, even if these were tagged
  before.
\end{warning}

A \textbf{single-particle cut} is defined for the observable of a
single particle, \eg a cut on its transverse momentum. 
The corresponding parameters, possible values, and default values are
summarised in \refta{tab:single_cuts1}. A
single-particle cut requires the tag \xmltag{jet_type} specifying the
single jet the cut is applied to, a minimum and/or a maximum value in
\xmltag{min_value} and/or \xmltag{max_value}, respectively, and the number of
required particles \xmltag{n_required} of this jet type in the final state. Optionally, \xmltag{jet_tag} and
\xmltag{jet_order} can be specified. Otherwise, these are assumed to
be zero, meaning that the cut applies to all untagged and unordered jets.
For a sensible cut, at least the tags \xmltag{jet_type} and either
\xmltag{min_value} or \xmltag{max_value} have to be provided.
The action of single-particle cuts is summarised in \refta{tab:single_cuts2}.
\begin{table}
      \centering
      \renewcommand\arraystretch{1.2}
      \rowcolors{2}{tablerowcolor}{}
      \begin{tabular}{lll}
        \toprule\rowcolor{tableheadcolor} 
        \textbf{Parameter} & \textbf{possible values} &
        \textbf{default value}  \\
        \midrule
      \xmltag{name} & string                   & none  \\
      \xmltag{jet_type} & $1,2,\ldots,50$                & $~0$  \\
      \xmltag{jet_tag} &  $0,1$                        & $~0$  \\
      \xmltag{jet_order} &  $0,1,\dots$                & $~0$  \\
      \xmltag{target}    &  $0,1$                      & $~0$  \\
      \xmltag{action}    &  $-1,0,1$                   & $-1$  \\
      \xmltag{order}    &  $0,1$                       & $~0$  \\
      \xmltag{n\_required}    &  $0,1,\ldots$          & $~0$  \\
      \xmltag{min\_value}    &  real number          & $0.0$  \\
      \xmltag{max\_value}    &  real number          & huge($0.0$) \\
      \bottomrule
      \end{tabular}
      \caption{\label{tab:single_cuts1} Parameters and possible values of
        single-particle cuts. For a cut of type \texttt{rapidity} or
        \texttt{pseudorapidity}, the default value of
        \xmltag{min\_value} is the negative of  \xmltag{max\_value}.}
\end{table}
\begin{table}
      \centering
      \renewcommand\arraystretch{1.2}
      \rowcolors{2}{tablerowcolor}{}
      \begin{tabular}{ccp{.7\textwidth}}
        \toprule\rowcolor{tableheadcolor} 
        \xmltag{target} & \xmltag{action} & \textbf{result}  \\
         $0$ &  $-1$ & remove event if less than \xmltag{n\_required} jets
         pass the cut \\
         $0$ &   $~1$ & remove event 
         if the cut variable for at least one jet is within \xmltag{min_value} and
\xmltag{max_value} (veto).
         \newline  (parameter \xmltag{n\_required} not allowed!) \\
         $1$ &  $~1$ & tag jet if it passes the cut and 
                       remove event if less than \xmltag{n\_required}
                       tagged jets pass the cut \\
         $1$ &  $~0$ & untag jet if it does not pass the cut and 
                       remove event if less than \xmltag{n\_required}
                       tagged jets pass the cut \\
         $1$ &  $-1$ & remove jet if it does not pass the cut and 
                       remove event if less than \xmltag{n\_required}
                       (tagged or untagged) jets pass the cut. Jets
                       are removed irrespective of their tagging status.\\
      \midrule
        \xmltag{order} &          & \textbf{result}  \\
         $1$ &   & order tagged jets according to $p_{\mathrm{T}}$ and
         add them to the list of ordered recombinants\\
      \bottomrule
      \end{tabular}
      \caption{\label{tab:single_cuts2} Action of single-particle cuts.}
\end{table}

Available single-particle cut types are: \nobreak
\begin{itemize}
\item \texttt{transverse\_momentum}, \quad $p_{\mathrm{T}}=\sqrt{p_x^2+p_y^2}$,
\item \texttt{energy}, \quad $E=p_0$, 
\item  \texttt{rapidity}, \quad $\displaystyle y = \frac12\ln\left(\frac{p_0+p_z}{p_0-p_z}\right)$,
\item  \texttt{pseudorapidity}, \quad $\displaystyle \eta = \frac12\ln\left(\frac{|\vec{p}|+p_z}{|\vec{p}|-p_z}\right)$,
\item  \texttt{theta\_angle}, \quad  $\displaystyle\frac{180}{\pi}\arccos\left(\frac{p_z}{|\vec{p}|}\right)$,
\item  \texttt{single\_invariant\_mass}, \quad $M=\sqrt{p^2}$.
\end{itemize}
Note that the cut
\texttt{single\_invariant\_mass} acts on the invariant mass of a
single jet.

In the example in \refli{lst:cuts2}, the first cut acts on 
untagged and unordered jets of \xmltag{jet_type}${}=1$ and requires 
a transverse-momentum of $\ptsub{\Pj}>30\GeV$. It tags jets that fulfil this criterion
and requires two tagged jets in the final state.  The third cut in
\refli{lst:cuts2}  is a rapidity cut on
tagged jets with  $|y_{\Pj}|<4.7\GeV$, which untags jets and orders
the remaining tagged jets. 

A \textbf{pair-particle cut} is defined for an observable of two
clusters, \eg their $\Delta R_{ij}$ distance or their invariant mass
$m_{ij}$.  A pair-particle cut requires two jet types via
\xmltag{jet_type_1} and \xmltag{jet_type_2}, possibly being the same
type, corresponding jet tags \xmltag{jet_tag_1} and
\xmltag{jet_tag_2}, and jet orders \xmltag{jet_order_1} and
\xmltag{jet_order_2}.  Jet tags and orders that are not
specified are assumed to be zero. The associated cut observable is evaluated
for all cluster pairs with the two jet types, jet tags, and jet orders
in the final state, excluding pairs of identical clusters. 
The complete set of parameters for pair-particle cuts is listed in
\refta{tab:pair_cuts1}, while the actions of the cuts are summarised in \refta{tab:pair_cuts2}.
The parameter \xmltag{n_required} is not allowed for \xmltag{target}${}=0$.
For \xmltag{target}${}=1$ or \xmltag{target}${}=2$, \xmltag{n_required}
applies to \xmltag{jet_type_1} or \xmltag{jet_type_2}, respectively. 
Tagging and ordering are restricted to  jets within \xmltag{jet_type_1} for \xmltag{target}${}=1$
and to jets within \xmltag{jet_type_2} for \xmltag{target}${}=2$.
\begin{table}
      \centering
      \renewcommand\arraystretch{1.2}
      \rowcolors{2}{tablerowcolor}{}
      \begin{tabular}{lll}
        \toprule\rowcolor{tableheadcolor} 
        \textbf{Parameter} & \textbf{possible values} &
        \textbf{default value}  \\
        \midrule
      \xmltag{name} & string                   & none  \\
      \xmltag{jet_type_1} & $1,2,\ldots,50$              & $~0$  \\
      \xmltag{jet_tag_1} &  $0,1$                        & $~0$  \\
      \xmltag{jet_order_1} &  $0,1,\dots$                & $~0$  \\
      \xmltag{jet_type_2} & $1,2,\ldots,50$                   & $~0$  \\
      \xmltag{jet_tag_2} &  $0,1$                        & $~0$  \\
      \xmltag{jet_order_2} &  $0,1,\dots$                & $~0$  \\
      \xmltag{target}    &  $0,1,2$                      & $~0$  \\
      \xmltag{action}    &  $-1,0,1$                   & $-1$  \\
      \xmltag{order_1}    &  $0,1$                       & $~0$  \\
      \xmltag{order_2}    &  $0,1$                       & $~0$  \\
      \xmltag{n\_required}    &  $0,1,\ldots$          & $~0$  \\
      \xmltag{min\_value}    &  real number          & $~0.0$  \\
      \xmltag{max\_value}    &  real number          & huge($0.0$) \\
      \bottomrule
      \end{tabular}
      \caption{\label{tab:pair_cuts1} Parameters and possible values of
        pair-particle cuts.}
\end{table}
\begin{table}
      \centering
      \renewcommand\arraystretch{1.2}
      \rowcolors{2}{tablerowcolor}{}
      \begin{tabular}{ccp{.7\textwidth}}
        \toprule\rowcolor{tableheadcolor} 
        \xmltag{target} & \xmltag{action} & \textbf{result}  \\
         $0$ &  $-1$ & remove event if at least one specified jet pair does not pass the cut\\
         $0$ &  $~1$ & remove event if at least one specified jet pair passes the cut (veto)\\
         $1$ &  $~1$ & tag jet  of type \xmltag{jet_type_1} if it passes the cut and 
                       remove event if less than \xmltag{n\_required}
                       tagged jets of type \xmltag{jet_type_1} pass the cut \\
         $1$ &  $~0$ & untag jet  of type \xmltag{jet_type_1} if it does not pass the cut and 
                       remove event if less than \xmltag{n\_required}
                       tagged jets  of type \xmltag{jet_type_1} pass the cut \\
         $1$ &  $-1$ & remove jet  of type \xmltag{jet_type_1} if it does not pass the cut and 
                       remove event if less than \xmltag{n\_required}
                       (tagged or untagged) jets of type \xmltag{jet_type_1} pass the cut \\
         $2$ &  $~1$ & tag jet  of type \xmltag{jet_type_2} if it passes the cut and 
                       remove event if less than \xmltag{n\_required}
                       tagged jets of type \xmltag{jet_type_2} pass the cut \\
         $2$ &  $~0$ & untag jet  of type \xmltag{jet_type_2} if it does not pass the cut and 
                       remove event if less than \xmltag{n\_required}
                       tagged jets  of type \xmltag{jet_type_2} pass the cut \\
         $2$ &  $-1$ & remove jet  of type \xmltag{jet_type_2} if it does not pass the cut and 
                       remove event if less than \xmltag{n\_required}
                       (tagged or untagged) jets of type \xmltag{jet_type_2} pass the cut \\
      \midrule
         \xmltag{target} & \xmltag{order_1} &    \textbf{result}  \\
         $1$ & $1$  & order tagged jets of type  \xmltag{jet_type_1}
         according to $p_{\mathrm{T}}$ and 
         add them to the list of ordered recombinants\\
      \midrule
         \xmltag{target} & \xmltag{order_2} &    \textbf{result}  \\
         $2$ & $1$  & order tagged jets of type  \xmltag{jet_type_2}
         according to $p_{\mathrm{T}}$ and 
         add them to the list of ordered recombinants\\
      \bottomrule
      \end{tabular}
      \caption{\label{tab:pair_cuts2} Action of pair-particle cuts.}
\end{table}

Available pair-particle cut types are: 
\begin{itemize}
\item \texttt{distance}, \quad $\Delta R_{12}=\sqrt{(y_1-y_2)^2 +
    (\Delta\phi)^2}$, \quad $\Delta\phi = \min(|\phi_1-\phi_2|,2\pi-|\phi_1-\phi_2|)$,
\item \texttt{pseudodistance}, \quad $\Delta r_{12}=\sqrt{(\eta_1-\eta_2)^2 + (\Delta\phi)^2}$,
\item  \texttt{pair\_invariant\_mass}, \quad $M_{12}=\sqrt{(p_1+p_2)^2}$,
\item  \texttt{pair\_transverse\_momentum}, 
\quad $p_{\mathrm{T},12}=\sqrt{(p_{x,1}+p_{x,2})^2+(p_{y,1}+p_{y,2})^2}$,
\item  \texttt{angular\_separation}, \quad
  $\displaystyle\Delta\theta_{12}=\frac{180}{\pi}\arccos\left(\frac{\vec{p}_1\cdot\vec{p}_2}{|\vec{p}_1||\vec{p}_2|}\right)$,
\item  \texttt{rapidity\_separation}, \quad $\Delta y_{12}=|y_1-y_2|$,
\item  \texttt{pseudorapidity\_separation}, \quad $\Delta \eta_{12}=|\eta_1-\eta_2|$,
\item  \texttt{rapidity\_product}, \quad $y_1\times y_2$.
\end{itemize}

The second cut defined in \refli{lst:cuts2} is a distance cut between
all positrons and tagged jets in the final state of $\Delta R>0.4$, 
which is an isolation criterion so that jets too close to the lepton
are removed from the list of jets.

A \textbf{multi-particle} cut involves an arbitrary number of
particles. In contrast to the single-particle cuts, \xmltag{jet_type}
and the optional parameters \xmltag{jet_tag} and \xmltag{jet_order}
are lists of integers. If the latter two lists are not specified or
shorter than the list \xmltag{jet_type}, they are complemented with
zero entries. The cut function depends on all specified jet entries,
an example being the invariant mass of the sum of all momenta of the
specified jets. The complete set of parameters for multi-particle cuts is listed in
\refta{tab:multi_cuts_1} and their actions in \refta{tab:multi_cuts_2}. 
For \xmltag{target}${}=0$, the action can be an
ordinary cut if \xmltag{action}${}=-1$ or a veto cut if
\xmltag{action}${}=1$. A \xmltag{target}${}>0$ is only allowed for
specific cases mentioned below.
The parameter \xmltag{n_required} is not supported for multi-particle cuts.
\begin{table}
      \centering
      \renewcommand\arraystretch{1.2}
      \rowcolors{2}{tablerowcolor}{}
      \begin{tabular}{llll}
        \toprule\rowcolor{tableheadcolor} 
        \textbf{Parameter} & \textbf{possible values} &
        \textbf{default value} & \textbf{remark}  \\
        \midrule
      \xmltag{name} & string                   & none  \\
      \xmltag{jet_type} & list of jet types                   & none &
      $n_{\mathrm{jet}} $ \\
      \xmltag{jet_tag} &  corresponding list of jet tags      & [0,\ldots,0]
      & $\le n_{\mathrm{jet}}$ \\
      \xmltag{jet_order} &   corresponding list of jet orders &
      [0,\ldots,0] 
      & $\le n_{\mathrm{jet}}$  \\
      \xmltag{target}    &  $0,1,\ldots,n_{\mathrm{jet}}$                      & $~0$ & $>0$
      only for specific cuts  \\
      \xmltag{action}    &  $-1,1$                   & $-1$   & $-1$
      only for  \xmltag{target} = 0 \\
      \xmltag{order}    &  $0,1$                       & $~0$  & only
      for specific cuts\\
      \xmltag{min\_value}    &  real number          & $~0.0$  &\\
      \xmltag{max\_value}    &  real number          & huge($0.0$) &\\
      \xmltag{mid\_value}    &  real number          & $~0.0$ & only
      for specific cuts\\
      \bottomrule
      \end{tabular}
      \caption{\label{tab:multi_cuts_1} Parameters and possible values of
        multi-particle cuts.}
\end{table}
\begin{table}
      \centering
      \renewcommand\arraystretch{1.2}
      \rowcolors{2}{tablerowcolor}{}
      \begin{tabular}{ccp{.7\textwidth}}
        \toprule\rowcolor{tableheadcolor} 
        \xmltag{target} & \xmltag{action} & \textbf{result}  \\
         $0$   &  $-1$ & remove event if cut not passed \\
         $0$   &  $~1$ & remove event if cut passed (veto)\\
         $n>0$ &  $~1$ & tag first $n$ jets of list \xmltag{jet_type} if
         the cut is passed and  remove event if cut not passed\\ 
      \midrule
        \xmltag{order} &          & \textbf{result}  \\
         $1$ &   & set order of jets tagged by this cut and append them 
         to the list of ordered recombinants (only
         for specific cuts)\\
      \bottomrule
      \end{tabular}
      \caption{\label{tab:multi_cuts_2} Action of multi-particle cuts.}
\end{table}

Available multi-particle cut types are: 
\begin{itemize}
\item \texttt{missing\_transverse\_momentum}, \quad
$\ptsub{\text{miss}} = \left|\sum_i \vec{p}_{\text{T},\nu_i}\right|$,
(sum over all neutrinos),
\item \texttt{missing\_energy}, \quad 
$E_{\text{miss}} = \left|\sum_i E_{\nu_i}\right|$,
(sum over all neutrinos),
\item  \texttt{invariant\_mass}, \quad $M = \sqrt{(\sum_i p_i)^2}$,
(sum over all specified jets),
\item  \texttt{mt\_jets\_neutrinos}, \quad
$M_\mathrm T = \sqrt{2\ptsub{\mathrm{jets}}p_{\mathrm{ T,miss}}
[1 - \cos  \Delta\phi(p_{\mathrm{jets}},p_{\mathrm{miss}})]}$, \\
\strut\hspace{10em} ($p_{\mathrm{jets}}={}$momentum sum of all specified jets),
\item  \texttt{zeppenfeld\_variable}, \quad
$\displaystyle z = \frac{2y_1 -y_2-y_3}{y_2-y_3}$,
\item  \texttt{jet\_number}, \quad  $N_{\mathrm{j}} = \sum_i 1$,
(sum over all specified jets),
\item  \texttt{max\_invariant\_mass}, \quad
$M_{\text{max}} = \max_{i_k}\sqrt{(\sum_k p_{i_k})^2}$,\\
\strut\hspace{10em}
(maximum over all allowed jets $i_k$ for each specified type $k$),
\item  \texttt{mid\_invariant\_mass}, \quad
$M = M_{\text{mid}}\pm\min_{i_k}|\sqrt{(\sum_k p_{i_k})^2}-M_{\text{mid}}|$,\\
\strut\hspace{10em}
(minimum over all allowed jets $i_k$ for each specified type $k$),
\item  \texttt{leading\_transverse\_momentum}, \quad
$\ptsub{\text{lead}} = \max_i \vec{p}_{\text{T},i}$,
(maximum over all specified jets).
\end{itemize}
The first two cuts apply to the vector sum of the transverse momenta and
energies of all neutrinos, respectively. No \xmltag{jet_type} input is needed, since
all jets of type 6 and 21--26 are considered as neutrinos.
The cut \texttt{invariant\_mass} applies to the invariant mass of all
jets of the list \xmltag{jet_type},
and \texttt{mt\_jets\_neutrinos} to the
transverse mass obtained from the sum of all neutrino momenta and the
sum of the momenta of all specified jets.
The latter quantity depends on  the azimuthal angle $\Delta\phi $ between the
sum of the jet momenta and the sum of the neutrino momenta.
The cut \texttt{zeppenfeld\_variable} acts on the Zeppenfeld
variable (or centrality), defined via the rapidities of the three(!) listed jets,
and the cut \texttt{jet\_number} restricts the total number of jets of the specified
types via the input values \xmltag{min_value} and \xmltag{max_value}.

Among all multi-particle cuts, the ones discussed in this paragraph support tagging and ordering.
The cut \texttt{max\_invariant\_mass} acts on the maximal invariant
mass calculated from all possible jets of the specified types, \eg
for \xmltag{jet\_type}${}= 1~5$ on the largest invariant mass of
all jet--electron pairs, while for \xmltag{jet\_type}${}= 1~1~1$ on the
 largest invariant mass of all combinations of three QCD jets. The cut
\texttt{mid\_invariant\_mass} works analogously, but applies to the 
invariant mass closest to the value indicated in
\xmltag{mid\_value}. These two cuts are particularly useful for tagging,
as they determine the jets that form the maximal invariant mass or
invariant mass closest to some given value.  For \xmltag{target}${}=
n$, they tag the first $n$ jets of the list \xmltag{jet\_type} 
that contribute to the invariant mass relevant for the
cut. Note that $n$ must be equal to or smaller than the number of specified jet types, which
 allows to tag all jets that form the invariant mass or only a
subset. Only one jet per entry in \xmltag{jet\_type} is
tagged. Finally, for the cut \texttt{leading\_transverse\_momentum}, the
entry with the leading transverse momentum is tagged, if  $n$ is greater than
or equal to the position of the corresponding jet type in the list
\xmltag{jet\_type}. The cut acts on the transverse momentum of the jet with the highest transverse momentum among the specified jet types. For
\xmltag{target}${}>0$, only \xmltag{action}${}=1$ is allowed. For
\xmltag{order}${}=1$, the
jets tagged by this cut are appended to the existing list of ordered jets.
This is also the case if the jets were already tagged but would
  be tagged by the multi-particle cut as well.
If no jets fulfilling the tagging criterion are found, the event is cut.

A \textbf{reconstruction cut} serves to define reconstructed objects,
\eg reconstructed neutrinos or resonances, so that it does not
necessarily act as a cut. If, however, values for \xmltag{max_value} and
\xmltag{min_value} are provided, the corresponding cut acts on the
first reconstructed object. The full set of parameters for a
reconstruction cut is listed in \refta{tab:reco_cuts_1}.
\begin{table}
      \centering
      \renewcommand\arraystretch{1.2}
      \rowcolors{2}{tablerowcolor}{}
      \begin{tabular}{llll}
        \toprule\rowcolor{tableheadcolor} 
        \textbf{Parameter} & \textbf{possible values} &
        \textbf{default value} 
        & \textbf{remark}  
        \\
        \midrule
      \xmltag{name} & string                   & none  \\
      \xmltag{jet_type_i} & list of jet types                   & none & $n_i$\\
      \xmltag{jet_tag_i} &  corresponding list of jet tags      & [0,\ldots,0]  & $\le n_i$\\
      \xmltag{jet_order_i} &   corresponding list of jet orders & [0,\ldots,0]  & $\le n_i$\\
      \xmltag{parameter_i} & list of parameters for the cut     & none & \\
      \xmltag{reconstructed_particle_i} & reconstructed particle name & none & \\      
      \xmltag{target}    &  $0,1,\ldots,n_i-1$                         & $~1$ & \\
      \xmltag{action}    &  $-1,0,1$                   & $~0$ &  \\
      \xmltag{order}    &  $0,1$                       & $~0$ &  \\
      \xmltag{n\_required}    &  $0,1,\ldots$          & $~1$ & \\
      \xmltag{min\_value}    &  real number          & $~0.0$ & \\
      \xmltag{max\_value}    &  real number          & huge($0.0$) & \\
      \bottomrule
      \end{tabular}
      \caption{\label{tab:reco_cuts_1} Parameters and possible values of
        reconstruction cuts.}
\end{table}
A reconstruction cut involves an arbitrary number of particles,
organised in (up to two) groups. As for the multi-particle cuts,
\xmltag{jet_type_i} and the optional parameters \xmltag{jet_tag_i} and
\xmltag{jet_order_i}, $i=1,2$, are lists of integers. If the latter
two lists are not specified or shorter than the list
\xmltag{jet_type_i}, they are complemented with zero entries.  While
the last (rightmost) entry in each list, denoted as $n_i$th entry in the
following, defines the reconstructed object(s), the other entries
specify the clusters used to define the reconstructed object(s) and are
selected as for the multi-particle cuts above. More precisely, the last entry of
\xmltag{jet_type_i} defines the jet type of the reconstructed object,
which can be either an existing type or a new type in the range $30$--$50$,
the last entry of
\xmltag{jet_tag_i} specifies whether the reconstructed object
is tagged, and the last entry of \xmltag{jet_order_i}
determines whether the reconstructed tagged object is ordered
and thus added to the list of ordered objects. The jet order of the
reconstructed object is initially equal to zero. For cases where more
than one object is reconstructed, a value of the last entry of
\xmltag{jet_order_i} larger than zero indicates how many of them should be
ordered.

The properties of the reconstructed object can be defined in two
alternative ways, either with the (real) values of
\xmltag{parameter_i} or with the (string) values of
\xmltag{reconstructed_particle_i}.  When indicating its name using
\xmltag{reconstructed_particle_i}, the reconstructed object inherits
the properties of an existing particle. Otherwise, its mass and width can be
defined via the first two entries of the list \xmltag{parameter_i}.

The cut parameters \xmltag{action} and
\xmltag{order} work as for the multi-particle cuts, but apply only to
the clusters used for reconstruction and not to the reconstructed
ones. Note also that only jets that are tagged by
reconstruction cuts are ordered by them. This is also the case if the
jets were already tagged but would be tagged by the reconstruction cut as well.
It is worth emphasising that, as for any other cut,
the clusters used for reconstruction will either be tagged or untagged
by a reconstruction cut (but not necessarily left in the same tagging state as before the cut).
The action of the reconstruction cuts is summarised in \refta{tab:reco_cuts_2}.
Depending on the cut, one or more reconstructed objects can be present.
The event is discarded, if less than
\xmltag{n_required} objects are reconstructed for the first
group \xmltag{jet_type_1}, independently of the value of the cut
observable, or if the latter fulfils the cut condition (see below).
Default values for the  cut parameters \xmltag{target}, \xmltag{action}, \xmltag{order}, and
\xmltag{n_required} are 1, 0, 0, and 1, meaning that the clusters used for
reconstruction are untagged and not ordered, and at least one recombined object must
be present.
\begin{table}
      \centering
      \renewcommand\arraystretch{1.2}
      \rowcolors{2}{tablerowcolor}{}
      \begin{tabular}{ccp{.7\textwidth}}
        \toprule\rowcolor{tableheadcolor} 
        \xmltag{target} & \xmltag{action} & \textbf{result on event
          and all but last jets specified by \xmltag{jet_type_i}}  \\
         $0$   &  $-1$ & remove event if cut not passed, no tagging/ordering \\
         $0$   &  $~1$ & remove event if cut passed (veto), no tagging/ordering\\
         $n>0$ &  $~1$ & tag all but last jets of list(s)
         \xmltag{jet_type_i} with $i\le n$ if
         the cut is passed and remove event if the cut is not passed\\ 
         $n>0$ &  $~0$ & untag all but last jets of list(s)
         \xmltag{jet_type_i} with $i\le n$ if
         the cut is passed and remove event if the cut is not passed\\ 
      \midrule
        \xmltag{order} &          & \textbf{result on
          all but last jets specified by \xmltag{jet_type_i}}  \\
         $1$ &   & set order of jets tagged by this cut and append them 
         to the list of ordered recombinants, if \xmltag{target}${}>0$\\
       \midrule
        \multicolumn{2}{l}{\xmltag{jet_type_i}$_{\mathrm{last}}$}   
& \textbf{result on last jet specified by \xmltag{jet_type_i}}, if \xmltag{target}${}>0$  \\
         $n$ &   & set type of reconstructed object to $n$ (use an
         existing particle type or a new one between 30 and 50)\\
       \midrule
        \multicolumn{2}{l}{\xmltag{jet_tag_i}$_{\mathrm{last}}$}  
        & \textbf{result on last jet specified by \xmltag{jet_type_i}}, if \xmltag{target}${}>0$  \\
         $1/0$ &   & tag/do not tag reconstructed object  \\
        \midrule
        \multicolumn{2}{l}{\xmltag{jet_order_i}$_{\mathrm{last}}$}  
        & \textbf{result on last jet specified by \xmltag{jet_type_i}}, if \xmltag{target}${}>0$  \\
         $1/0$ &   & order/do not order reconstructed objects, \ie add them to the
         list of tagged jets \\
         $n>0$ &   & generate new list of $n$ ordered objects, \\ 
               &   & only for cut
         \texttt{transverse\_momentum\_ordering} \\
        \midrule
        \multicolumn{2}{l}{\xmltag{n\_required}}  
        & \textbf{result on last jet specified by \xmltag{jet_type_1}}  \\
         $n$ &   &  require at least $n$ reconstructed objects for the
         first group \\ 
        &&  (not necessarily passing the cut condition) \\
    \bottomrule
      \end{tabular}
      \caption{\label{tab:reco_cuts_2} Action of reconstruction
        cuts. Only the listed combinations of \xmltag{target} and \xmltag{action} are allowed.
      }
\end{table}

\begin{warning}{Valid values for \xmltag{jet_type_i} created in
    the reconstruction cuts}   
  The new jet types appearing within the reconstruction cuts must correspond
  to existing jet types of \mocanlo as given in
  \refta{tab:ParticleNames} or new ones
  within the range $30$--$50$.
\end{warning}

{\sloppy
Available reconstruction cut types are: 
\begin{itemize}
\item \texttt{single\_neutrino\_pt\_reconstruction}, 
\item \texttt{staggered\_max\_pt\_tagging}, 
\item \texttt{two\_resonances\_likelihood\_reconstruction},
\item \texttt{resonance\_mass\_reconstruction}, \quad
\item \texttt{resonance\_trivial\_reconstruction}, \quad
\item \texttt{cluster\_cloning}, \quad
\item \texttt{transverse\_momentum\_ordering}.
\end{itemize}
Only the cut \texttt{two\_resonances\_likelihood\_reconstruction} requires
two lists of jet types, \linebreak 
\xmltag{jet_type_1} and \xmltag{jet_type_2}, while
all other reconstruction cuts expect one list  \xmltag{jet_type_1} (together
with corresponding list(s) for the tags \xmltag{jet_tag_i}, \xmltag{jet_order_i},
\xmltag{parameter_i}, \linebreak and \xmltag{reconstructed_particle_i}). 
}

{\sloppy
The cut \texttt{single\_neutrino\_pt\_reconstruction} allows to
reconstruct a massless object, more precisely its longitudinal
momentum $p_1^z$,
from the transverse component of the momentum $p_1$ of the first entry
in list \xmltag{jet_type_1} and the momenta $p_i$ of the $n_1-2$ entries of
the same list. The equation
$$
m_1^2 = \left(\sum_{i=1}^{n_1-1} p_i \right)^2
$$
is solved for the unknown $p_1^z$, where $m_1$ is the mass of the
cluster specified via the first entry of \xmltag{parameter_1}
or alternatively via \xmltag{reconstructed_particle_1}.
The reconstructed object, with the reconstructed longitudinal
momentum $p_1^z$, is put in the $n_1$th entry of \xmltag{jet_type_1}. If two
solutions exist, reconstructed objects are generated for both, the
first entry being the one with smaller longitudinal momentum. If the
solutions are complex, the real part is selected.  No cut is applied
if at least one reconstructed neutrino is found. 

The cut \texttt{staggered\_max\_pt\_tagging} creates a new jet type
with exactly one jet in its recombinant list from the list of all jet types specified by the
first $n_1-1$ entries of \xmltag{jet_type_1}. The new object is the
one with largest transverse momentum of the first type in
\xmltag{jet_type_1}. If this list is empty, the one with largest
transverse momentum of the second type in \xmltag{jet_type_1} is
searched. This is continued until one object is found or all $n_1-1$
types have been searched. The cut acts on
the transverse momentum of the resulting object.

The cut \texttt{two\_resonances\_likelihood\_reconstruction}
reconstructs two resonances from the objects in the first $n_1-1$ and
$n_2-1$ entries of the lists \xmltag{jet_type_1} and
\xmltag{jet_type_2}, respectively. From all allowed clusters with
fitting tags and orders, those objects are selected that maximise the
likelihood
$$
\mathcal{L}= \frac{1}{(M_1^2 - m_1^2)^2 + (m_1\Gamma_1)^2} \:
\frac{1}{(M_2^2 - m_2^2)^2 + (m_2\Gamma_2)^2}\,,
$$
where 
$$
M_1^2=\left(\sum_{i_1=1}^{n_1-1} p_{i_1}\right)^2, \qquad
M_2^2=\left(\sum_{i_1=1}^{n_2-1} p_{i_2}\right)^2,
$$
and $m_i$ and $\Gamma_i$ are the masses and widths of the two
resonances specified via the first two entries of \xmltag{parameter_i}
or via \xmltag{reconstructed_particle_i}. For \xmltag{target}${}=1$,
the tagging applies to all clusters that reconstruct the first
resonance, for \xmltag{target}${}=2$, it applies to all clusters
that reconstruct the two resonances. The tagging of the resonances is
steered by the values of the last entries in \xmltag{jet_tag_i}.  A
possible cut acts on the invariant mass of the reconstructed resonance
in the first list \xmltag{jet_type_1}.

The cuts \texttt{resonance\_mass2\_reconstruction} and \texttt{resonance\_mass\_reconstruction}
recon\-struct a resonance from the objects in the first $n_1-1$
entries of the list \xmltag{jet_type_1}. From all allowed clusters with
fitting tags and orders, those objects are selected that minimise the
distances
$$
\Delta= |M^2_1 - m^2_1| \quad\text{ or }\quad
\Delta= |M_1 - m_1|,
$$
respectively, where 
$$
M_1^2=\left(\sum_{i_1=1}^{n_1-1} p_{i_1}\right)^2,
$$
and $m_1$ is the mass of the resonance specified via the first
entry of \xmltag{parameter_1} or via \linebreak
\xmltag{reconstructed_particle_1}. Tagging applies to all clusters
that reconstruct the resonance. A possible cut acts on the invariant
mass of the reconstructed resonance upon setting the \xmltag{min\_value} and/or \xmltag{max\_value}.

The cut \texttt{resonance\_trivial\_reconstruction} reconstructs an
object from those in the first $n_1-1$ entries of the list
\xmltag{jet_type_1}. For each entry, the first allowed cluster with
fitting tags and orders is taken. Tagging applies to all clusters that
reconstruct the resonance. A possible cut acts on the invariant mass
of the reconstructed resonance.

The cut \texttt{cluster\_cloning} clones the objects in the first 
$n_1-1$ entries of the list \xmltag{jet_type_1} into the new jet 
type. For each entry, all allowed clusters with
fitting tags and orders are taken. No cut is applied. This cut does not
change the tagging status of the input jets as opposed to the other reconstruction cuts.
It can be used to combine objects with different jet types into one
new jet type to be used in subsequent cuts, histograms or scales.

The cut \texttt{transverse\_momentum\_ordering} serves to order
objects of different jet types according to their transverse momenta.
Up to the number of  objects specified by the $n_1$th entry of
\xmltag{jet_order_1} fitting to the first $n_1-1$ entries of
\xmltag{jet_type_1} are copied into a list for a new jet type (defined
by this cut) and ordered by their transverse momenta. 
This can for instance be used to order leptons of different types
(electrons, positrons, muons, \ldots) according to their transverse
momenta. A possible cut acts on the transverse momentum of the
selected object with largest transverse momentum. This cut does not
change the tagging of the input jets.

Examples for the use of reconstruction cuts are provided in \refli{lst:reco_cuts}.
\begin{figure}
\begin{lstlisting}[language=XML,caption=Examples for reconstruction
  cuts. ,label=lst:reco_cuts]
<cuts id="tzj">
    <cut type="two_resonances_likelihood_reconstruction ">
        <name>         top_Z_tagging </name>                  
        <jet_type_1>   16 6 2 7      </jet_type_1>     
        <jet_tag_1>    0  0 0 1      </jet_tag_1>     
        <jet_order_1>  0  0 0 0      </jet_order_1>
        <reconstructed_particle_1> t </reconstructed_particle_1>
        <jet_type_2>   4 5 8         </jet_type_2>
        <jet_tag_2>    0 0 1         </jet_tag_2>
        <jet_order_2>  0 0 0         </jet_order_2>
        <reconstructed_particle_2> z </reconstructed_particle_2>
        <n_required>   1             </n_required>
        <target>       2             </target>
        <action>       1             </action>
        <order>        1             </order>
    </cut>
    ...
    <cut type="transverse_momentum_ordering"> 
        <name>        lepton_pt_sort </name>
        <jet_type_1>   4  5 16 32    </jet_type_1>
        <jet_tag_1>    1  1  1  1    </jet_tag_1>
        <jet_order_1>  1  1  1  3    </jet_order_1>
        <n_required>   1             </n_required>
        <target>       0             </target>
        <order>        0             </order>
    </cut>
</cuts>
\end{lstlisting}
\end{figure}
The first cut, named \texttt{top\_Z\_tagging}, reconstructs a top
quark (\xmltag{jet_type}${}=7$) from all untagged positive
muons (16), neutrinos (6), and bottom jets (2), and a Z~boson (8) from
all untagged electrons (4) and positrons (5), and tags and orders all
objects that reconstruct the top quark and the Z boson. The parameters
for the masses and widths of the top quark and
Z~boson are taken from the values for the corresponding particles.
The second cut, named \texttt{lepton\_pt\_sort}, sorts all tagged electrons,
positrons and positively charged muons with \xmltag{jet_order} equal to 1
according to their transverse momenta and puts the first three objects
into the \xmltag{jet_type} 32.

The typical use of the cut scheme for a VBS process with jet tagging can be found in \refli{lst:cuts2}.
The first cut of type \texttt{transverse\_momentum} acts on all jets
(\xmltag{jet_type}${}=1$) and tags them if their transverse momentum is above
$30\GeV$. The following cut of type \texttt{distance} acts on tagged jets
and untags them if their distance to positrons (\xmltag{jet_type}${}=4$) is below
$0.4$. The third cut of type \texttt{rapidity} acts on tagged jets,
untags them if their rapidity is outside the interval $[-4.7,4.7]$,
and orders the tagged jets. Finally, the cut of type
\texttt{invariant\_mass} acts on tagged jets with specified order, \ie the
two leading transverse-momentum jets, and discards
events with $M_{\Pj_1\Pj_2}<400\GeV$. Note that the order of the cuts
is crucial.

\subsection{Fixing scales via the cut card}
\label{se:scale}

Dynamical scales can be defined via the cut card
(\texttt{cut\_card.xml}) by using generic scale types generated from 
selected final-state jets. To this end, the tag
\xmltag{dynamical_scale_type} (in the \xmltag{run_parameters} section of the \texttt{param\_card.xml})
must be set to 1 and the definition
of scales must be provided in the cut card.  
This requires a corresponding tag
<\texttt{scales id="\ldots"\/}> in \texttt{run\_card.xml}.  If this
tag is not found, the code will stop with an error message.

The central scale can be constructed by multiplicatively combining
several scale sums~$l$, which in turn are sums of
individual scales $k$.
The implemented types for individual scales are listed in
\refta{tab:ScaleTypes} and the corresponding parameters in \refta{tab:scale}.
\begin{table}
      \centering
      \renewcommand\arraystretch{1.2}
      \rowcolors{2}{tablerowcolor}{}
      \begin{tabular}{ll}
      \toprule\rowcolor{tableheadcolor} 
      \textbf{observable} & \textbf{type} \\
      \midrule
      $\left(\prod_{i=1}^n p_{\text{T},i}\right)^{1/n}$  & \texttt{transverse\_momentum\_product}\\
      $(\sum_{i=1}^n  p^m_{\text{T},i})^{1/m}$     & \texttt{transverse\_momentum\_sum}  \\
      $\max_i  p_{\text{T},i}$             & \texttt{transverse\_momentum\_maximum}  \\
      $\left(\prod_{i=1}^n E_{\text{T},i}\right)^{1/n}$ & \texttt{transverse\_energy\_product} \\
      $(\sum_{i=1}^n E^m_{\text{T},i})^{1/m}$              & \texttt{transverse\_energy\_sum}  \\
      $\max_i  E_{\text{T},i}$             & \texttt{transverse\_energy\_maximum}  \\
      $\left(\prod_{i=1}^n M_{\text{T},i}\right)^{1/n}$ & \texttt{transverse\_mass\_product} \\
      $(\sum_{i=1}^n  M^m_{\text{T},i})^{1/m}$              & \texttt{transverse\_mass\_sum}  \\
      $\max_i  M_{\text{T},i}$             & \texttt{transverse\_mass\_maximum}  \\
      $\left(\prod_{i=1}^n M_{i}\right)^{1/n}$          & \texttt{invariant\_mass\_product} \\
      $\left(\sum_{i=1}^n  M^m_{i}\right)^{1/m}$        & \texttt{invariant\_mass\_sum}  \\
      $\max_i  M_{i}$                      & \texttt{invariant\_mass\_maximum}  \\
      $\left(\prod_{i=1}^n m_{i}\right)^{1/n}$          & \texttt{particle\_mass\_product} \\
      $\left(\sum_{i=1}^n  m^m_{i}\right)^{1/m}$        & \texttt{particle\_mass\_sum}  \\
      \bottomrule
      \end{tabular}
      \caption{\label{tab:ScaleTypes} Implemented individual scale types. The
        final state jets $i$ are selected via the input tags
        \xmltag{jet_type}, \xmltag{jet_tag}, and
        \xmltag{jet_order}. The \xmltag{power} of the scale is denoted
        my $m$.}
\end{table}%
\begin{table}
      \centering
      \renewcommand\arraystretch{1.2}
      \rowcolors{2}{tablerowcolor}{}
      \begin{tabular}{llll}
        \toprule\rowcolor{tableheadcolor} 
        \textbf{Parameter} & \textbf{possible values} &
        \textbf{default value}  \\
        \midrule
      \xmltag{name} & string                   & none  \\
      \xmltag{jet_type} & list of jet types                   & none  \\
      \xmltag{jet_tag} &  corresponding list of jet tags      & none  \\
      \xmltag{jet_order} &   corresponding list of jet orders & none  \\
      \xmltag{jet_mass} & real number     & 0.0  \\
      \xmltag{jet_particle} & particle name     & none  \\
      \xmltag{offset} & real number     & 0.0  \\
      \xmltag{factor} & real number     & 1.0  \\
      \xmltag{power}  & real number     & 1.0  \\
      \xmltag{weight} & real number     & 1.0  \\
      \bottomrule
      \end{tabular}
      \caption{\label{tab:scale} Parameters of an individual scale
        type (some parameters are not available for all types).}
\end{table}%
We define the transverse energy for a momentum $p$ with transverse
part $p_{\text{T}}$ via
\begin{equation}\label{eq:ET}
E_{\text{T}} = \sqrt{p^2+p_{\text{T}}^2}
\end{equation}
and the transverse mass via
\begin{equation}\label{eq:MT}
M_{\text{T}} = \sqrt{m^2+p_{\text{T}}^2}\,,
\end{equation}
where $m$ is a mass associated to the corresponding jet (see below). The
individual scales are calculated by taking the geometric average, the sum, or the
maximum of the corresponding observable for a selection of final-state
objects specified via the tags \xmltag{jet_type}, \xmltag{jet_tag},
and \xmltag{jet_order}${}>0$. 
Note that the jets are selected according to the relevant tagging
scheme as described at the end of \refse{se:histograms}.

The jets can be associated a mass indicated via the tag
\xmltag{jet_mass} or the mass of the particle specified via the tag
\xmltag{jet_particle}, which can take the values in column~2
of \refta{tab:ParticleNames}. In addition, the resulting quantity can
be increased by an offset \xmltag{offset} (typically obtained from the
masses of certain particles) and multiplied by a factor
\xmltag{factor}.  An individual scale of type \texttt{\textit{xxx}\_product}, where \texttt{xxx} refers to the observables in Table~\ref{tab:ScaleTypes}, is calculated
via
$$
scale_k = \textit{factor}_k * \left(\prod_{i=1}^n (observable_{k,i})^{1/n} + \textit{offset}_k\right)
$$
and an individual scale of type \texttt{\textit{xxx}\_sum} via
$$
scale_k = \textit{factor}_k * \left[\left(\sum_{i=1}^n
(observable_{k,i})^{\textit{power}_k}\right)^{1/{\textit{power}_k}}
+ \textit{offset}_k\right],
$$
where \textit{power} is fixed by the tag \xmltag{power},
and an  individual scale of type \texttt{\textit{xxx}\_max} via
$$
scale_k = \textit{factor}_k * (\max_i observable_{k,i} + \textit{offset}_k).
$$

Individual scales can be combined in a scale sum indicated by the tag
\xmltag{scale_sum} as
$$
scale\_sum_l = \left[\sum_k (scale_k)^{\textit{power}_l}\right]^{1/{\textit{power}_l}},
$$
where \xmltag{power} is the parameter of the scale sum $l$ (see \refta{tab:scale_sum}).
\begin{table}
      \centering
      \renewcommand\arraystretch{1.2}
      \rowcolors{2}{tablerowcolor}{}
      \begin{tabular}{llll}
        \toprule\rowcolor{tableheadcolor} 
        \textbf{Parameter} & \textbf{possible values} &
        \textbf{default value}  \\
        \midrule
      \xmltag{power}  & real number     & 1  \\
      \xmltag{weight} & real number     & 1  \\
      \bottomrule
      \end{tabular}
      \caption{\label{tab:scale_sum} Parameters of a scale sum.}
\end{table}
A scale sum can receive a further parameter \xmltag{weight},
which is relevant for the calculation of the final scale
according to \refeq{eq:final_scale}.
A \xmltag{scale} that is not part of a \xmltag{scale_sum} environment
is considered as a scale sum with just this scale. The
\xmltag{weight} of this scale is passed to the corresponding
\xmltag{scale_sum}. The tag \xmltag{weight} of the individual scale is ignored for a
\xmltag{scale} within a \xmltag{scale_sum}. 

The final scale is combined from all scale sums defined in the input card
via
\begin{equation}\label{eq:final_scale}
final\_scale = global\_factor * \left[\prod_l scale\_sum_l^{weight_l}\right]^{\frac{1}{\sum_l weight_l}},
\end{equation}
where $global\_factor$ is the value of the tag \xmltag{global_factor} of the complete
list of scales.

If \xmltag{global_factor} and the \xmltag{weight} of all
\xmltag{scale_sum} subsections contain two entries each, the first set
of entries
is used to construct the factorisation scale and the second one the
renormalisation scale.
Thus, \xmltag{scale_sum} sections with \xmltag{weight} w1 0
\xmltag{/weight} define the factorisation scale, while the ones with
\xmltag{weight} 0 w2 \xmltag{/weight} define the renormalisation
scale. Sections with \xmltag{weight} w1 w2 \xmltag{/weight} enter
the definition of both scales.

An example for the definition of a scale used for
$\Pp\Pp\to\Pt\bar\Pt\Pb\bar\Pb\to\mu^-\bar\nu_\mu\Pe^+\nu_\Pe\bar{\Pb}\Pb\bar\Pb\Pb$
is given in \refli{lst:scales}.
\begin{figure}
\begin{lstlisting}[language=XML,caption=Example for scale definition
  via \texttt{cut\_card.xml}. ,label=lst:scales]
<scales id="ttxbbx">
    <global_factor>  .5d0       </global_factor>
    <scale_sum>
    <weight>          1.0       </weight>
    <scale type="transverse_energy_sum">
        <name>  transverse_energy_bottom_pair </name>
        <jet_type>      2       </jet_type>
        <jet_tag>       0       </jet_tag>
        <jet_order>     0       </jet_order>
        <factor>        1       </factor>
    </scale>
    </scale_sum>
    <scale_sum>
    <weight>          1.0       </weight>
    <scale type="transverse_momentum_sum">
        <name>  missing_transverse_momentum </name>
        <jet_type>     31       </jet_type>
        <jet_tag>       1       </jet_tag>
        <jet_order>     1       </jet_order>
        <factor>        1       </factor>
    </scale>
    <scale type="transverse_energy_sum">
        <name>  transverse_energy_charged_fermions </name>
        <jet_type>      4 17 2  </jet_type>
        <jet_tag>       0 0  0  </jet_tag>
        <jet_order>     0 0  0  </jet_order>
        <factor>        1       </factor>
        <offset>      346       </offset>
    </scale>
    </scale_sum>
</scales>
\end{lstlisting}
\end{figure}
This defines the scale 
\begin{equation}
\label{eq:scale}
 \mu_0 = \frac{1}{2}\Biggl[
\Biggl( \sum_{i={\mathrm j_\Pb}} E_{\rT,i}\Biggr)
\Biggl( p^{\mathrm{miss}}_{\rT} +
    \sum_{i=\Pe^+,\mu^-,{\mathrm j_\Pb}}
    E_{\rT,i}+ 346\GeV \Biggr)\Biggr]^{1/2},
\end{equation}
where $\mathrm j_\Pb$ denotes all bottom and antibottom jets.
The missing transverse momentum $p^{\mathrm{miss}}_\rT$ is defined via the reconstruction cut in
\refli{lst:missing}.
\begin{figure}
\begin{lstlisting}[language=XML,caption=Definition of missing momentum
  via a reconstruction cut in \texttt{cut\_card.xml}. ,label=lst:missing]
    <cut type="resonance_trivial_reconstruction">
        <name> missing_momentum </name>-
        <jet_type_1>    6  6 31 </jet_type_1>
        <jet_tag_1>     0  0  1 </jet_tag_1>
        <jet_order_1>   0  0  1 </jet_order_1>
        <target>        1       </target>
        <action>        0       </action>
        <order>         0       </order>
    </cut>
\end{lstlisting}
\end{figure}
The top-quark mass is accounted for by the offset
$2m_\mathrm{t}=346\GeV$. Alternatively, it can be taken into account by
adding an extra scale as shown in \refli{lst:particle_mass_scale} to
the second \xmltag{scale_sum} in \refli{lst:scales}.
\begin{figure}
\begin{lstlisting}[language=XML,caption=Definition of scale from mass
  of top particle times 2. ,label=lst:particle_mass_scale]
    <scale type="particle_mass_sum">
        <name>  particle_mass_top </name>
        <jet_type>      7      </jet_type>
        <jet_tag>       1      </jet_tag>
        <jet_order>     1      </jet_order>
        <jet_particle>  t      </jet_particle>
        <factor>        2      </factor>
    </scale>
\end{lstlisting}
\end{figure}

Another example used for $\Pp\Pp\to\Pe^+\Pe^-\gamma$
is shown in \refli{lst:scales_eexa}
\begin{figure}
\begin{lstlisting}[language=XML,caption=Example for scale definition
  via \texttt{cut\_card.xml}. ,label=lst:scales_eexa]
<scales id="eexa">
    <global_factor> 0.7071067811865475d0 </global_factor>
    <scale_sum>
        <weight> 1.0 </weight>
        <power> 2.0 </power>
        <scale type="transverse_momentum_sum">
            <name> transverse_momenta_squared </name>
            <jet_type>     23 3 1  </jet_type>
            <jet_tag>       0 0 0  </jet_tag>
            <jet_order>     0 0 0  </jet_order>
            <factor>        1      </factor>
            <power>         2      </power>
        </scale>
        <scale type="particle_mass_sum">
            <name> particle_mass_squared </name>
            <jet_type>     23      </jet_type>
            <jet_tag>       0      </jet_tag>
            <jet_order>     0      </jet_order>
            <jet_particle>  z      </jet_particle>
            <factor>        1      </factor>
        </scale>
    </scale_sum>
</scales>
\end{lstlisting}
\end{figure}
and defines the scale
\begin{equation}
\label{eq:scale_eexa}
 \mu_0 = \frac{1}{\sqrt{2}}\Biggl[
\MZ^2+ \sum_{i=\PZ,\gamma,\Pj} p_{\rT,i}^2\Biggr]^{1/2}.
\end{equation}

\subsection{Consistency checks of input}

\mocanlo performs some consistency checks on the input. When reading
the cuts and scales, it checks, in particular, whether the input for
\xmltag{jet_type}, \xmltag{jet_tag}, and \xmltag{jet_order} matches
with the final state of the considered process or the objects
constructed during the recombination and cut procedure. 
By default, the code stops if it detects a potential
inconsistency, which is useful to correctly set up the input cards.
At their own risk, the stop can be switched off for some warnings by the user by adding
\begin{lstlisting}[language=XML]
<ignore_stop>  true </ignore_stop>
\end{lstlisting}
in the \xmltag{cut} or \xmltag{scale} environment.
Moreover,  by using the tag
\begin{lstlisting}[language=XML]
<ignore_input_check>  true </ignore_input_check>
\end{lstlisting}
all checks for a particular \xmltag{cut} or
\xmltag{scale} can be disabled. However, this is not recommended, as it usually leads to run-time errors.

\section[Listing observables with the \texttt{\large plot\_card.xml}]
{Listing observables with the \texttt{\Large plot\_card.xml}}
\label{se:histograms}

The plot card \texttt{plot\_card.xml} contains the section
\xmltag{histograms}, where histograms of observables are listed. 
Individual histograms are defined in \xmltag{histogram} sections,
with their type specified by the attribute \texttt{type}.
The histograms operate on momenta
after cuts, \ie on recombined and/or reconstructed objects called generically jets.

The implemented histogram types are listed in \refta{tab:HistogramTypes}.
\begin{table}
      \centering
      \renewcommand\arraystretch{1.2}
      \renewcommand\arraystretch{1.19}
      \rowcolors{2}{tablerowcolor}{}
      \begin{tabular}{lll}
      \toprule\rowcolor{tableheadcolor} 
      \textbf{category} & \textbf{observable} & \textbf{type} \\
      \midrule
      single-jet based  &
      $\sum_n E_n$               & \texttt{energy       }  \\
      single-jet based  &
      $(\sum_n p_n)_\rT$         & \texttt{transverse\_momentum       }  \\
      single-jet based  &
      $\sum_n|p_{\rT_n}|$        & \texttt{transverse\_momentum\_scalar\_sum}  \\
      single-jet based &
      $\sum_nE_{\rT n}$          & \texttt{transverse\_energy         }  \\
      single-jet based  &
      $\sum_nM_{\rT n}$          & \texttt{transverse\_mass           }  \\
      single-jet based  &
      $\sqrt{ (\sum_n p_n)^2}$   & \texttt{invariant\_mass            }  \\
      single-jet based  &
      $y(\sum_np_n)$             & \texttt{rapidity                  }   \\
      single-jet based  &
      $\eta(\sum_np_n)$          & \texttt{pseudorapidity          }   \\
      single-jet based  &
      $\cos\theta(\sum_np_n)$    & \texttt{cosine\_angle          }   \\
      single-jet based  &
      $\tilde{M}_{\rT}$          & \texttt{transverse\_mass\_neutrino}  \\
      \midrule 
      multi-jet based &
      $\Delta R_{ij}$            & \texttt{distance                  }   \\
      multi-jet based &
      $\Delta r_{ij}$            & \texttt{pseudodistance            }   \\
      multi-jet based &
      $\phi_{ij}$                & \texttt{azimuthal\_angle\_separation} \\
      multi-jet based &
      $\cos\theta_{ij,k}$        & \texttt{cosine\_angle\_separation   } \\
      multi-jet based &
      $y_{ij}$                   & \texttt{rapidity\_separation       }  \\
      multi-jet based &
      $\eta_{ij}$                & \texttt{pseudorapidity\_separation  }  \\
      multi-jet based &
      $p_{\rT,j}/p_{\rT,i}$      & \texttt{transverse\_momentum\_ratio  }  \\
      multi-jet based &
      $\cos\theta^*_{ij,k}$      & \texttt{cosine\_decay\_angle     } \\
      multi-jet based &
      $\phi^*_{ijkl,m}$          & \texttt{azimuthal\_angle\_planes } \\
      multi-jet based &
      $z_{ijk}$                  & \texttt{z3rap                      }  \\
      \midrule
      overall     &
      $H_\text{T}$               & \texttt{total\_transverse\_energy   } \\
      overall     &
      $\sqrt{\hat{s}}$           & \texttt{cms\_energy                }  \\
      \bottomrule
      \end{tabular}
      \caption{\label{tab:HistogramTypes} Implemented histogram types.}
\end{table}
{\sloppy
The parameters of a histogram are summarised in \refta{tab:histogram}.
\begin{table}
      \centering
      \renewcommand\arraystretch{1.2}
      \rowcolors{2}{tablerowcolor}{}
      \begin{tabular}{lll}
        \toprule\rowcolor{tableheadcolor} 
        \textbf{Parameter} & \textbf{possible values} &
        \textbf{default value} \\
        \midrule
      \xmltag{name} & string                   & none  \\
      \xmltag{jet_type} & list of jet types              & none \\
      \xmltag{jet_tag} &  corresponding list of jet tags      & [0,\ldots,0] \\
      \xmltag{jet_order} &   corresponding list of jet orders & [0,\ldots,0] \\
      \xmltag{x_min} & real number     & histogram-specific  \\
      \xmltag{x_max} & real number     & histogram-specific  \\
      \xmltag{bin_width} & real number & histogram-specific  \\
      \xmltag{x_scale_up} & real number  & 1.0  \\
      \xmltag{x_scale_down} & real number & 1.0  \\
      \xmltag{abs} & true/false & false  \\
      \xmltag{use_tagging} & 0,1,2     & 1  \\
      \xmltag{latex_title}             & string    &  none  \\
      \xmltag{latex\_observable}       & latex\_string & none      \\
      \xmltag{plot_xmax} & real number & \xmltag{x_max}  \\
      \xmltag{plot_xmin} & real number & \xmltag{x_min}  \\
      \xmltag{plot_ymax} & real number & none  \\
      \xmltag{plot_ymin} & real number & none  \\
      \xmltag{plot_kmax} & real number & none  \\
      \xmltag{plot_kmin} & real number & none  \\
      \xmltag{logscale}  & true/false  & histogram-specific \\
      \bottomrule
      \end{tabular}
      \caption{\label{tab:histogram} Parameters of a histogram.}
\end{table}
\refli{lst:transverse_momentum_plot} shows as an example the definition of a
histogram of the transverse momentum of a bottom-jet pair.
\begin{lstlisting}[language=XML,caption=Transverse-momentum histogram
  in \texttt{plot\_card.xml}. ,label=lst:transverse_momentum_plot,float]
    <histogram type="transverse_momentum">
        <name>             "bb12"                              </name>
        <jet_type>         2 2                                 </jet_type>
        <jet_tag>          1 1                                 </jet_tag>
        <jet_order>        1 2                                 </jet_order>
        <x_min>            0d0                                 </x_min>
        <x_max>            13000.0                             </x_max>
        <bin_width>        20.0                                </bin_width>
        <latex_title>      "Transverse Momentum b-Jet Pair"    </latex_title>
        <latex_observable> "p_{\text{T},\text{b}_1\text{b}_2}" </latex_observable>
        <plot_xmin>        0.0                                 </plot_xmin>
        <plot_xmax>        400.0                               </plot_xmax>
        <plot_ymin>        1d-4                                </plot_ymin>
        <plot_ymax>        0.01                                </plot_ymax>
        <plot_kmin>        0.6                                 </plot_kmin>
        <plot_kmax>        2.2                                 </plot_kmax>
        <logscale>         true                                </logscale>
    </histogram>
\end{lstlisting}
The parameter \xmltag{name} is used for naming the resulting
data file, \eg for the example in
\refli{lst:transverse_momentum_plot} it reads
\texttt{histogram\_transverse\_momentum\_bb12.dat}.
The tags \xmltag{jet_type},
\xmltag{jet_tag}, and \xmltag{jet_order} indicate which transverse
momentum to plot. The arrays of integers in \xmltag{jet_type} specify
the types of jets, those in \xmltag{jet_tag} their tagging status and
those in \xmltag{jet_order} the position of the jets in the ordered
lists for \xmltag{jet_type} as defined by the tagging scheme (see
below). While \xmltag{jet_type} and
\xmltag{jet_order} are mandatory for all jet-based histograms, \xmltag{jet_tag}
is assumed to be zero if not present.  
The definitions in
\refli{lst:transverse_momentum_plot} correspond to add the momenta of the
first and second tagged and ordered b jets and to bin the transverse momentum of this sum,
labelled as $\ptsub{\Pb_1\Pb_2}$. The tags \xmltag{x_min} and
\xmltag{x_max} fix the observable's range to bin and
\xmltag{bin_width} the bin width to be used.

The tags \xmltag{x_scale_up}, \xmltag{x_scale_down}, and
\xmltag{abs} are optional and modify the default settings of the
histogram. If \xmltag{abs} is set to \texttt{true}, the absolute value
of the observable is binned. With \xmltag{x_scale_up} and
\xmltag{x_scale_down}, the observable is multiplied and divided by
the respective input number before filling the bins. By default,
observables with mass dimension are measured in GeV and angles in
degrees. In the latter case, setting \xmltag{x_scale_down} to 180 and \xmltag{x_scale_up}
to $3.1415$ will display the distributions in radians.

The remaining tags are
only passed through the histogram files to gnuplot.  The strings given
in \xmltag{latex_title} and \xmltag{latex_observable} are
used as title and observable label of the
histogram in the \LaTeX\ output. The other tags define the visible plot ranges for
the $x$ axis via \xmltag{plot_xmin} and \xmltag{plot_xmax}, the ranges
of the $y$ axis via \xmltag{plot_ymin} and \xmltag{plot_ymax}, while
\xmltag{plot_kmin} and \xmltag{plot_kmax} define the $y$ axis range of
the $k$ factor  in relative plots. The tag \xmltag{logscale} allows to
specify whether to use a logarithmic scale in the plot.}

For the single-jet-based histogram types, the lists \xmltag{jet_type},
\xmltag{jet_tag}, and \xmltag{jet_order} can have an arbitrary number of entries (same
number for all lists) and the observables of all partons that match
the corresponding entries are summed up as indicated in
\refta{tab:HistogramTypes}. The transverse energy $E_\rT$ and the
transverse mass $M_\rT$ are defined as in \refeq{eq:ET} and
\refeq{eq:MT}, respectively.  In the
histogram of type \texttt{transverse\_mass\_neutrino},
the momenta of jets of type neutrino are summed separately from
all other entries,
resulting in two momenta $p_\nu$ and $p_{\mathrm{vis}}$. From these
momenta the transverse mass is calculated as
$$
\tilde{M}_{\rT} = \sqrt{(p_{\rT,\mathrm{vis}}+p_{\rT,\nu})^2 - (\vec{p}_{\rT,\mathrm{vis}}+\vec{p}_{\rT,\nu})^2}.
$$

For multi-jet-based histogram types with two indices $ij$, exactly two entries
are expected in each of these lists. In those cases all observables
are computed from momenta in the laboratory system. For the observables
with indices $ij,k$ either two or three entries can be given.
If the third entry is missing for \texttt{cosine\_angle\_separation}, $\cos\theta_{ij}$ 
is calculated in the laboratory system.
If the third entry is missing for \texttt{cosine\_decay\_angle}, $\cos\theta^*_{ij}$, \ie
the decay angle of parton $i$ in the rest frame of parton $j$, is computed by first boosting
the momentum $i$ in the frame of $j$, and evaluating in this frame the angle of $i$ with respect to the boost direction.
If the third argument $k$ is present, in both cases all momenta are
first boosted to the rest frame of parton $k$, and thereafter the
angles are calculated as specified above.
The histogram \texttt{azimuthal\_angle\_planes} works similarly and
produces a distribution in the angle between two planes, defined by
the momenta of the first two pairs of the specified jets. Without a fifth
jet specified, this angle is calculated in the laboratory frame,
otherwise it is calculated in the rest system of the fifth jet.
Finally, the Zeppenfeld variable (or centrality) is defined as 
$$
z_{ijk} = \frac{2 y_i - y_j - y_k}{2(y_j-y_k)}
$$
and requires exactly three entries.

For the histograms \texttt{total\_transverse\_energy} and
\texttt{cms\_energy} the tags \xmltag{jet_type}, \xmltag{jet_tag}, and
\xmltag{jet_order} are not allowed.
The former histogram 
yields the sum of the transverse
energies of all final-state objects, including neutrinos (missing
energy) and bremsstrahlung objects. The histogram \texttt{cms\_energy}
yields the total partonic CM energy.

Histograms can be based on three different tagging schemes, which are
selected using the tag \xmltag{use_tagging}, which either applies to all histograms
(when specified in the section \xmltag{histograms}) or to 
individual histograms (when specified within the
\xmltag{histogram} definitions).} The default is
\xmltag{use_tagging}${}=1$.

For \xmltag{use_tagging}${}=0$, tagging is not used, \ie
\xmltag{jet_tag} should not be present. All jets are taken from the
lists of \emph{recombinants} of the corresponding \xmltag{jet\_type}, 
and \xmltag{jet_order} refers to the
corresponding position in this list.  The ordering in the list of \emph{recombinants}
corresponds to an ordering in transverse momenta.

\begin{warning}{Use of \xmltag{use_tagging}${}=0$}
The tag \xmltag{jet_order} refers to the position in the list of
recombinants, regardless of whether some of the recombinants have been cut.
Thus, this option should only be used for simple cases and if no tagging is required.
\end{warning}

For \xmltag{use_tagging}${}=1$ (default), the tagging information of the cut
routines is consistently used. Specific jets \emph{must} be tagged and
ordered and only ordered jets can be addressed via the tag
\xmltag{jet_order}.  \emph{This might require the use of extra cuts
  that take care of the tagging.} For \xmltag{jet_order}${}>0$, the jets
are taken from the list of \emph{ordered recombinants}, and \xmltag{jet_order}
refers to the corresponding position in this list.  
For \xmltag{jet_order}${}=0$, the jets are taken from the list of
\emph{recombinants} after discarding those entries
whose tag does not match
\xmltag{jet_tag} or with \xmltag{jet_order}${}\ne0$. For histograms in the
single-jet category, all matching jets are taken into account with \xmltag{jet_order}${}=0$.
This is useful if extra NLO jets, \ie jets that may or may not be
present in the final state, are to be included.  For histograms in
the multi-jet category, \xmltag{jet_order}${}=0$ is not allowed.

For \xmltag{use_tagging}${}=2$, the tagging information of the cut
routines is used in a different way. If  \xmltag{jet_tag} is missing
in the histogram definition, the jets are assumed not to be tagged.
Furthermore, all tagged jets must also be ordered.
For \xmltag{jet_tag}${}>0$, the jets are taken from the list of
\emph{ordered recombinants}.  For \xmltag{jet_tag}${}=0$, the jets are
taken from the list of \emph{recombinants} after discarding those
entries with non-zero \xmltag{jet_tag}. In both cases,
\xmltag{jet_order} refers to the corresponding entry in this list and
not necessarily to the corresponding \xmltag{jet_order} as defined by
the cuts. For
histograms that can take lists of jets, all matching jets are taken
into account. This scheme allows to access untagged jets in
\emph{recombinants} via \xmltag{jet_order}.

A further example is shown in \refli{lst:mjj-vbs-histo}, which is understood
together with the cuts defined in \refli{lst:cuts2}, which define the two
tagged jets of the process according to the \texttt{jet\_rapidity} cut. 
\begin{lstlisting}[language=XML,caption={Histogram for the invariant mass of the two tagged jets, which are defined by the \texttt{invariant\_mass\_tag\_jets} cut of \refli{lst:cuts2}.},label=lst:mjj-vbs-histo,float]
    <histogram type="invariant_mass">
        <name>             tag_jets                     </name>
        <jet_type>         1 1                          </jet_type>
        <jet_order>        1 2                          </jet_order>
        <jet_tag>          1 1                          </jet_tag>
        <x_min>            500                          </x_min>
        <x_max>            4000                         </x_max>
        <bin_width>        50                           </bin_width>
        <latex_title>      Invariant mass tag jets      </latex_title>
        <latex_observable> M_{\mathrm{j}_1\mathrm{j}_2} </latex_observable>
        <plot_xmin>        500                          </plot_xmin>
        <plot_xmax>        4000                         </plot_xmax>
        <plot_ymin>        0.00001                      </plot_ymin>
        <plot_ymax>        0.1                          </plot_ymax>
        <plot_kmin>        0.5                          </plot_kmin>
        <plot_kmax>        1.6                          </plot_kmax>
        <logscale>         true                         </logscale>
    </histogram>
\end{lstlisting}
The histogram bins the invariant mass of the two tagged jets,
calculated from the first and second momenta (\xmltag{jet_order} = 1
2) of the ordered recombinants.

As for cuts and scales, \mocanlo checks the input for
\xmltag{jet_type}, \xmltag{jet_tag}, and \xmltag{jet_order} also for
histograms. By setting the tag 
\begin{lstlisting}[language=XML]
<ignore_stop>  true </ignore_stop>
\end{lstlisting}
in the \xmltag{histogram} environment, one can prevent the code from stopping, 
and if one sets
\begin{lstlisting}[language=XML]
<ignore_input_check>  true </ignore_input_check>
\end{lstlisting}
all checks are switched off for the corresponding histogram.
This, however, might lead to run-time errors.

\section{User-defined input features}
\label{se:user-defined}

{\sloppy
Despite the general structure and flexibility of the cards,
there might still be cuts, scales, or histograms that cannot be
obtained from the input in the cards.
These objects can be defined in dedicated Fortran files, which are named
\texttt{user\_defined\_cuts.F90}, \texttt{user\_defined\_scales.F90},
and 
\linebreak
\texttt{user\_defined\_histograms.F90},
respectively.}

Templates for such files can be found in the folder
\begin{lstlisting}[language=Bash]
/path/to/mocanlo/src/mocanlo/user_defined_inputs
\end{lstlisting}
Depending on the user's needs, one, two, or all of them must be copied from that location
to the folder of the process to be computed, and specifically in a subfolder named
\texttt{user\_defined}.
As an example, we consider the process \vbsProcessPol~with polarised intermediate bosons, to be found in
the folder \texttt{/path/to/process/vbs/wz/pp\_wz\_ew\_ew\_pol}.
Once the necessary files have been copied, the process directory will look like%
\dirtree{%
  .1 pp\_wz\_ew\_ew\_pol/ .
  .2 cards/.
  .3 \ldots .
  .2 user\_defined/ .
  .3 user\_defined\_cuts.F90 .
  .3 user\_defined\_scales.F90 .
  .3 user\_defined\_histograms.F90 .
}%
Then, the files must be modified accordingly, as described below. Finally, to link these files
when compiling the code, the script \texttt{compile\_mocanlo} must be run
with the (relative or absolute) path of the process folder as an argument, like
\begin{lstlisting}[language=Bash]
 ./compile_mocanlo /path/to/processes/vbs/wz/pp_wz_ew_ew_pol
\end{lstlisting}

The modification of a user-defined file always requires three
steps: specifying the numbers of new features that one wants to implement, setting a template for
all of them, and setting up the procedure that contains their actual definitions.

We start showing how a new cut acting on a Zeppenfeld-like variable can be defined for \vbsProcessPol{}
by modifying the file \texttt{user\_defined\_cuts.F90}. The content of the file is reported
in \refli{lst:vbs-user-cuts}.

\begin{lstlisting}[language=Fortran,basicstyle=\ttfamily\scriptsize,
caption={Example of a \texttt{user\_defined\_cuts.F90} file for the
  process \vbsProcessPol}. ,captionpos=b,label=lst:vbs-user-cuts]
  #include "helpers.h"

  module user_defined_cuts
    use user_defined_cut2
    implicit none
   
    ! !!!!!!!!!!!!!!!!!!!!!!!!!!!!!!!!!!!!!!!!!!
    ! (1) Set number of user-defined cuts
    ! !!!!!!!!!!!!!!!!!!!!!!!!!!!!!!!!!!!!!!!!!!  
    integer :: n_user_cuts = 1

    type, extends(mc_user_defined_cut) :: mc_zep_cut
    contains
      procedure :: calc_observable => mc_zep_cut_get_value
    end type mc_zep_cut

  contains

    ! !!!!!!!!!!!!!!!!!!!!!!!!!!!!!!!!!!!!!!!!!!
    ! (2) Add template to list
    ! !!!!!!!!!!!!!!!!!!!!!!!!!!!!!!!!!!!!!!!!!!
    subroutine init_user_cuts_templ(user_template,index)
      class(mc_user_defined_cut), pointer :: user_template
      integer :: index
    
      if(index==1)then
        allocate(mc_zep_cut :: user_template)
        call user_template%init_template(label="zep_cut") ! <- This is the cut type 
      end if
    
    end subroutine init_user_cuts_templ
    
    ! !!!!!!!!!!!!!!!!!!!!!!!!!!
    ! (3) Define new cuts
    ! !!!!!!!!!!!!!!!!!!!!!!!!!!
    function mc_zep_cut_get_value(this, particles, sorted_clusters2) result(cut_applied)
      class(mc_zep_cut), intent(in) :: this
      logical :: cut_applied
      class(mc_particle), dimension(:), intent(inout) :: particles
      type(mc_recombinant2_list), dimension(:), intent(inout), target :: sorted_clusters2
      real(kind=dp) :: zeppenfeld_variable
      real(kind=dp) :: lep_rap, j1_rap, j2_rap
      integer :: jet_type, lep_type, i, n_tagged_jets, i_order
      logical :: ordered_jets
      
      ignore(particles) ! Act on sorted clusters, not at truth level
      cut_applied = .false.
      
      ! This cut is applied on a Zeppenfeld variable. This cut only operates
      ! on tagged jets that must be ordered. If less than 2 tagged jets are found,
      ! the cut is applied. If at least 2 tagged jets are found, but they are not ordered,
      ! this cut orders them and puts them in the list of ordered_recombinants.
      ! Only at this point the observable is computed.
      
      ! (1) Check if at least two tagged light jets exist
      jet_type = 1 ! Light jets
      n_tagged_jets = 0
      ordered_jets = .true.
      do i = 1, sorted_clusters2(jet_type)%n_recombinants
        if (sorted_clusters2(jet_type)%recombinants(i)%ptr%get_jet_tag() == 1) then
          n_tagged_jets = n_tagged_jets + 1
          if(sorted_clusters2(jet_type)%recombinants(i)%ptr%get_jet_order() == 0) then
            ordered_jets = .false.
          end if
        end if
      end do
      if(n_tagged_jets < 2) then
        cut_applied = .true.
        return
      end if
      
      ! (2) If needed, order the tagged jets and add them to list of ordered recombinants
      i_order = 0
      if(.not.ordered_jets) then
        do i = 1, sorted_clusters2(jet_type)%n_recombinants
          if (sorted_clusters2(jet_type)%recombinants(i)%ptr%get_jet_tag() == 1) then
            i_order = i_order + 1
            sorted_clusters2(jet_type)%n_ordered_recombinants = i_order
            ! Only the variable 'order' of the recombinants is set, since in this
            ! example we assume that the objects in the sorted_clusters2 list are
            ! already ordered in their transverse momentum
            call sorted_clusters2(jet_type)%recombinants(i)%ptr%set_jet_order(i_order)
            sorted_clusters2(jet_type)%ordered_recombinants(i_order) = &
              sorted_clusters2(jet_type)%recombinants(i)
          else
            call sorted_clusters2(jet_type)%recombinants(i)%ptr%set_jet_order(0)
          end if
        end do
      end if
      
      ! (3) Compute variable
      lep_type = 4 ! Positron
      lep_rap = sorted_clusters2(lep_type)%recombinants(1)%ptr%rapidity()
      j1_rap = sorted_clusters2(jet_type)%ordered_recombinants(1)%ptr%rapidity()
      j2_rap = sorted_clusters2(jet_type)%ordered_recombinants(2)%ptr%rapidity()
      
      zeppenfeld_variable = ( lep_rap - 0.5 * (j1_rap + j2_rap) ) / abs(j1_rap - j2_rap)
      
      if(abs(zeppenfeld_variable) > 0.4) cut_applied = .true.
    
    end function mc_zep_cut_get_value

  end module user_defined_cuts

\end{lstlisting}

First, the variable \texttt{n\_user\_cuts} is set to one, \ie the number of new cuts that we want
to add. Right below, a new type \texttt{mc\_zep\_cut} is declared, inheriting from the
type \texttt{mc\_user\_defined\_cut}. The new type just needs to define its own
\texttt{calc\_observable} procedure,
which in the example here is renamed \texttt{mc\_zep\_cut\_get\_value}.

As a second step, the if-statement in the subroutine \texttt{init\_user\_cuts\_templ} has to be adapted.
Specifically, for each new cut type an else-if scope must be added with the condition \texttt{index==n},
where \texttt{n} is an integer number indexing the different cuts. In the corresponding scope,
a template for the cut has to be allocated and initialised. For a user-defined cut, this just requires
to set a label. This label defines the type with which the cut becomes accessible from the
\texttt{cut\_card.xml}:
\begin{lstlisting}[language=xml,captionpos=b]
    <cut type="zep_cut">
        <name>  my_zep_cut </name>
    </cut>  
\end{lstlisting}
Note that, as for any other cut, the position of the user-defined cut in the sequence of cuts
in the \texttt{cut\_card.xml} is important.

Finally, the function \texttt{mc\_zep\_cut\_get\_value} must be defined. The function is expected to
return a logical variable \texttt{cut\_applied} that tells whether the cut must be applied or not.
Moreover, it always operates on \texttt{sorted\_clusters2}, which
collects the reconstructed objects for all jet types possibly
resulting from the recombination chain
and the action of previous cuts. Each of these jet types is a component of the \texttt{sorted\_clusters2}
array. Depending on the process, each component can collect multiple objects of the same type.
Such objects are accessible via two different lists (for further context, see discussion at
the beginning of \refse{se:cut-scheme}): 
\begin{itemize}
\item \textbf{\texttt{sorted\_clusters2(jet\_type)\%recombinants(i)\%ptr}} is the list of
  \emph{recombinants} containing  \texttt{sorted\_clusters2(jet\_type)\%n\_recombinants} objects
  of type \texttt{jet\_type}. This is the only way of accessing objects if they have not been ordered
  by the action of a cut via the tag \xmltag{order}. 
  One can inquire whether an object in the list has been tagged or ordered
  by checking the value of \texttt{get\_jet\_tag()} and \texttt{get\_jet\_order()}, respectively.
\item \sloppy
  \textbf{\texttt{sorted\_clusters2(jet\_type)\%ordered\_recombinants(i)\%ptr}} is the list of
  \emph{ordered recombinants} containing  \texttt{sorted\_clusters2(jet\_type)\%n\_ordered\_recombinants}
  objects. This list for a given \texttt{jet\_type} is only filled and accessible once the
  corresponding objects have been ordered by a previous cut.
  The order of an object corresponds to the position \texttt{i} in the list. 
\end{itemize}
{\sloppy
Accessing the member \texttt{ptr} of the recombinant and ordered-recombinant lists allows
to perform a set of operations on particle momenta via some  pre-defined methods, like computing
the transverse momentum or the rapidity using
\texttt{ptr\%rapidity()} and \texttt{ptr\%transverse\_momentum()}, respectively.
A more comprehensive list of available operations can be found by inspecting the
file \texttt{/path/to/mocanlo/src/mocanlo/math/lorentz.f90}.
}

After this basic introduction, the example reported in \refli{lst:vbs-user-cuts} can be
 understood as follows. The user-defined cut first checks if at least two tagged light jets (\texttt{jet\_type==1})
exist. If not, the event is cut away. Otherwise, the function \texttt{mc\_zep\_cut\_get\_value} proceeds
filling the list of \texttt{ordered\_recombinants}, in case the jets were not already ordered by a
previous cut. Finally, a Zeppenfeld variable among the positron, the
leading, and the subleading jet is
computed, and the event is cut away if this variable exceeds a hard-coded value.

Defining a new dynamical scale via the \texttt{user\_defined\_scales.F90} file requires to go through
very similar steps as for the definition of a new cut. The user is expected to introduce
a new scale type inheriting from \texttt{mc\_scale} having its own \texttt{calc} procedure.
As for the cuts, the latter has access to the particle momenta via the list of
\texttt{sorted\_clusters2}. One should keep in mind that when a scale is computed, the objects
collected in \texttt{sorted\_clusters2} are the ones that went through the complete sequence of
cuts of \texttt{cut\_card.xml}. It is the responsibility of the user to make sure that the objects used
in the construction of the scale really exist and are well defined. The actual scale definition
must be part of an appropriate \texttt{calc} procedure that returns the scale as a real variable.
The user-defined scale can then be used either as an independent scale
or as part of a \texttt{scale\_sum} definition (see \refse{se:scale}). It can be called from the input card
like
\begin{lstlisting}[language=xml,captionpos=b]
<scales id="wzscatt">
    <scale_sum>
        ...
        <scale type="pt_user_scale">
            <name> my_scale  </name>
        </scale>
        ...  
    </scale_sum>
    ...
</scales>
\end{lstlisting}
where \texttt{pt\_user\_scale} is the label used to initialise the scale template.

To introduce new histograms, the workflow is pretty similar up to some small differences that
are illustrated in \refli{lst:vbs-user-histos}.
\begin{lstlisting}[language=Fortran,basicstyle=\ttfamily\scriptsize,
caption={Example of a \texttt{user\_defined\_histograms.F90} file for
  the process \vbsProcessPol}. ,captionpos=b. ,label=lst:vbs-user-histos]
  module user_defined_histograms
    use observable_histogram
    implicit none
    
    ! !!!!!!!!!!!!!!!!!!!!!!!!!!!!!!!!!!!!!!!!!!
    ! (1) Set number of user-defined histograms
    ! !!!!!!!!!!!!!!!!!!!!!!!!!!!!!!!!!!!!!!!!!!  
    integer :: n_user_histos = 2
    
    type, extends(mc_observable_histogram) :: mc_normalised_pt_difference_histogram
     contains
       procedure :: calc => mc_normalised_pt_difference_histogram_calc
    end type mc_normalised_pt_difference_histogram
    type, extends(mc_observable_histogram) :: mc_zep_histogram
     contains
       procedure :: calc => mc_zep_histogram_calc
    end type mc_zep_histogram       
    ...
    
  contains

    ! !!!!!!!!!!!!!!!!!!!!!!!!!!!!!!!!!!!!!!!!!!
    ! (2) Add template to list
    ! !!!!!!!!!!!!!!!!!!!!!!!!!!!!!!!!!!!!!!!!!!
    subroutine init_user_histos_templ(user_template,index)
      class(mc_observable_histogram), pointer :: user_template
      integer :: index
      
      if(index==1)then
        allocate (mc_normalised_pt_difference_histogram :: user_template)
        call user_template%init_template( &
            label="normalised_pt_difference", &
            latex_unit="", &
            x_min=0d0, &
            x_max=1d0, &
            bin_width=0.01d0, &
            plot_xmin=0d0, &
            plot_xmax=1d0, &
            logscale=.true. &
            )
      end if(index==2)then
        allocate (mc_zep_histogram :: user_template)
        ...
      end if
    end subroutine init_user_histos_templ
    
    ! !!!!!!!!!!!!!!!!!!!!!!!!!!
    ! (3) Define new histograms  
    ! !!!!!!!!!!!!!!!!!!!!!!!!!!
    function mc_normalised_pt_difference_histogram_calc(this, partonic_process) &
     result(observable)
      class(mc_normalised_pt_difference_histogram) :: this
      type(mc_partonic_process), intent(in) :: partonic_process
      real(kind=dp) :: observable
      type(mc_recombinant) :: p_j1, p_j2
      real(kind=dp) :: pt_j1, pt_j2
        
      p_j1 = partonic_process%sorted_clusters2(1)%ordered_recombinants(1)%ptr
      p_j2 = partonic_process%sorted_clusters2(1)%ordered_recombinants(2)%ptr
    
      pt_j1 = p_j1%transverse_momentum()
      pt_j2 = p_j2%transverse_momentum()
    
      observable = ( pt_j1 - pt_j2 ) / ( pt_j1 + pt_j2 )
    end function mc_normalised_pt_difference_histogram_calc

    ...
    
end module user_defined_histograms

\end{lstlisting}

In the example above, two new histograms are introduced. The first thing to note is the
definition of the template for the new types. In this case, the template contains a label 
together with an additional set of entries setting default values (that could always be 
overwritten from the input card, if needed) for the histogram ranges, binning and so on.
The second difference is found in the \texttt{calc} procedure. There, the particle momenta
must now be obtained through the \texttt{partonic\_process} argument, that in turn provides access to
the list of \texttt{sorted\_clusters2}, already described above. 

Once a histogram is defined,
it can be used in the \texttt{plot\_card.xml} as any other distribution:
\begin{lstlisting}[language=xml,captionpos=b]
<histograms id="wzscatt">
    ...
    <histogram type="normalised_pt_difference">
        <name>             "aptj"                         </name>
        <latex_title>      "Jet pt normalised difference" </latex_title>
        <latex_observable> "A_{p_{\text{T}, j}}"          </latex_observable>
    </histogram>
    ...  
</histograms>
\end{lstlisting}
where in the example above
the default values for all additional features of the histograms will be used as defined in the allocation of the template for the new type.

\section{Conclusions}
\label{se:conclusions}

This manual describes the usage of \mocanlo, a Monte Carlo integration
code designed for arbitrary scattering processes at high-energy colliders.  It
allows the calculation of both NLO QCD and EW corrections as long as
external particles that can give rise to real
radiation are massless. Non-radiating final-state particles can be
massive, and collinear singularities arising from initial-state
electrons and muons can be regulated by their masses.
For the calculation of tree-level
and one-loop matrix elements, the program \recola is used, which in turn
relies on \collier for the loop integrals.
Infrared singularities are
treated with the Catani--Seymour dipole formalism and its extensions,
whereby all light fermions are assumed to be massless.
Integrated and differential results are
obtained upon combining different contributions like the Born, the virtual,
the subtracted real and the integrated dipoles for a standard NLO calculation.

\mocanlo has already been used for a large number of scattering
processes at the LHC, including processes with up to eight
particles in the final state. It can also deal with processes at
lepton--lepton and lepton--proton colliders, as well as
with ultra-peripheral collisions of photons in heavy-ion collisions. It supports
the helicity selection for intermediate weak bosons and for incoming leptons.

In the future \mocanlo could be extended in several respects.
While \mocanlo provides results for varied renormalisation and
factorisation scales in an efficient way, it would be nice 
to also evaluate uncertainties from the
parton-distributions functions efficiently.
The treatment of processes with polarised resonances, which is currently
subject to some limitations, is expected to be progressively extended and generalised.
Presently, we are working on the matching of \mocanlo predictions to parton showers via the
\textsc{MC@NLO} method.  Also,  support for $1\to n$ decay processes
and $2\to 1$ processes might be considered as an additional feature.

\section*{Acknowledgements}
The first version of \mocanlo was devised and implemented by Robert
Feger.  Notable contributions were made by Jean-Nicolas Lang, Stefan
Rode, Timo Schmidt, and Christopher Schwan.  We are indebted to Jean-Nicolas
Lang and Sandro Uccirati for maintaining \recola.
We thank Jose Luis Hernando Ariza and Jo Reimer for testing \mocanlo.

This work is supported by the German Federal Ministry for
Research, Technology and Space (BMFTR) under contracts no.\ 05H12WWE, 05H15WWCA1,
05H18WWCA1, 05H21WWCAA and the German Research Foundation (DFG) under
reference numbers DE 623/6-1, DE 623/6-2, DE 623/8-1, and through the Research Training Group RTG2044.
It was also supported by the state of Baden-W\"urttemberg through bwHPC and the DFG through grant No.\ INST 39/963-1 FUGG (bwForCluster NEMO).
This research has received funding from the European Research Council (ERC) under the EU Horizon 2020 Research and Innovation Programme (grant agreement no.\ 683211).
The research of DL has also been partially supported by
the Italian Ministry of Universities and Research (MUR) under the FIS grant
(CUP: D53C24005480001, FLAME).
GP is supported by the EU Horizon Europe Research and Innovation programme under the Marie-Sk\l{}odowska Curie Action (MSCA)
``POEBLITA - POlarised Electroweak Bosons at the LHC with Improved Theoretical Accuracy'', grant agreement no.\ 101149251 (CUP H45E2300129000).

\appendix

\section{Validated processes}
\label{se:validated_processes}

This section contains a list of calculations that have been carried out with
\mocanlo, together with the respective references. The corresponding
input cards are provided in the subfolders of the folder \texttt{validated\_processes/}.

\input{validated_processes.tex}

\bibliographystyle{JHEPmod}
\bibliography{mocanlo}

\end{document}

%% file: macros.tex
\def\refeq#1{\mbox{Eq.~(\ref{#1})}}

\def\refta#1{\mbox{Table~\ref{#1}}}
\def\refli#1{\mbox{Listing~\ref{#1}}}
\def\reftas#1{\mbox{Tables~\ref{#1}}}
\def\refse#1{\mbox{Sec.~\ref{#1}}}
\def\refapp#1{\mbox{Appendix~\ref{#1}}}
\def\citere#1{\mbox{Ref.~\cite{#1}}}
\def\citeres#1{\mbox{Refs.~\cite{#1}}}

\newcommand{\ri}{\mathrm i}

\def\be{\begin{equation}}
\def\ee{\end{equation}}

\newcommand{\rd}{\ensuremath{\text{d}}\xspace}

\newcommand{\PH}{\ensuremath{\text{H}}\xspace}
\newcommand{\Pj}{\ensuremath{\text{j}}\xspace}
\newcommand{\Pp}{\ensuremath{\text{p}}\xspace}
\newcommand{\Pe}{\ensuremath{\text{e}}\xspace}
\newcommand{\Pb}{\ensuremath{\text{b}}\xspace}
\newcommand{\Pq}{\ensuremath{\text{q}}\xspace}
\newcommand{\Pt}{\ensuremath{\text{t}}\xspace}
\newcommand{\Pu}{\ensuremath{\text{u}}\xspace}
\newcommand{\Pd}{\ensuremath{\text{d}}\xspace}
\newcommand{\Ps}{\ensuremath{\text{s}}\xspace}
\newcommand{\Pc}{\ensuremath{\text{c}}\xspace}
\newcommand{\Pg}{\ensuremath{\text{g}}\xspace}

\newcommand{\PW}{\ensuremath{\text{W}}\xspace}
\newcommand{\PZ}{\ensuremath{\text{Z}}\xspace}

\newcommand{\MW}{\ensuremath{M_\PW}\xspace}

\newcommand{\MZ}{\ensuremath{M_\PZ}\xspace}

\newcommand{\GeV}{\ensuremath{\,\text{GeV}}\xspace}

\newcommand{\alphas}{\ensuremath{\alpha_\text{s}}\xspace}
\newcommand{\order}[1]{\ensuremath{\mathcal{O}{\left(#1\right)}}\xspace}

\newcommand{\GF}{\ensuremath{G_\mu}}

\newcommand{\ptsub}[1]{\ensuremath{p_{\text{T},#1}}\xspace}

\newcommand{\mur}{\ensuremath{\mu_\text{R}}\xspace}
\newcommand{\muf}{\ensuremath{\mu_\text{F}}\xspace}

\newcommand{\ttbarProcess }{\ensuremath{\Pp\Pp\to\Pe^+\nu_\Pe\Pb \mu^-\bar{\nu}_\mu\bar{\Pb}}\xspace}
\newcommand{\fullProcess  }{\ensuremath{\Pp\Pp\to\Pe^+\nu_\Pe\Pb \mu^-\bar{\nu}_\mu\bar{\Pb}}\xspace}
\newcommand{\ggProcess    }{\ensuremath{\Pg\Pg\to\Pe^+\nu_\Pe\Pb \mu^-\bar{\nu}_\mu\bar{\Pb}}\xspace}

\newcommand{\vbsProcess    }{\ensuremath{\Pp\Pp\to\Pe^+\nu_\Pe \mu^-\bar{\nu}_\mu\Pj\Pj\xspace}}
\newcommand{\vbsProcessPol    }{\ensuremath{\Pp\Pp\to\Pe^+\Pe^-\mu^-\bar{\nu}_\mu\Pj\Pj\xspace}}

\newcommand{\recola}{{\scshape Recola}\xspace}
\newcommand{\mocanlo}{{\scshape MoCaNLO}\xspace}
\newcommand{\MCG}{{\scshape MCG}\xspace}
\newcommand{\collier}{{\scshape Collier}\xspace}
\newcommand{\sherpa}{{\scshape Sherpa}\xspace}
\newcommand{\powheg}{{\scshape Powheg-Box}\xspace}
\newcommand{\pythia}{{\scshape Pythia}\xspace}
\newcommand{\herwig}{{\scshape Herwig}\xspace}
\newcommand{\madgraph}{{\scshape MadGraph5\_aMC@NLO}\xspace}
\newcommand{\rT}{{\mathrm{T}}}
\newcolumntype{.}{D{.}{.}{-1}}
\newcolumntype{d}[1]{D{.}{.}{#1}}

\renewcommand{\vec}[1]{\mathbf{#1}}
\colorlet{tableoverheadcolor}{gray!37.5}
\colorlet{tableheadcolor}{gray!25}
\colorlet{tablerowcolor}{gray!12.5}

\newlength{\width}
\newlength{\height}

\providecommand*\eg{e.g.\ }
\providecommand*\ie{i.e.\ }

\renewcommand{\Re}{\mathop{\mathrm{Re}}\nolimits}

\newcommand{\MSbar}{\ensuremath{\overline{{\text{MS}}\xspace}}}

\def\draftdate{\relax}
\def\mda{\relax}
\def\mua{\relax}
\def\mla{\relax}
\def\draft{
\def\thtystars{******************************}
\def\sixtystars{\thtystars\thtystars}
\typeout{}
\typeout{\sixtystars**}
\typeout{* Draft mode!
         For final version remove \protect\draft\space in source file *}
\typeout{\sixtystars**}
\typeout{}
\def\draftdate{\today}
\def\mua{\marginpar[\boldmath\hfil$\uparrow$]%
                   {\boldmath$\uparrow$\hfil}\color{black}%
                    \typeout{marginpar: $\uparrow$}\ignorespaces}
\def\mda{\color{red}\marginpar[\boldmath\hfil$\downarrow$]%
                   {\boldmath$\downarrow$\hfil}%
                    \typeout{marginpar: $\downarrow$}\ignorespaces}
\def\mla{\marginpar[\boldmath\hfil$\rightarrow$]%
                   {\boldmath$\leftarrow $\hfil}%
                    \typeout{marginpar: $\leftrightarrow$}\ignorespaces}
\def\Mua{\marginpar[\boldmath\hfil$\Uparrow$]%
                   {\boldmath$\Uparrow$\hfil}\color{black}%
                    \typeout{marginpar: $\uparrow$}\ignorespaces}
\def\Mda{\color{red}\marginpar[\boldmath\hfil$\Downarrow$]%
                   {\boldmath$\Downarrow$\hfil}%
                    \typeout{marginpar: $\downarrow$}\ignorespaces}
\def\Mla{\marginpar[\boldmath\hfil\textcolor{red}{$\Rightarrow$}]%
                   {\boldmath\textcolor{red}{$\Leftarrow $}\hfil}%
                    \typeout{marginpar: $\leftrightarrow$}\ignorespaces}
\overfullrule 5pt
\oddsidemargin -15mm
\marginparwidth 29mm
\def\Red##1{{\color{red}{##1}}}
\def\ADm##1{{\color{red}{##1}}}
\def\AD##1{{\color{red}{AD: ##1}}}
\def\DL##1{{\color{magenta}{DL: ##1}}}
\def\GP##1{{\color{orange}{GP: ##1}}}
\def\SL##1{{\color{blue}{SL: ##1}}}
\def\MP##1{{\color{darkgreen}{MP: ##1}}}
}


%% file: validated_processes.tex
\subsection{Top--antitop production and associated production}

\begin{enumerate}
 \item \texttt{pp\_epvemumvmxbbx\_qcd}\\
 The cards reproduce the calculation of NLO QCD corrections to off-shell tt production in \citere{Denner:2012yc}.
 \item \texttt{pp\_epvemumvmxbbx\_ew}\\
 The cards have been set up for the calculation of NLO EW corrections
 for off-shell tt production in \citere{Denner:2016jyo}.
 \item \texttt{pp\_epvemumvmxbbx\_ew\_DPA\_ww}\\
 The cards have been set up for the calculation of NLO EW corrections for off-shell tt production in \citere{Denner:2016jyo} in the WW DPA for the virtual part.
 \item \texttt{pp\_epvemumvmxbbx\_ew\_DPA\_tt}\\
 The cards have been set up for the calculation of NLO EW corrections for off-shell tt production in \citere{Denner:2016jyo} in the tt DPA for the virtual part.
 \item \texttt{pp\_epvemumvmxbbxh\_qcd}\\
 The cards have been set up for the calculation of NLO QCD corrections for off-shell ttH production in \citere{Denner:2016wet}. Please also cite \citere{Denner:2015yca}.
\item \texttt{pp\_epvemumvmxbbxh\_ew}\\
 The cards have been set up for the calculation of NLO EW corrections for off-shell ttH production in \citere{Denner:2016wet}.
\item \texttt{pp\_epvemumvmxbbxh\_ew\_DPA\_ww}\\
 The cards have been set up for the calculation of NLO EW corrections for off-shell ttH production in \citere{Denner:2016wet} in the WW DPA for the virtual part.
 \item \texttt{pp\_epvemumvmxbbxh\_ew\_DPA\_tt}\\
   The cards have been set up for the calculation of NLO EW corrections for off-shell ttH production in \citere{Denner:2016wet} in the tt DPA for the virtual part.
 \item \texttt{pp\_ttw}\\
   The cards have been set up for the calculation of the NLO corrections
   of orders $\mathcal{O}(\alphas^3\alpha^6)$, $\mathcal{O}(\alphas^2\alpha^7)$, and $\mathcal{O}(\alphas\alpha^8)$ to off-shell $\Pt\overline{\Pt}\PW^+$ production in the fully leptonic decay channel
   of \citeres{Denner:2020hgg,Denner:2021hqi}.
 \item \texttt{pp\_ttz}\\
   The cards have been set up for the calculation of the full set of LO and NLO corrections for off-shell  $\Pt\overline{\Pt}\PZ$ production in \citere{Denner:2023eti}.
  \item \texttt{pp\_ttbb}\\
  The cards have been set up for the calculation of the NLO QCD corrections for off-shell  $\Pt\overline{\Pt}\Pb\overline{\Pb}$ production in \citere{Denner:2020orv}.
\end{enumerate}

\subsection{Single-top production}

\begin{enumerate}
 \item \texttt{pp\_tzj}\\
 The cards have been set up for the calculation of NLO QCD and EW corrections to
 associated tZj production in \citere{Denner:2022fhu}.
\end{enumerate}

\subsection{Vector-boson scattering}

\begin{enumerate}
 \item \texttt{pp\_ssww\_ew\_ew}\\
 The cards have been set up for the calculation of $\mathcal{O}\!\left(\alpha^7 \right)$ corrections to same-sign WW scattering in \citere{Biedermann:2017bss}. 
 Please also cite \citere{Biedermann:2016yds} for the $\PW^+\PW^+$ signature.
 \item \texttt{pp\_ssww\_ew\_qcd}\\
 The cards have been set up for the calculation of $\mathcal{O}\!\left(\alpha^6 \alphas \right)$ corrections to same-sign WW scattering in \citere{Biedermann:2017bss} for the $\PW^+\PW^+$ signature.
 \item \texttt{pp\_ssww\_qcd\_ew}\\
 The cards have been set up for the calculation of $\mathcal{O}\!\left(\alpha^5 \alphas^2 \right)$ corrections to same-sign WW scattering in \citere{Biedermann:2017bss} for the $\PW^+\PW^+$ signature.
 \item \texttt{pp\_ssww\_qcd\_qcd}\\
 The cards have been set up for the calculation of $\mathcal{O}\!\left(\alpha^4 \alphas^3 \right)$ corrections to same-sign WW scattering in \citere{Biedermann:2017bss} for the $\PW^+\PW^+$ signature.
 \item \texttt{pp\_ssww\_ew\_ew\_minus\_minus}\\
 The cards have been set up for the calculation of $\mathcal{O}\!\left(\alpha^7 \right)$ corrections to same-sign WW scattering in \citere{Biedermann:2017bss} for the $\PW^-\PW^-$ signature.
 Please also cite Refs.~\cite{Biedermann:2016yds,Chiesa:2019ulk}.
 \item \texttt{pp\_ssww\_ew\_qcd\_minus\_minus}\\
 The cards have been set up for the calculation of $\mathcal{O}\!\left(\alpha^6 \alphas \right)$ corrections to same-sign WW scattering in \citere{Biedermann:2017bss}  for the $\PW^-\PW^-$ signature.
 \item \texttt{pp\_ssww\_ew\_ew\_pol}\\
   The cards have been set up for the calculation of $\mathcal{O}\!\left(\alpha^7 \right)$ corrections to polarised $\PW^+\PW^+$ scattering in \citere{Denner:2024tlu}.    
\item \texttt{pp\_wz\_ew\_ew}\\
 The cards have been set up for the calculation of $\mathcal{O}\!\left(\alpha^7 \right)$ corrections to $\PW^+\PZ$ scattering in \citere{Denner:2019tmn}. 
 \item \texttt{pp\_wz\_ew\_qcd}\\
 The cards have been set up for the calculation of $\mathcal{O}\!\left(\alpha^6 \alphas \right)$ corrections to $\PW^+\PZ$ scattering in \citere{Denner:2019tmn}.
 \item \texttt{pp\_wz\_ew\_ew\_minus}\\
 The cards have been set up for the calculation of $\mathcal{O}\!\left(\alpha^7 \right)$ corrections to $\PW^-\PZ$ scattering in \citere{Denner:2019tmn}. 
 \item \texttt{pp\_wz\_ew\_qcd\_minus}\\
   The cards have been set up for the calculation of $\mathcal{O}\!\left(\alpha^6 \alphas \right)$ corrections to $\PW^-\PZ$ scattering in \citere{Denner:2019tmn}.
 \item \texttt{pp\_wz\_ew\_ew\_pol}\\
   The cards have been set up for the calculation of $\mathcal{O}\!\left(\alpha^7 \right)$ corrections to polarised $\PW^+\PZ$ scattering in \citere{Denner:2025xdz}.    
 \item \texttt{pp\_zz\_ew\_ew}\\
 The cards have been set up for the calculation of $\mathcal{O}\!\left(\alpha^7 \right)$ corrections to ZZ scattering in \citere{Denner:2021hsa}. 
 Please also cite \citere{Denner:2020zit}.
 \item \texttt{pp\_zz\_ew\_qcd}\\
 The cards have been set up for the calculation of $\mathcal{O}\!\left(\alpha^6 \alphas \right)$ corrections to ZZ scattering in \citere{Denner:2021hsa}.
 Please also cite \citere{Denner:2020zit}.
 \item \texttt{pp\_zz\_qcd\_ew}\\
 The cards have been set up for the calculation of $\mathcal{O}\!\left(\alpha^5 \alphas^2 \right)$ corrections to ZZ scattering in \citere{Denner:2021hsa}.
 \item \texttt{pp\_zz\_qcd\_qcd}\\
 The cards have been set up for the calculation of $\mathcal{O}\!\left(\alpha^4 \alphas^3 \right)$ corrections to ZZ scattering in \citere{Denner:2021hsa}.
 \item \texttt{pp\_wpwm\_ew\_ew}\\
 The cards have been set up for the calculation of $\mathcal{O}\!\left(\alpha^7 \right)$ corrections to $\PW^+\PW^-$ scattering in \citere{Denner:2022pwc}. 
 \item \texttt{pp\_wpwm\_ew\_qcd}\\
 The cards have been set up for the calculation of $\mathcal{O}\!\left(\alpha^6 \alphas \right)$ corrections to $\PW^+\PW^-$ scattering in \citere{Denner:2022pwc}.
 
\end{enumerate}

\subsection{Triboson production}

\begin{enumerate}
 \item \texttt{pp\_www\_ew\_ew}\\
 The cards have been set up for the calculation of $\mathcal{O}\!\left(\alpha^7 \right)$ corrections to WWW production with one W boson decaying hadronically in \citere{Denner:2024ufg}.
 Please also cite \citere{Biedermann:2016yds}.
 \item \texttt{pp\_www\_ew\_qcd}\\
 The cards have been set up for the calculation of $\mathcal{O}\!\left(\alpha^6 \alphas \right)$ corrections to WWW production with one W boson decaying hadronically in \citere{Denner:2024ufg}.
 \item \texttt{pp\_www\_qcd\_ew}\\
 The cards have been set up for the calculation of $\mathcal{O}\!\left(\alpha^5 \alphas^2 \right)$ corrections to WWW production with one W boson decaying hadronically in \citere{Denner:2024ufg}.
 \item \texttt{pp\_www\_qcd\_qcd}\\
 The cards have been set up for the calculation of $\mathcal{O}\!\left(\alpha^4 \alphas^3 \right)$ corrections to WWW production with one W boson decaying hadronically in \citere{Denner:2024ufg}.
\item \texttt{wzv\_semi\_leptonic}\\
  The folder collects the cards for the different contributions to WVZ production, with V being either a W or a Z boson decaying hadronically, with the setup described in \citere{Denner:2024ndl}.
  The subfolder \texttt{pp\_wvw\_ew\_ew} contains the $\mathcal{O}\!\left(\alpha^6 \right)$ and the $\mathcal{O}\!\left(\alpha^7\right)$ corrections, while the subfolder \texttt{pp\_wvw\_ew\_qcd} 
  the $\mathcal{O}\!\left(\alpha^6 \alphas \right)$ corrections. The LO contributions at $\mathcal{O}\!\left(\alpha^5 \alphas\right)$ and $\mathcal{O}\!\left(\alpha^4 \alphas^2\right)$
  can be found in \texttt{pp\_wvw\_LO\_as1\_a5} and \texttt{pp\_wvw\_LO\_as2\_a4}, respectively. Cards to reproduce the results for the bottom- and photon-induced channels
  at the perturbative orders described in \citere{Denner:2024ndl} are
   located in the subfolder \texttt{pp\_wvw\_bottoms} and \texttt{pp\_wvw\_photons}, respectively.
\end{enumerate}

\subsection{Diboson production}
\begin{enumerate}
 \item \texttt{pp\_ww\_pol}\\
 The cards have been set up for the calculation of NLO EW corrections to inclusive $\PW^+\PW^-$ production at the LHC with polarised bosons in the fully leptonic channel \cite{Denner:2023ehn}.
 Please also cite \citere{Denner:2020bcz}.
 \item \texttt{pp\_wz\_pol}\\
 The cards have been set up for the calculation of NLO QCD and NLO EW corrections to inclusive $\PW^+\PZ$ production at the LHC with polarised bosons in the fully leptonic channel \cite{Pelliccioli:2025com}.
 Please also cite \citere{Denner:2020eck}.
 \item \texttt{pp\_zz\_pol}\\
 The cards have been set up for the calculation of NLO QCD and NLO EW corrections, as well as the loop-induced $\Pg\Pg$ contribution, to inclusive $\PZ\PZ$ production at the LHC with polarised bosons in the fully leptonic channel \cite{Carrivale:2025mjy}.
 Please also cite \citere{Denner:2021csi}.
\end{enumerate}

\subsection{Higgs production}

\begin{enumerate}
 \item \texttt{vbf\_h}\\
 It contains the cards of \citere{Barone:2025jey} for Higgs production via vector-boson fusion (VBF) in one of the setups considered in the reference.
 It provides the NLO QCD and NLO EW corrections in the full computation to VBF in addition to the subleading contributions of orders $\mathcal{O}\!\left(\alphas^2 \alpha^2\right)$, 
 $\mathcal{O}\!\left(\alphas^3 \alpha^2\right)$, and $\mathcal{O}\!\left(\alphas^4 \alpha\right)$.
 Please also cite \citere{Dreyer:2020xaj} when relevant, as indicated in the README files.
 \item \texttt{vbf\_hh}\\
 It contains cards for di-Higgs production via VBF from \citere{Dreyer:2020xaj} at NLO QCD and NLO EW.
 In addition, cards for an inclusive set-up are also available for NLO EW corrections.
 In particular, these are used by the LHC Higgs Cross Section Working Group as canonical predictions for VBF.
\end{enumerate}

\subsection{Ultra-peripheral collisions}

\begin{enumerate}
 \item \texttt{tautau}\\
 The cards have been set up for one check of the calculation of EW corrections to tau-pair production in UPC~\cite{Dittmaier:2025ikh}, namely on-shell massless tau fermions leptons.
\end{enumerate}


%% file: mocanlo_users_guide.bbl
\providecommand{\href}[2]{#2}\begingroup\raggedright\begin{thebibliography}{10}

\bibitem{Alwall:2014hca}
J.~Alwall, et~al., {\itshape {The automated computation of tree-level and
  next-to-leading order differential cross sections, and their matching to
  parton shower simulations}},  {\em JHEP} {\bfseries 07} (2014) 079,
  [\href{http://arxiv.org/abs/1405.0301}{{\ttfamily arXiv:1405.0301}}].

\bibitem{Alioli:2010xd}
S.~Alioli, P.~Nason, C.~Oleari, and E.~Re, {\itshape {A general framework for
  implementing NLO calculations in shower Monte Carlo programs: the POWHEG
  BOX}},  {\em JHEP} {\bfseries 06} (2010) 043,
  [\href{http://arxiv.org/abs/1002.2581}{{\ttfamily arXiv:1002.2581}}].

\bibitem{Sherpa:2024mfk}
{\bfseries Sherpa} Collaboration, E.~Bothmann et~al., {\itshape {Event
  generation with Sherpa 3}},  {\em JHEP} {\bfseries 12} (2024) 156,
  [\href{http://arxiv.org/abs/2410.22148}{{\ttfamily arXiv:2410.22148}}].

\bibitem{Bewick:2023tfi}
G.~Bewick et~al., {\itshape {Herwig 7.3 release note}},  {\em Eur. Phys. J. C}
  {\bfseries 84} (2024) 1053,
  [\href{http://arxiv.org/abs/2312.05175}{{\ttfamily arXiv:2312.05175}}].

\bibitem{Bierlich:2022pfr}
C.~Bierlich et~al., {\itshape {A comprehensive guide to the physics and usage
  of PYTHIA 8.3}},  {\em SciPost Phys. Codeb.} {\bfseries 2022} (2022) 8,
  [\href{http://arxiv.org/abs/2203.11601}{{\ttfamily arXiv:2203.11601}}].

\bibitem{Berends:1994pv}
F.~A. Berends, R.~Pittau, and R.~Kleiss, {\itshape {All electroweak four
  fermion processes in electron - positron collisions}},  {\em Nucl. Phys. B}
  {\bfseries 424} (1994) 308--342,
  [\href{http://arxiv.org/abs/hep-ph/9404313}{{\ttfamily hep-ph/9404313}}].

\bibitem{Denner:1999gp}
A.~Denner, S.~Dittmaier, M.~Roth, and D.~Wackeroth, {\itshape {Predictions for
  all processes $e^+ e^-\to 4\,\mathrm{fermions}+\gamma$}},  {\em Nucl. Phys.}
  {\bfseries B560} (1999) 33--65,
  [\href{http://arxiv.org/abs/hep-ph/9904472}{{\ttfamily hep-ph/9904472}}].

\bibitem{Roth:1999kk}
M.~Roth, {\itshape {Precise predictions for four fermion production in electron
  positron annihilation}},  {PhD thesis, ETH Z\"{u}rich}, 1999.

\bibitem{Dittmaier:2002ap}
S.~Dittmaier and M.~Roth, {\itshape {LUSIFER: A LUcid approach to six FERmion
  production}},  {\em Nucl. Phys.} {\bfseries B642} (2002) 307--343,
  [\href{http://arxiv.org/abs/hep-ph/0206070}{{\ttfamily hep-ph/0206070}}].

\bibitem{Actis:2012qn}
S.~Actis, A.~Denner, L.~Hofer, A.~Scharf, and S.~Uccirati, {\itshape {Recursive
  generation of one-loop amplitudes in the Standard Model}},  {\em JHEP}
  {\bfseries 04} (2013) 037, [\href{http://arxiv.org/abs/1211.6316}{{\ttfamily
  arXiv:1211.6316}}].

\bibitem{Actis:2016mpe}
S.~Actis, et~al., {\itshape {RECOLA: REcursive Computation of One-Loop
  Amplitudes}},  {\em Comput. Phys. Commun.} {\bfseries 214} (2017) 140--173,
  [\href{http://arxiv.org/abs/1605.01090}{{\ttfamily arXiv:1605.01090}}].

\bibitem{Denner:2016kdg}
A.~Denner, S.~Dittmaier, and L.~Hofer, {\itshape {COLLIER: a fortran-based
  Complex One-Loop LIbrary in Extended Regularizations}},  {\em Comput. Phys.
  Commun.} {\bfseries 212} (2017) 220--238,
  [\href{http://arxiv.org/abs/1604.06792}{{\ttfamily arXiv:1604.06792}}].

\bibitem{Catani:1996vz}
S.~Catani and M.~Seymour, {\itshape {A general algorithm for calculating jet
  cross-sections in NLO QCD}},  {\em Nucl. Phys. B} {\bfseries 485} (1997)
  291--419, [\href{http://arxiv.org/abs/hep-ph/9605323}{{\ttfamily
  hep-ph/9605323}}]. [Erratum: \textit{Nucl. Phys. B} \textbf{510} (1998)
  503--504].

\bibitem{Nagy:1998bb}
Z.~Nagy and Z.~Tr{\'o}cs{\'a}nyi, {\itshape {Next-to-leading order calculation
  of four jet observables in electron positron annihilation}},  {\em Phys. Rev.
  D} {\bfseries 59} (1999) 014020,
  [\href{http://arxiv.org/abs/hep-ph/9806317}{{\ttfamily hep-ph/9806317}}].
  [Erratum: Phys.Rev.D 62, 099902 (2000)].

\bibitem{Dittmaier:1999mb}
S.~Dittmaier, {\itshape {A general approach to photon radiation off fermions}},
   {\em Nucl. Phys. B} {\bfseries 565} (2000) 69--122,
  [\href{http://arxiv.org/abs/hep-ph/9904440}{{\ttfamily hep-ph/9904440}}].

\bibitem{Catani:2002hc}
S.~Catani, S.~Dittmaier, M.~H. Seymour, and Z.~Tr{\'o}cs{\'a}nyi, {\itshape
  {The dipole formalism for next-to-leading order QCD calculations with massive
  partons}},  {\em Nucl. Phys. B} {\bfseries 627} (2002) 189--265,
  [\href{http://arxiv.org/abs/hep-ph/0201036}{{\ttfamily hep-ph/0201036}}].

\bibitem{Dittmaier:2008md}
S.~Dittmaier, A.~Kabelschacht, and T.~Kasprzik, {\itshape {Polarized QED
  splittings of massive fermions and dipole subtraction for non-collinear-safe
  observables}},  {\em Nucl. Phys.} {\bfseries B800} (2008) 146--189,
  [\href{http://arxiv.org/abs/0802.1405}{{\ttfamily arXiv:0802.1405}}].

\bibitem{Basso:2015gca}
L.~Basso, S.~Dittmaier, A.~Huss, and L.~Oggero, {\itshape {Techniques for the
  treatment of IR divergences in decay processes at NLO and application to the
  top-quark decay}},  {\em Eur. Phys. J.} {\bfseries C76} (2016) 56,
  [\href{http://arxiv.org/abs/1507.04676}{{\ttfamily arXiv:1507.04676}}].

\bibitem{Frixione:1998jh}
S.~Frixione, {\itshape {Isolated photons in perturbative QCD}},  {\em Phys.
  Lett. B} {\bfseries 429} (1998) 369--374,
  [\href{http://arxiv.org/abs/hep-ph/9801442}{{\ttfamily hep-ph/9801442}}].

\bibitem{Glover:1993xc}
E.~W.~N. Glover and A.~G. Morgan, {\itshape {Measuring the photon fragmentation
  function at LEP}},  {\em Z. Phys. C} {\bfseries 62} (1994) 311--322.

\bibitem{Denner:2010ia}
A.~Denner, S.~Dittmaier, T.~Gehrmann, and C.~Kurz, {\itshape {Electroweak
  corrections to hadronic event shapes and jet production in ${e}^+{e}^-$
  annihilation}},  {\em Nucl. Phys.} {\bfseries B836} (2010) 37--90,
  [\href{http://arxiv.org/abs/1003.0986}{{\ttfamily arXiv:1003.0986}}].

\bibitem{Denner:2014bna}
A.~Denner, S.~Dittmaier, M.~Hecht, and C.~Pasold, {\itshape {NLO QCD and
  electroweak corrections to $W+\gamma$ production with leptonic W-boson
  decays}},  {\em JHEP} {\bfseries 04} (2015) 018,
  [\href{http://arxiv.org/abs/1412.7421}{{\ttfamily arXiv:1412.7421}}].

\bibitem{Denner:2019zfp}
A.~Denner, S.~Dittmaier, M.~Pellen, and C.~Schwan, {\itshape {Low-virtuality
  photon transitions $\gamma^*\to f\bar f$ and the photon-to-jet conversion
  function}},  {\em Phys. Lett. B} {\bfseries 798} (2019) 134951,
  [\href{http://arxiv.org/abs/1907.02366}{{\ttfamily arXiv:1907.02366}}].

\bibitem{Denner:2005fg}
A.~Denner, S.~Dittmaier, M.~Roth, and L.~H. Wieders, {\itshape {Electroweak
  corrections to charged-current $e^+ e^-\to 4\,$ fermion processes: Technical
  details and further results}},  {\em Nucl. Phys. B} {\bfseries 724} (2005)
  247--294, [\href{http://arxiv.org/abs/hep-ph/0505042}{{\ttfamily
  hep-ph/0505042}}]. [Erratum: \textit{Nucl.Phys. B} \textbf{854}, 504--507
  (2012)].

\bibitem{Denner:2006ic}
A.~Denner and S.~Dittmaier, {\itshape {The complex-mass scheme for perturbative
  calculations with unstable particles}},  {\em Nucl. Phys. B Proc. Suppl.}
  {\bfseries 160} (2006) 22--26,
  [\href{http://arxiv.org/abs/hep-ph/0605312}{{\ttfamily hep-ph/0605312}}].

\bibitem{Denner:2019vbn}
A.~Denner and S.~Dittmaier, {\itshape {Electroweak Radiative Corrections for
  Collider Physics}},  {\em Phys. Rept.} {\bfseries 864} (2020) 1--163,
  [\href{http://arxiv.org/abs/1912.06823}{{\ttfamily arXiv:1912.06823}}].

\bibitem{Stuart:1991xk}
R.~G. Stuart, {\itshape {Gauge invariance, analyticity and physical observables
  at the $Z^0$ resonance}},  {\em Phys. Lett. B} {\bfseries 262} (1991)
  113--119.

\bibitem{Aeppli:1993rs}
A.~Aeppli, G.~J. van Oldenborgh, and D.~Wyler, {\itshape {Unstable particles in
  one loop calculations}},  {\em Nucl. Phys. B} {\bfseries 428} (1994)
  126--146, [\href{http://arxiv.org/abs/hep-ph/9312212}{{\ttfamily
  hep-ph/9312212}}].

\bibitem{Denner:2000bj}
A.~Denner, S.~Dittmaier, M.~Roth, and D.~Wackeroth, {\itshape {Electroweak
  radiative corrections to $e^+ e^-\to W W\to 4\,$fermions in double pole
  approximation: The RACOONWW approach}},  {\em Nucl. Phys. B} {\bfseries 587}
  (2000) 67--117, [\href{http://arxiv.org/abs/hep-ph/0006307}{{\ttfamily
  hep-ph/0006307}}].

\bibitem{Denner:1997ia}
A.~Denner, S.~Dittmaier, and M.~Roth, {\itshape {Non-factorizable photonic
  corrections to $e^+ e^- \to WW \to$ four fermions}},  {\em Nucl. Phys. B}
  {\bfseries 519} (1998) 39--84,
  [\href{http://arxiv.org/abs/hep-ph/9710521}{{\ttfamily hep-ph/9710521}}].

\bibitem{Accomando:2004de}
E.~Accomando, A.~Denner, and A.~Kaiser, {\itshape {Logarithmic electroweak
  corrections to gauge-boson pair production at the LHC}},  {\em Nucl. Phys. B}
  {\bfseries 706} (2005) 325--371,
  [\href{http://arxiv.org/abs/hep-ph/0409247}{{\ttfamily hep-ph/0409247}}].

\bibitem{Dittmaier:2015bfe}
S.~Dittmaier and C.~Schwan, {\itshape {Non-factorizable photonic corrections to
  resonant production and decay of many unstable particles}},  {\em Eur. Phys.
  J. C} {\bfseries 76} (2016) 144,
  [\href{http://arxiv.org/abs/1511.01698}{{\ttfamily arXiv:1511.01698}}].

\bibitem{Denner:2015yca}
A.~Denner and R.~Feger, {\itshape {NLO QCD corrections to off-shell top-antitop
  production with leptonic decays in association with a Higgs boson at the
  LHC}},  {\em JHEP} {\bfseries 11} (2015) 209,
  [\href{http://arxiv.org/abs/1506.07448}{{\ttfamily arXiv:1506.07448}}].

\bibitem{Denner:2016jyo}
A.~Denner and M.~Pellen, {\itshape {NLO electroweak corrections to off-shell
  top-antitop production with leptonic decays at the LHC}},  {\em JHEP}
  {\bfseries 08} (2016) 155, [\href{http://arxiv.org/abs/1607.05571}{{\ttfamily
  arXiv:1607.05571}}].

\bibitem{Denner:2016wet}
A.~Denner, J.-N. Lang, M.~Pellen, and S.~Uccirati, {\itshape {Higgs production
  in association with off-shell top-antitop pairs at NLO EW and QCD at the
  LHC}},  {\em JHEP} {\bfseries 02} (2017) 053,
  [\href{http://arxiv.org/abs/1612.07138}{{\ttfamily arXiv:1612.07138}}].

\bibitem{Denner:2017kzu}
A.~Denner and M.~Pellen, {\itshape {Off-shell production of top-antitop pairs
  in the lepton+jets channel at NLO QCD}},  {\em JHEP} {\bfseries 02} (2018)
  013, [\href{http://arxiv.org/abs/1711.10359}{{\ttfamily arXiv:1711.10359}}].

\bibitem{Denner:2020hgg}
A.~Denner and G.~Pelliccioli, {\itshape {NLO QCD corrections to off-shell
  $\text{t}{}\bar{\text{t}}\text{W}{}^+$ production at the LHC}},  {\em JHEP}
  {\bfseries 11} (2020) 069, [\href{http://arxiv.org/abs/2007.12089}{{\ttfamily
  arXiv:2007.12089}}].

\bibitem{Denner:2020orv}
A.~Denner, J.-N. Lang, and M.~Pellen, {\itshape {Full NLO QCD corrections to
  off-shell ttbb production}},  {\em Phys. Rev. D} {\bfseries 104} (2021)
  056018, [\href{http://arxiv.org/abs/2008.00918}{{\ttfamily
  arXiv:2008.00918}}].

\bibitem{Denner:2021hqi}
A.~Denner and G.~Pelliccioli, {\itshape {Combined NLO EW and QCD corrections to
  off-shell $\text {t} \overline{\text {t}}\text {W} $ production at the LHC}},
   {\em Eur. Phys. J. C} {\bfseries 81} (2021) 354,
  [\href{http://arxiv.org/abs/2102.03246}{{\ttfamily arXiv:2102.03246}}].

\bibitem{Denner:2023eti}
A.~Denner, D.~Lombardi, and G.~Pelliccioli, {\itshape {Complete NLO corrections
  to off-shell $ \textrm{t}\overline{\textrm{t}}\textrm{Z} $ production at the
  LHC}},  {\em JHEP} {\bfseries 09} (2023) 072,
  [\href{http://arxiv.org/abs/2306.13535}{{\ttfamily arXiv:2306.13535}}].

\bibitem{Denner:2023grl}
A.~Denner, M.~Pellen, and G.~Pelliccioli, {\itshape {NLO QCD corrections to
  off-shell top{\textendash}antitop~production with semi-leptonic decays at
  lepton colliders}},  {\em Eur. Phys. J. C} {\bfseries 83} (2023) 353,
  [\href{http://arxiv.org/abs/2302.04188}{{\ttfamily arXiv:2302.04188}}].

\bibitem{Czakon:2022khx}
M.~Czakon, A.~Mitov, M.~Pellen, and R.~Poncelet, {\itshape {A detailed
  investigation of W+c-jet at the LHC}},  {\em JHEP} {\bfseries 02} (2023) 241,
  [\href{http://arxiv.org/abs/2212.00467}{{\ttfamily arXiv:2212.00467}}].

\bibitem{Biedermann:2017yoi}
B.~Biedermann, et~al., {\itshape {Automation of NLO QCD and EW corrections with
  Sherpa and Recola}},  {\em Eur. Phys. J.} {\bfseries C77} (2017) 492,
  [\href{http://arxiv.org/abs/1704.05783}{{\ttfamily arXiv:1704.05783}}].

\bibitem{Brauer:2020kfv}
S.~Br{\"a}uer, A.~Denner, M.~Pellen, M.~Sch{\"o}nherr, and S.~Schumann,
  {\itshape {Fixed-order and merged parton-shower predictions for WW and WWj
  production at the LHC including NLO QCD and EW corrections}},  {\em JHEP}
  {\bfseries 10} (2020) 159, [\href{http://arxiv.org/abs/2005.12128}{{\ttfamily
  arXiv:2005.12128}}].

\bibitem{Dreyer:2020xaj}
F.~A. Dreyer, A.~Karlberg, J.-N. Lang, and M.~Pellen, {\itshape {Precise
  predictions for double-Higgs production via vector-boson fusion}},  {\em Eur.
  Phys. J. C} {\bfseries 80} (2020) 1037,
  [\href{http://arxiv.org/abs/2005.13341}{{\ttfamily arXiv:2005.13341}}].

\bibitem{Barone:2025jey}
G.~Barone et~al., {\itshape {Higgs production via vector-boson fusion at the
  LHC}},  \href{http://arxiv.org/abs/2507.22574}{{\ttfamily arXiv:2507.22574}}.

\bibitem{Denner:2024ufg}
A.~Denner, M.~Pellen, M.~Sch{\"o}nherr, and S.~Schumann, {\itshape {Tri-boson
  and WH production in the W$^{+}$W$^{+}$jj channel: predictions at full NLO
  accuracy and beyond}},  {\em JHEP} {\bfseries 08} (2024) 043,
  [\href{http://arxiv.org/abs/2406.11516}{{\ttfamily arXiv:2406.11516}}].

\bibitem{Denner:2024ndl}
A.~Denner, D.~Lombardi, S.~L.~P. Chavez, and G.~Pelliccioli, {\itshape {NLO
  corrections to triple vector-boson production in final states with three
  charged leptons and two jets}},  {\em JHEP} {\bfseries 09} (2024) 187,
  [\href{http://arxiv.org/abs/2407.21558}{{\ttfamily arXiv:2407.21558}}].

\bibitem{Biedermann:2016yds}
B.~Biedermann, A.~Denner, and M.~Pellen, {\itshape {Large electroweak
  corrections to vector-boson scattering at the Large Hadron Collider}},  {\em
  Phys. Rev. Lett.} {\bfseries 118} (2017) 261801,
  [\href{http://arxiv.org/abs/1611.02951}{{\ttfamily arXiv:1611.02951}}].

\bibitem{Biedermann:2017bss}
B.~Biedermann, A.~Denner, and M.~Pellen, {\itshape {Complete NLO corrections to
  W$^{+}$W$^{+}$ scattering and its irreducible background at the LHC}},  {\em
  JHEP} {\bfseries 10} (2017) 124,
  [\href{http://arxiv.org/abs/1708.00268}{{\ttfamily arXiv:1708.00268}}].

\bibitem{Denner:2019tmn}
A.~Denner, S.~Dittmaier, P.~Maierh{\"o}fer, M.~Pellen, and C.~Schwan, {\itshape
  {QCD and electroweak corrections to WZ scattering at the LHC}},  {\em JHEP}
  {\bfseries 06} (2019) 067, [\href{http://arxiv.org/abs/1904.00882}{{\ttfamily
  arXiv:1904.00882}}].

\bibitem{Pellen:2019ywl}
M.~Pellen, {\itshape {Exploring the scattering of vector bosons at LHCb}},
  {\em Phys. Rev. D} {\bfseries 101} (2020) 013002,
  [\href{http://arxiv.org/abs/1908.06805}{{\ttfamily arXiv:1908.06805}}].

\bibitem{Chiesa:2019ulk}
M.~Chiesa, A.~Denner, J.-N. Lang, and M.~Pellen, {\itshape {An event generator
  for same-sign W-boson scattering at the LHC including electroweak
  corrections}},  {\em Eur. Phys. J. C} {\bfseries 79} (2019) 788,
  [\href{http://arxiv.org/abs/1906.01863}{{\ttfamily arXiv:1906.01863}}].

\bibitem{Denner:2020zit}
A.~Denner, R.~Franken, M.~Pellen, and T.~Schmidt, {\itshape {NLO QCD and EW
  corrections to vector-boson scattering into ZZ at the LHC}},  {\em JHEP}
  {\bfseries 11} (2020) 110, [\href{http://arxiv.org/abs/2009.00411}{{\ttfamily
  arXiv:2009.00411}}].

\bibitem{Denner:2021hsa}
A.~Denner, R.~Franken, M.~Pellen, and T.~Schmidt, {\itshape {Full NLO
  predictions for vector-boson scattering into Z bosons and its irreducible
  background at the LHC}},  {\em JHEP} {\bfseries 10} (2021) 228,
  [\href{http://arxiv.org/abs/2107.10688}{{\ttfamily arXiv:2107.10688}}].

\bibitem{Denner:2022pwc}
A.~Denner, R.~Franken, T.~Schmidt, and C.~Schwan, {\itshape {NLO QCD and EW
  corrections to vector-boson scattering into W$^+$W$^-$ at the LHC}},  {\em
  JHEP} {\bfseries 06} (2022) 098,
  [\href{http://arxiv.org/abs/2202.10844}{{\ttfamily arXiv:2202.10844}}].

\bibitem{Denner:2024xul}
A.~Denner, D.~Lombardi, and C.~Schwan, {\itshape {Double-pole approximation for
  leading-order semi-leptonic vector-boson scattering at the LHC}},  {\em JHEP}
  {\bfseries 08} (2024) 146, [\href{http://arxiv.org/abs/2406.12301}{{\ttfamily
  arXiv:2406.12301}}].

\bibitem{Denner:2022fhu}
A.~Denner, G.~Pelliccioli, and C.~Schwan, {\itshape {NLO QCD and EW corrections
  to off-shell tZj production at the LHC}},  {\em JHEP} {\bfseries 10} (2022)
  125, [\href{http://arxiv.org/abs/2207.11264}{{\ttfamily arXiv:2207.11264}}].

\bibitem{Denner:2020bcz}
A.~Denner and G.~Pelliccioli, {\itshape {Polarized electroweak bosons in
  W$^+$W$^-$ production at the LHC including NLO QCD effects}},  {\em JHEP}
  {\bfseries 09} (2020) 164, [\href{http://arxiv.org/abs/2006.14867}{{\ttfamily
  arXiv:2006.14867}}].

\bibitem{Denner:2020eck}
A.~Denner and G.~Pelliccioli, {\itshape {NLO QCD predictions for
  doubly-polarized WZ production at the LHC}},  {\em Phys. Lett. B} {\bfseries
  814} (2021) 136107, [\href{http://arxiv.org/abs/2010.07149}{{\ttfamily
  arXiv:2010.07149}}].

\bibitem{Denner:2021csi}
A.~Denner and G.~Pelliccioli, {\itshape {NLO EW and QCD corrections to
  polarized ZZ production in the four-charged-lepton channel at the LHC}},
  {\em JHEP} {\bfseries 10} (2021) 097,
  [\href{http://arxiv.org/abs/2107.06579}{{\ttfamily arXiv:2107.06579}}].

\bibitem{Denner:2022riz}
A.~Denner, C.~Haitz, and G.~Pelliccioli, {\itshape {NLO QCD corrections to
  polarized diboson production in semileptonic final states}},  {\em Phys. Rev.
  D} {\bfseries 107} (2023) 053004,
  [\href{http://arxiv.org/abs/2211.09040}{{\ttfamily arXiv:2211.09040}}].

\bibitem{Denner:2023ehn}
A.~Denner, C.~Haitz, and G.~Pelliccioli, {\itshape {NLO EW corrections to
  polarised W$^+$W$^-$ production and decay at the LHC}},  {\em Phys. Lett. B}
  {\bfseries 850} (2024) 138539,
  [\href{http://arxiv.org/abs/2311.16031}{{\ttfamily arXiv:2311.16031}}].

\bibitem{Grossi:2024jae}
M.~Grossi, G.~Pelliccioli, and A.~Vicini, {\itshape {From angular coefficients
  to quantum observables: a phenomenological appraisal in di-boson systems}},
  {\em JHEP} {\bfseries 12} (2024) 120,
  [\href{http://arxiv.org/abs/2409.16731}{{\ttfamily arXiv:2409.16731}}].

\bibitem{Denner:2024tlu}
A.~Denner, C.~Haitz, and G.~Pelliccioli, {\itshape {NLO EW and QCD corrections
  to polarised same-sign WW scattering at the LHC}},  {\em JHEP} {\bfseries 11}
  (2024) 115, [\href{http://arxiv.org/abs/2409.03620}{{\ttfamily
  arXiv:2409.03620}}].

\bibitem{Carrivale:2025mjy}
C.~Carrivale et~al., {\itshape {Precise Standard-Model predictions for
  polarised Z-boson pair production and decay at the LHC}},  {\em Eur. Phys. J.
  C} {\bfseries 85} (2025) 1342,
  [\href{http://arxiv.org/abs/2505.09686}{{\ttfamily arXiv:2505.09686}}].

\bibitem{DelGratta:2025xjp}
M.~Del~Gratta, et~al., {\itshape {Z-boson quantum tomography at next-to-leading
  order}},  {\em JHEP} {\bfseries 02} (2026) 056,
  [\href{http://arxiv.org/abs/2509.20456}{{\ttfamily arXiv:2509.20456}}].

\bibitem{Pelliccioli:2025com}
G.~Pelliccioli and R.~Poncelet, {\itshape {Precise predictions for joint
  polarisation fractions in WZ production at the LHC}},
  \href{http://arxiv.org/abs/2510.25898}{{\ttfamily arXiv:2510.25898}}.

\bibitem{Denner:2025xdz}
A.~Denner, R.~Franken, C.~Haitz, D.~Lombardi, and G.~Pelliccioli, {\itshape
  {Electroweak corrections to doubly polarised WZ scattering at the LHC}},
  \href{http://arxiv.org/abs/2510.26462}{{\ttfamily arXiv:2510.26462}}.

\bibitem{Buckley:2014ana}
A.~Buckley, et~al., {\itshape {LHAPDF6: parton density access in the LHC
  precision era}},  {\em Eur. Phys. J. C} {\bfseries 75} (2015) 132,
  [\href{http://arxiv.org/abs/1412.7420}{{\ttfamily arXiv:1412.7420}}].

\bibitem{Shao:2022cly}
H.-S. Shao and D.~d'Enterria, {\itshape {gamma-UPC: automated generation of
  exclusive photon-photon processes in ultraperipheral proton and nuclear
  collisions with varying form factors}},  {\em JHEP} {\bfseries 09} (2022)
  248, [\href{http://arxiv.org/abs/2207.03012}{{\ttfamily arXiv:2207.03012}}].

\bibitem{Vermaseren:1994je}
J.~A.~M. Vermaseren, {\itshape {Axodraw}},  {\em Comput. Phys. Commun.}
  {\bfseries 83} (1994) 45--58.

\bibitem{Binosi:2003yf}
D.~Binosi and L.~Theu{\ss}l, {\itshape {JaxoDraw: A Graphical user interface
  for drawing Feynman diagrams}},  {\em Comput. Phys. Commun.} {\bfseries 161}
  (2004) 76--86, [\href{http://arxiv.org/abs/hep-ph/0309015}{{\ttfamily
  hep-ph/0309015}}].

\bibitem{Melnikov:1993np}
K.~Melnikov and O.~I. Yakovlev, {\itshape {Top near threshold: all $\alpha_S$
  corrections are trivial}},  {\em Phys. Lett.} {\bfseries B324} (1994)
  217--223, [\href{http://arxiv.org/abs/hep-ph/9302311}{{\ttfamily
  hep-ph/9302311}}].

\bibitem{Fadin:1993dz}
V.~S. Fadin, V.~A. Khoze, and A.~D. Martin, {\itshape {Interference radiative
  phenomena in the production of heavy unstable particles}},  {\em Phys. Rev.}
  {\bfseries D49} (1994) 2247--2256.

\bibitem{Fadin:1993kt}
V.~S. Fadin, V.~A. Khoze, and A.~D. Martin, {\itshape {How suppressed are the
  radiative interference effects in heavy instable particle production?}},
  {\em Phys. Lett.} {\bfseries B320} (1994) 141--144,
  [\href{http://arxiv.org/abs/hep-ph/9309234}{{\ttfamily hep-ph/9309234}}].

\bibitem{Denner:1998rh}
A.~Denner, S.~Dittmaier, and M.~Roth, {\itshape {Further numerical results on
  nonfactorizable corrections to $e^+ e^-\to W^+ W^-\to\,$four fermions}},
  {\em Phys. Lett. B} {\bfseries 429} (1998) 145--150,
  [\href{http://arxiv.org/abs/hep-ph/9803306}{{\ttfamily hep-ph/9803306}}].

\bibitem{Bardin:1988xt}
D.~{\relax Yu}. Bardin, A.~Leike, T.~Riemann, and M.~Sachwitz, {\itshape
  {Energy-dependent width effects in $e^+ e^-$ annihilation near the Z-boson
  pole}},  {\em Phys. Lett.} {\bfseries B206} (1988) 539--542.

\bibitem{Bertone:2022ktl}
V.~Bertone, et~al., {\itshape {Improving methods and predictions at high-energy
  $e^{+}e^{-}$ colliders within collinear factorisation}},  {\em JHEP}
  {\bfseries 10} (2022) 089, [\href{http://arxiv.org/abs/2207.03265}{{\ttfamily
  arXiv:2207.03265}}].

\bibitem{Beenakker:1996kt}
W.~Beenakker et~al., {\itshape {$W W$ cross-sections and distributions}},  in
  {\em {Physics at LEP2}} ({G. Altarelli, T.~Sj\"ostrand, F.~Zwirner}, ed.),
  vol.~1, (Geneva), pp.~79--139, CERN, 1996.
\newblock \href{http://arxiv.org/abs/hep-ph/9602351}{{\ttfamily
  hep-ph/9602351}}.
\newblock {CERN-96-01}.

\bibitem{Nagy:2003tz}
Z.~Nagy, {\itshape {Next-to-leading order calculation of three jet observables
  in hadron hadron collision}},  {\em Phys. Rev. D} {\bfseries 68} (2003)
  094002, [\href{http://arxiv.org/abs/hep-ph/0307268}{{\ttfamily
  hep-ph/0307268}}].

\bibitem{James:1993np}
F.~James, {\itshape {RANLUX: A FORTRAN implementation of the high quality
  pseudorandom number generator of Luscher}},  {\em Comput. Phys. Commun.}
  {\bfseries 79} (1994) 111--114. [Erratum: \textit{Comput.Phys.Commun. 97}
  \textbf{357} (1996)].

\bibitem{Cahn:1990jk}
R.~N. Cahn and J.~D. Jackson, {\itshape {Realistic equivalent photon yields in
  heavy ion collisions}},  {\em Phys. Rev. D} {\bfseries 42} (1990) 3690--3695.

\bibitem{Vidovic:1992ik}
M.~Vidovic, M.~Greiner, C.~Best, and G.~Soff, {\itshape {Impact parameter
  dependence of the electromagnetic particle production in ultrarelativistic
  heavy ion collisions}},  {\em Phys. Rev. C} {\bfseries 47} (1993) 2308--2319.

\bibitem{Kleiss:1994qy}
R.~Kleiss and R.~Pittau, {\itshape {Weight optimization in multichannel Monte
  Carlo}},  {\em Comput. Phys. Commun.} {\bfseries 83} (1994) 141--146,
  [\href{http://arxiv.org/abs/hep-ph/9405257}{{\ttfamily hep-ph/9405257}}].

\bibitem{Cacciari:2008gp}
M.~Cacciari, G.~P. Salam, and G.~Soyez, {\itshape {The anti-$k_t$ jet
  clustering algorithm}},  {\em JHEP} {\bfseries 04} (2008) 063,
  [\href{http://arxiv.org/abs/0802.1189}{{\ttfamily arXiv:0802.1189}}].

\bibitem{Dokshitzer:1997in}
Y.~L. Dokshitzer, G.~D. Leder, S.~Moretti, and B.~R. Webber, {\itshape {Better
  jet clustering algorithms}},  {\em JHEP} {\bfseries 08} (1997) 001,
  [\href{http://arxiv.org/abs/hep-ph/9707323}{{\ttfamily hep-ph/9707323}}].

\bibitem{Wobisch:1998wt}
M.~Wobisch and T.~Wengler, {\itshape {Hadronization corrections to jet
  cross-sections in deep inelastic scattering}},  in {\em {Workshop on Monte
  Carlo Generators for HERA Physics (Plenary Starting Meeting)}}, pp.~270--279,
  4, 1998.
\newblock \href{http://arxiv.org/abs/hep-ph/9907280}{{\ttfamily
  hep-ph/9907280}}.

\bibitem{Catani:1993hr}
S.~Catani, Y.~L. Dokshitzer, M.~H. Seymour, and B.~R. Webber, {\itshape
  {Longitudinally invariant $k_\perp$-clustering algorithms for hadron hadron
  collisions}},  {\em Nucl. Phys. B} {\bfseries 406} (1993) 187--224.

\bibitem{Ellis:1993tq}
S.~D. Ellis and D.~E. Soper, {\itshape {Successive combination jet algorithm
  for hadron collisions}},  {\em Phys. Rev. D} {\bfseries 48} (1993)
  3160--3166, [\href{http://arxiv.org/abs/hep-ph/9305266}{{\ttfamily
  hep-ph/9305266}}].

\bibitem{Glover:1993he}
E.~W.~N. Glover and A.~G. Morgan, {\itshape {Soft gluon radiation in photon
  plus single jet events at LEP}},  {\em Phys. Lett. B} {\bfseries 324} (1994)
  487--491.

\bibitem{Buskulic:1995au}
{\bfseries ALEPH} Collaboration, D.~Buskulic et~al., {\itshape {First
  measurement of the quark to photon fragmentation function}},  {\em Z. Phys.}
  {\bfseries C69} (1996) 365--378.

\bibitem{Denner:2009gj}
A.~Denner, S.~Dittmaier, T.~Kasprzik, and A.~M{\"u}ck, {\itshape {Electroweak
  corrections to W+jet hadroproduction including leptonic W-boson decays}},
  {\em JHEP} {\bfseries 08} (2009) 075,
  [\href{http://arxiv.org/abs/0906.1656}{{\ttfamily arXiv:0906.1656}}].

\bibitem{Denner:2011vu}
A.~Denner, S.~Dittmaier, T.~Kasprzik, and A.~M{\"u}ck, {\itshape {Electroweak
  corrections to dilepton + jet production at hadron colliders}},  {\em JHEP}
  {\bfseries 06} (2011) 069, [\href{http://arxiv.org/abs/1103.0914}{{\ttfamily
  arXiv:1103.0914}}].

\bibitem{Denner:2014ina}
A.~Denner, L.~Hofer, A.~Scharf, and S.~Uccirati, {\itshape {Electroweak
  corrections to lepton pair production in association with two hard jets at
  the LHC}},  {\em JHEP} {\bfseries 01} (2015) 094,
  [\href{http://arxiv.org/abs/1411.0916}{{\ttfamily arXiv:1411.0916}}].

\bibitem{Denner:2015fca}
A.~Denner, S.~Dittmaier, M.~Hecht, and C.~Pasold, {\itshape {NLO QCD and
  electroweak corrections to $Z+\gamma$ production with leptonic Z-boson
  decays}},  {\em JHEP} {\bfseries 02} (2016) 057,
  [\href{http://arxiv.org/abs/1510.08742}{{\ttfamily arXiv:1510.08742}}].

\bibitem{Keshavarzi:2018mgv}
A.~Keshavarzi, D.~Nomura, and T.~Teubner, {\itshape {Muon $g-2$ and
  $\alpha(M_Z^2)$: a new data-based analysis}},  {\em Phys. Rev. D} {\bfseries
  97} (2018) 114025, [\href{http://arxiv.org/abs/1802.02995}{{\ttfamily
  arXiv:1802.02995}}].

\bibitem{Denner:2012yc}
A.~Denner, S.~Dittmaier, S.~Kallweit, and S.~Pozzorini, {\itshape {NLO QCD
  corrections to off-shell top-antitop production with leptonic decays at
  hadron colliders}},  {\em JHEP} {\bfseries 10} (2012) 110,
  [\href{http://arxiv.org/abs/1207.5018}{{\ttfamily arXiv:1207.5018}}].

\bibitem{Dittmaier:2025ikh}
S.~Dittmaier, T.~Engel, J.~L.~H. Ariza, and M.~Pellen, {\itshape {Electroweak
  corrections to $\tau^+\tau^-$ production in ultraperipheral heavy-ion
  collisions at the LHC}},  {\em JHEP} {\bfseries 08} (2025) 051,
  [\href{http://arxiv.org/abs/2504.11391}{{\ttfamily arXiv:2504.11391}}].

\end{thebibliography}\endgroup
